\documentclass[11pt]{article}
\pdfoutput=1
\usepackage[T1]{fontenc} 
\usepackage{bbm}
\usepackage{amsmath}
\usepackage{amssymb}
\usepackage{amsfonts}
\usepackage{mathrsfs}
\usepackage[pdftex]{graphicx}
\usepackage{axodraw}
\usepackage{color}
\usepackage{cancel}
\newcommand{\be}{\begin{equation}}
\newcommand{\ee}{\end{equation}}
\newcommand{\bea}{\begin{eqnarray}}
\newcommand{\eea}{\end{eqnarray}}

\begin{document}
%\begin{titlepage}
{\vspace*{-2cm}
%\phantom{hep-ph/yymmnnn}
%\hfill{LOCAL-PREPRINT-NUMBER}
\flushright{ULB-TH/11-26}
\vspace{-4mm}
%\flushright{NSF-KITP-08-79}
\vskip 1.5cm}
\begin{center}
{\Large \bf The Four Basic Ways of Creating Dark Matter Through a Portal\\
\vspace{3mm}
}
\vspace{1mm}
\end{center}
\vskip 0.5cm
\begin{center}

{\large Xiaoyong Chu$^a$, Thomas Hambye$^{a,b}$ and Michel H.G. Tytgat$^{a,}$}\footnote{xiaoyong.chu@ulb.ac.be,thambye@ulb.ac.be,mtytgat@ulb.ac.be}
\\
\vskip .7cm
$^a$Service de Physique Th\'eorique,\\
\vspace{0.7mm}
Universit\'e Libre de Bruxelles, 1050 Brussels, Belgium\\
\vspace{2.0mm}
$^b$Departamento de F\'isica Te\'orica, Universidad Aut\'onoma de Madrid and\\ 
Instituto de F\'isica Te\'orica IFT-UAM/CSIC, Cantoblanco, 28049 Madrid, Spain 
 \end{center}
\vskip 0.5cm

\begin{abstract}
We consider the possibility that along the thermal history of the Universe, dark matter (DM) would have been created from Standard Model particles, either through a kinetic mixing portal to an extra $U(1)$ gauge field, or through the Higgs portal. Depending solely on the DM particle mass, on the portal and on the DM hidden sector interaction, we show how the observed DM relic density can be obtained. There are four possible freeze-in/reannihilation/freeze-out regimes, which together result in a simple characteristic relic density phase diagram, with the shape of a ``Mesa". 
In the case of the kinetic mixing portal, we show that, unlike other freeze-in scenarios discussed in the literature, the freeze-in regime can
%can
 be probed by forthcoming DM direct detection experiments. 
These results are well representative {of} any scenario where a DM hidden sector would be created out of the Standard Model {sector}.

 \vspace{2mm}
\end{abstract}
\setcounter{footnote}{0}
\vskip2truecm

\section{Introduction}

There is an all series of gravitational evidences for the existence of dark matter (DM). In particular, from observations of the Cosmic Microwave Background  anisotropies and large-scale structures, it appears that a large fraction of the energy density of the Universe is made of DM, with $\Omega_{DM} =0.229 \pm 0.015$ \cite{Komatsu:2010fb}.
To explain such a proportion, the WIMP candidates are particularly interesting because, by producing  thermally the DM from the thermal bath of the Universe, through DM pair creation and annihilation, they lead to a relic density which depends only on the masses and interactions involved at the time of the decoupling of these processes. Moreover it turns out that for interactions of order the electroweak ones, or more generally for couplings of order unity, they lead to the observed relic density for a DM mass of the order of the electroweak scale. This offers the possibility of DM production at colliders, on top of the possibilities of direct and indirect detection.

Another simple mechanism to produce the DM relic density, which was pushed forward recently \cite{McDonald:2001vt,Hall:2009bx,Cheung:2010gj,Cheung:2010gk}, is the freeze-in mechanism. {Here the relic density, instead of resulting from the thermal freeze-out of the DM annihilation process, stems solely from the DM creation processes, for instance through annihilation into DM particles, $A A\rightarrow DM DM$, or through a decay process $A\rightarrow DM B$.} These processes, {that are} assumed to be out of thermal equilibrium, {freeze} when the temperature $T$ drops below the mass of the source particle $A$ or DM particle, {\em i.e.}~when the production rate gets Boltzmann suppressed.
{While being out-of-equilibrium}, this {mechanism} is {actually} also thermal in the sense that the source particle $A$ is assumed to be in thermal equilibrium, so that, here too, the number of particles produced depends only on the masses and couplings involved in the DM creation process.
A difference though with the usual freeze-out scenario is that the relic density may also receive a contribution from a "primordial" density, which is not washed-out by any process since the DM creation process is out of thermal equilibrium. However there exists a whole regime for which the DM density produced from the Standard Model (SM) sector can be the dominant one.\footnote{If both sectors are feebly connected, one could argue that the inflaton is likely to belong to only one of the sector, so that it couple dominantly to one sector and reheating occurs dominantly in this sector.} This is the possibility  we will consider in the present work. {Another important} difference with freeze-out is that the {characteristic} coupling required to produce the relic density through freeze-in is tiny, {typically} about 8-10 orders of magnitude below {that} required by freeze-out, implying that the DM basically lies in a hidden sector that is feebly coupled to the SM {one}. {Hence it is {\em a priori} very difficult to probe experimentally the freeze-in scenario, either at colliders or from direct/indirect detection.} There is nevertheless one way one could consider to test this mechanism in specific scenarios \cite{Hall:2009bx,Cheung:2010gj,Cheung:2010gk}{: if freeze-in proceeds through $A \rightarrow DM B$ decay (assuming typically that $A$ and DM particles are both odd under a $Z_2$ symmetry so that DM is stable) and if, in the visible sector, $A$ has substantial coupling to the SM particles, then $A$ particles may be abundantly produced at colliders. Their slow} decay to DM from the feeble interaction could then be probed from the search of long-lived particles at colliders. To be efficiently probed, this {scenario}  requires that $B$ itself is substantially coupled  to SM particles (being typically charged or colored), so that only DM  {particles are} feebly coupled to all other particles. 

In the following we show that there is another possibility of testing a DM relic density created through freeze-in, {\em i.e.}~direct detection. The scenario we consider is {quite} different. {In particular} we do not assume any new particles in the visible sector{, but the ones from the SM. This means that in our framework the A particle is a SM particle, which implies that  the predicted DM relic density depends neither on the mass of any new hypothetical particles in the visible sector, nor on the interactions these particles  may have with the rest of the visible sector.} {The DM} relic density depends only on the DM particle mass, on the connector interaction and on any interaction which may be {relevant} within the hidden sector.
One class of models we consider is based on the {possible} existence of {a light mediator particle}. We {will show} that such a light mediator interaction, as tiny as the one required for {DM production through} freeze-in, could be probed by direct detection, because the elastic cross section with nuclei, which exhibits a collinear divergence, is strongly enhanced.

The light mediator DM model we consider in the following as an example is very simple. {It involves, in addition to the SM (the visible sector), a hidden sector composed of  a single particle, which is assumed to be charged under a new $U(1)'$ gauge symmetry.}
{This hidden sector is coupled to the SM sector through kinetic mixing, ${\cal L} \owns -\frac{1}{2} \varepsilon F_Y^{\mu\nu} F'_{\mu\nu}$ (see Ref.~\cite{Pospelov:2007mp,Chun:2010ve} for a "secluded" DM structure of this type with a massive $\gamma'$, see also the related model in Ref.~\cite{Feldman:2006wd}). Thus DM  may interact with the SM particles through a massless mediator, or a light mediator if the $U(1)'$ gauge group is slightly broken (either through the Higgs or St\"uckelberg mechanisms).} {On top of the connector interaction, this model involves QED-like interactions between the $\gamma'$ and the charged DM particle in the hidden sector.}
{Another possible motivation for this model is that there are not that many simple mechanisms available to stabilize DM (see e.g.~\cite{Hambye:2010zb}). Among these, one of the simplest possibilities is to consider the lightest particle charged under an extra, unbroken gauge symmetry. Hence the model we invoke here may also be seen as the simplest and yet generic representative of a larger class of models. In particular we expect that the features we will discuss should hold for many  models of the same kind.} {A further motivation is that, although} the cosmological and astrophysical consequences of such a new long range {gauge interaction}  have been considered extensively in Refs.~\cite{Ackerman:2008gi,Feng:2009mn,Feng:2008mu}, {this has been done} basically without considering the {possibility of} kinetic mixing connecting both sectors.

Beyond the fact that the freeze-in mechanism {may} be {testable within} this {specific} scenario, the main, {more general} purpose of our work is to determine the various ways a viable DM relic density can be obtained in {the context of particle creation through a portal}.
To this end we {have explored} the full parameter space of this model.
If both the kinetic mixing  and the hidden sector interactions are feeble, the DM relic density can be calculated by simply counting the number of DM particle produced through the connector. This is precisely the freeze-in scenario. {However}, if instead {either one} or both {of the}
 interactions are {stronger, the DM relic density is affected by the thermalization of
the corresponding processes.} 
We show {that} this leads to a simple characteristic "Mesa" phase diagram for the relic density as a function {of the connector and hidden sector couplings}. {As we explain in details this phase diagram displays four distinct} phases: (i) freeze-in, (ii) simultaneous freezing of the (in-equilibrium) hidden sector interactions and the (out-of-equilibrium) connector, {\em i.e.}~reannihilation, (iii) freeze-out of the hidden sector interaction, and (iv) freeze-out of {connector} interaction. An interesting feature of this diagram is that{, in each phase, the relic density is essentially characterized by  only one interaction, either the connector  or that in the hidden sector}.

In order to address the generality of this classification, we consider the creation of DM through another simple portal between the SM and a hidden DM sectors, the so-called Higgs portal. We show that DM production from SM particles through the Higgs portal leads to a similar simple characteristic phase diagram which also has characteristic ``Mesa'' shape. This holds even though the mediator, {\em i.e.}~the Higgs boson, is massive and may even be heavier than the DM particle, unlike for the kinetic mixing where we consider the opposite limit of a mediator (much) lighter than the DM candidate. In this sense the "Mesa" diagram 
is a generic feature of models within which DM is created out of SM particles through a portal (or more generally from particles that are in thermal equilibrium with the SM sector when they create the DM particles).

\bigskip
{The plan of this article is as follows.
In Section 2 we first compute the transfer of energy from the visible sector to the hidden sector in the case of the kinetic mixing portal. This is necessary to compute the DM relic density in all regimes, except in the freeze-in one}. We then compute the DM number density produced. {In Section 3 we present and discuss in details the corresponding phase diagram obtained and give analytical approximations for the relic density. In Section 4 we discuss the possibility of direct detection signal through kinetic mixing, and the prospect for testing the freeze-in mechanism. In Section 5 we discuss the compatibility of the relic abundance and direct detection signatures obtained with the various existing cosmological and astrophysical constraints on DM interacting through long range interactions. 
In Section 6 we present our results for the Higgs portal, and discuss in details the similarities and differences between the cases of massive and massless mediators. Finally, we summarize our results in the last section.}

%%%%%%%%%%%%%%%%%%%%%%%%%%%%%%%%%%%%%%%%%%%%%%%
\section{DM number density and energy transfer from the visible to the  hidden sector through kinetic mixing}

It is remarkable that the addition of the simplest gauge structure one can think of, that is to say of a $U(1)'$ gauge structure with a single charged particle, leads to a viable DM candidate \cite{Ackerman:2008gi,Feng:2009mn,Feng:2008mu}.
In the following we will consider a charged fermion (the DM candidate), 
\begin{equation}
{\cal L} ={\cal L}_{SM} +  \bar{\psi^\prime} (i\!\not \hspace{-1.2mm}D^\prime -m_\psi) \psi^\prime \,,
\label{qedprime}
\end{equation}
with $D^\prime_\mu=\partial_\mu + i e' A^\prime_\mu$, but the results would be essentially the same for a scalar particle (except in this case for a possible additional Higgs portal interaction between 2 DM particles and 2 Higgs doublets, which will be considered in Section 6).
This QED$'$ model contains, on top of the SM particles, an extra massless gauge boson $\gamma'$ {(the hidden photon)}, and an extra fermion, that we call "$e'$" (singlet of the SM gauge group), and nothing else. 
This set up  can be coupled to the SM {in an unique way through} the kinetic mixing portal \cite{Holdom:1985ag,Foot:1991kb}
\begin{equation}
{\cal L} \owns  - \frac{\epsilon}{2} F_Y^{\mu\nu} F'_{\mu\nu} \,.
\label{Lkinetic}
\end{equation}
where $F_Y^{\mu\nu}$ is the hypercharge field strength. 
Consequently this model Lagrangian involves only three new parameters: $\alpha'=e'^2/4\pi$, $\epsilon$ and $m_{e'}=m_{DM}$.
The non-canonical kinetic term implies a mixing between the visible and hidden sector photons, $\gamma$ and $\gamma'$, as well as between the $Z$ boson and $\gamma'$.
For all purposes it is convenient to work in a basis where the kinetic terms are canonical which can be obtained from a non-unitary transformation. {We use}
\begin{equation}
\begin{pmatrix}
\gamma'_\mu\\
\gamma_\mu\\
Z_\mu
\end{pmatrix}=
\begin{pmatrix}
1   &0  & 0\\
0          &\cos\theta_\epsilon &\sin\theta_\epsilon\\
0 &-\sin\theta_\epsilon  &\cos\theta_\epsilon
\end{pmatrix}
\begin{pmatrix}
1   &\epsilon    &0\\
      0      &\sqrt{1-\epsilon^2}  &0\\
0                     &0  &1
\end{pmatrix}
%\begin{pmatrix}
%1    &0    &0\\
%0    &\cos\theta_W  &-\sin\theta_W\\
%0    &\sin\theta_W &\cos\theta_W
%\end{pmatrix}
\begin{pmatrix}
A'_\mu\\
B_\mu\\
W^3_\mu
\end{pmatrix},
\label{KMtransfo}
\end{equation}
where $\theta_W$ is the Weinberg angle and the $\theta_\epsilon$ mixing angle satisfies $\tan\theta_\epsilon=\tan\theta_W/\sqrt{1-\epsilon^2}$. 

{We then have the following couplings to the SM and DM currents}
\begin{equation}
{\cal L}\owns -e_{EM} J_{EM}^\mu\, \gamma_\mu + e^\prime \frac{\varepsilon\cos \theta_\varepsilon }{\sqrt{1-\varepsilon^2}}J^\mu_{DM}\, \gamma_\mu - e' J^\mu_{DM} \gamma'_\mu  - e^\prime {\varepsilon\sin\theta_\varepsilon\over \sqrt{1 - \varepsilon^2}} J_{DM}^\mu Z_\mu - g {\cos\theta_W\over \cos\theta_\varepsilon} J_Z^\mu Z_\mu\,,
\end{equation}
where we have defined 
$$e_{EM} = {e \cos\theta_\varepsilon\over \cos\theta_W\sqrt{1-\varepsilon^2}} = {e \over \sqrt{1- \varepsilon^2 \cos^2\theta_W}}$$
 and with $J^\mu_{SM}$ and $J^\mu_{DM}$ the corresponding $U(1)_{em}$ and $U(1)'$ currents. 

{As the visible and hidden photons are degenerate, there is some arbitrariness in the definition of the massless fields. Concretely, we have adopted a basis in which the hidden photon does not couple to the SM particles.}\footnote{
The basis of Eq.(\ref{KMtransfo}) is convenient for our purpose, as we focus on the possibility of creating DM through the kinetic portal. In particular we consider that the Universe contains initially only SM degrees of freedom, among which are ordinary visible photons. By definition the visible photon is the spin-one massless particle that couples, say, to the electron. In this basis, it is also clear that hidden radiation ({\em i.e.}~extra degrees of freedom) may be produced only through the production of DM particles. We could have worked in a different basis though. In particular, instead of the transformation of Eq.~(\ref{KMtransfo}), which gives $\gamma$ couplings to both $e$ and $e'$, and a $\gamma'$ coupling to only $e'$, we could have as well done a transformation which gives $\gamma'$ couplings to both $e$ and $e'$, but a $\gamma$ coupling only to $e$.
Naively we would then expect that the Compton scatterings $e e \rightarrow \gamma \gamma'$ and  $e \gamma \rightarrow e \gamma'$ play also a role, but this is not the case \cite{Holdom:1985ag}. 
In any basis, the actual visible photon is by definition the state that couples to electrically charged SM particles. If this state is represented by a mixture, then the amplitudes must be added coherently. Hence, in this basis too, it is clear that our initial condition for the Universe contains only (essentially) visible radiation if there is initially no (resp. negligible) DM. 
Physically, it is also clear that the energy transferred to the hidden sector must involve the creation of a pair of DM particles, a process which is manifestly basis independent.}
% (and similarly for the other SM fermions replacing their masses and electric charges accordingly).
Hence the massless connector between the SM and hidden sector is the ordinary photon. {This choice makes clear the fact that  the dominant process to create hidden particles or to transfer energy from the visible sector to the hidden sector is through the creation of DM pairs, see Fig. \ref{feyngraph}.
\begin{figure}[!t]
\centering
\includegraphics[height=2.8cm]{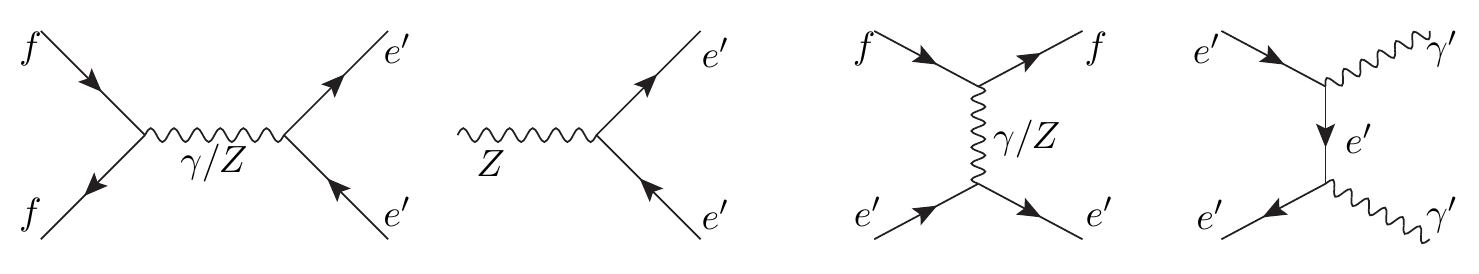}
\caption{Processes that are relevant for the production of DM and thermalization of the hidden sector through kinetic mixing. We work in the basis in which the DM is millicharged so that the hidden photons, $\gamma^\prime$, only couple to DM and not the SM degrees of freedom (see text).}
\label{feyngraph}
\end{figure}  
%Since the visible photon is massless, t
%This generically takes place through annihilation of charged SM particles.  
In this basis, by far the most important process to do so is DM creation from SM fermions through the photon, or Z boson, in the s-channel, $f \bar f \rightarrow e^\prime \bar{e^\prime}$,  see Eq.(\ref{sigmaf}) of Appendix A. There is also the $W^+W^-\rightarrow DM DM$ process but it turns out that it has always a small effect, even in regimes for which $\sqrt{s} \ge 2m_W$.
%because the $\gamma$ and $Z$  channels interfere destructively while, for heavy DM, $m_{DM} > m_Z$, annihilations of SM fermion pairs into a photon or a Z give similar contributions which interfere constructively. %Note also that, as we will see, that the $Z$ channel is negligible for $m_{DM} \ll m_Z$. There is however an intermediate regime within which the $Z$ boson plays an important role as we will see. 
}

{Note that, while in all the plots we  take into account all contributions, in this section and in the next section our discussions will be focused on the case where the production of DM is dominated by non-resonant scattering processes. For instance, for $f\bar{f}\rightarrow e'\bar{e}'$ processes, we put apart the contribution from  $Z\rightarrow e'\bar{e}'$ decay. 
These discussions therefore apply to the case of $m_{DM}>m_Z/2$ case, where obviously there is no $Z\rightarrow e'\bar{e}'$ decay. It also applies to $m_{DM}\lesssim 1$~GeV. Indeed, in the latter case the $Z$ decay contribution can be neglected because, as we will see, the production from the $\gamma$ mediated $f\bar{f}\rightarrow e'\bar{e}'$ scattering is enhanced at low temperature (being maximum at $T\sim \hbox{Max}[m_{DM},m_f]$), whereas the production from the $Z$ decay is Boltzmann suppressed at these temperatures. 
There is however an intermediate mass range, $1\,\hbox{GeV}\lesssim m_{DM} \lesssim m_Z/2$, for which the $Z\rightarrow e'\bar{e}'$ decay may dominate the DM production. In this case, it is technically more convenient to express all Boltzmann equations directly in terms of the decay width rather than to hide it in the on-shell part of the scattering contributions. 
This we have done in section 6 for the Higgs portal because, in this case, the Higgs decay to a DM particle pair (if allowed, thus for $m_{DM}<m_h/2$) always dominates the production of DM. For a better comprehension of results within this intermediate mass range  we therefore refer to section 6, where all equations apply in the same way, as well as to Appendix D, where a few supplementary subtleties concerning the interplay between decay and scattering in the case of the kinetic mixing portal are discussed in details.}

%n particular if $m_{DM} \lesssim m_Z$, the decay $Z \rightarrow e^\prime \bar{e^\prime}$ of on-shell $Z$ bosons in DM may give the dominant channel for the production of DM, at least in some of the regimes we will study in the sequel. As this process is only relevant for a specific range of DM candidates, and since this scenario is very similar to the Higgs portal, we will postpone the detailled discussion of the contribution of $Z$-decay to DM creation to Section 6.

{Note also that, {\em a priori}, we would expect the scattering $f e^\prime \rightarrow f e^\prime$ of DM particles with SM fermions to be also relevant for the transfer of energy between the visible and the invisible sector. Ev
However the behaviour of this t-channel process is very different from that of the s-channel processes.}
 To begin with, the energy transfer in this t-channel  process is {\em a priori} sub-dominant compared to that from the s-channel processes.
Moreover this process involves a hidden sector particle in the initial state, whose number density with respect to the one of the visible sector particles is suppressed by a factor of $\xi^3$, with $\xi\equiv T'/T$ the hidden-to-visible sectors temperature ratio. However, the t-channel has a collinear divergent behaviour at low-energy transfers which could  compensate for the two suppression effects. 
Nevertheless, we have checked that it is not the case and that the t-channel processes can be safely neglected in the calculation of the energy transfer between the visible and the hidden sector, as we will do in the sequel.

%\begin{figure}[b]
%\begin{center}
%\begin{picture}(410,73)(0,0)
%%%%%%%%%%%%%%%%%%%%
%%%%%%%%
%\ArrowLine(0,60)(30,30)
%\ArrowLine(30,30)(0,0)
%\Photon(30,30)(70,30){2}{6}
%\ArrowLine(70,30)(100,60)
%\ArrowLine(100,0)(70,30)
%\Text(50,22)[]{$\gamma'$}
%\Text(-3,55)[]{$f$}
%\Text(103,55)[]{$e'$}
%\Text(-3,7)[]{$f$}
%\Text(103,8)[]{$e'$}
%%%%%%%%
%\ArrowLine(130,60)(160,30)
%\ArrowLine(160,30)(130,0)
%\Photon(160,30)(200,30){2}{6}
%\DashArrowLine(200,30)(230,60){5}
%\DashArrowLine(230,0)(200,30){5}
%\Text(180,22)[]{$\gamma'$}
%\Text(128,55)[]{$f$}
%\Text(235,54)[]{$\phi$}
%\Text(128,7)[]{$f$}
%\Text(235,7)[]{$\phi$}
%%%%%%%%%%%%%%%%%%%%%%%%%
%\ArrowLine(260,60)(290,44)
%\ArrowLine(290,44)(320,60)
%\Photon(290,44)(290,16){2}{6}
%\ArrowLine(260,0)(290,16)
%\ArrowLine(290,16)(320,0)
%\Text(258,55)[]{$f$}
%\Text(325,55)[]{$f$}
%\Text(259,8)[]{$e'$}
%\Text(326,8)[]{$e'$}
%\Text(300,29)[]{$\gamma'$}
%%%%%%%%%%%%%
%\ArrowLine(350,60)(380,44)
%\ArrowLine(380,44)(410,60)
%\Photon(380,44)(380,16){2}{6}
%\DashArrowLine(350,0)(380,16){5}
%\DashArrowLine(380,16)(410,0){5}
%\Text(347,55)[]{$f$}
%\Text(414,55)[]{$f$}
%\Text(348,7)[]{$\phi$}
%\Text(415,7)[]{$\phi$}
%\Text(391,29)[]{$\gamma'$}
%%%%%
%\end{picture}
%\end{center}
%\caption{The pair creation and elastic processes for a hidden sector fermion $e'$ or scalar $\phi$. $f$ stands for a charged SM fermion (or a $W$).}
%\label{figpair}
%\end{figure}

\bigskip
Since the processes connecting the visible and hidden sector discussed above depend only on the combination
\begin{equation}
\kappa\equiv \hat{\epsilon} \sqrt{\alpha'/\alpha}\,,
\end{equation}
in what follows, we will express all results in term of this parameter, $\kappa$, which we call the connector parameter, and of $\alpha'$, the equivalent of the fine structure constant in the hidden sector. {The later coupling controls the  process $e^\prime \bar{e^\prime} \leftrightarrow \gamma^\prime \gamma^\prime$ that takes place in the hidden sector, see Eq.(\ref{sigma1}) of Appendix A.}

To study the evolution of the number density of DM particles, 
we will solve a simple Boltzmann equation in presence of both the connector processes, $SM_i SM_i \leftrightarrow DM DM$ with $i$ the various SM species, and the hidden sector process, $e' \bar{e}' \leftrightarrow  \gamma' \gamma'$. It is convenient to write this Boltzmann equation in terms of the so-called $\gamma \equiv \Gamma n_{eq}$ reaction densities, defined in Appendix B. It takes the form
\begin{equation}
s z H \frac{dY}{dz}= \sum_i \gamma^i_{connect}
\Big(1-\frac{Y^2}{Y_{eq}^{2}(T)}\Big)+\gamma_{HS}   \Big(1-\frac{Y^2}{Y_{eq}^{2}(T')}\Big) \,,
\label{generalboltzmann1}
\end{equation}
with $s$ the entropy density, $H$ the Hubble constant and $Y$ the DM number density to entropy density ratio, $Y\equiv n_{e'}/s$. Note that $Y_{DM}=2Y$ since DM is charged.
Equivalently, this Boltzmann equation may be expressed  in terms of the usual thermally averaged cross sections, $\langle \sigma v\rangle$, with $\gamma= \langle \sigma v\rangle n_{eq}^2\equiv \Gamma n_{eq},$
\begin{equation}
z \frac{H}{s} \frac{dY}{dz}=  \sum_i \langle \sigma_{connect} v \rangle_i
(Y_{eq}^{2}(T)-Y^2)+\langle \sigma_{HS} v \rangle (Y_{eq}^{2}(T')-Y^2) \,.
\label{generalboltzmann2}
\end{equation}
Note that, in either form of the Boltzmann equation, we need to distinguish $Y_{eq}(T)=n_{eq}(T)/s$ from $Y_{eq}(T')=n_{eq}(T')/s$. These abundances involve the DM equilibrium number densities expressed as a function of $T$ and $T'$ respectively, while $s$ is the total entropy density (visible plus hidden sectors).
The first one, $Y_{eq}(T)$, parametrizes the number of SM particles participating in the $SM_i SM_i \rightarrow DM DM$ processes. The second one, $Y_{eq}(T^\prime)$, parametrizes the number of $\gamma'$ that are  participating in the $\gamma' \gamma' \rightarrow e' \bar{e}' $ processes. This distinction is important, as  in the following we will consider situations for which $T'\ll T$, corresponding to $Y_{eq}(T^\prime)\ll Y_{eq}(T)$.

If the DM particles never thermalize
with the hidden photons, which, as we will see, actually happens if the combination $\kappa\alpha'$ is small,
%connector interaction $\kappa$ and hidden sector coupling $\alpha'$ are so small that the DM particles never thermalize with the hidden photons, 
the Boltzmann equation for the DM abundance simplifies, 
\begin{equation}
s z H \frac{dY}{dz}=\sum_i \gamma_{connect}^i 
\Big(1-\frac{Y^2}{Y_{eq}^{2}(T)}\Big).
\end{equation}
In other words, the number of DM particles is driven by the balance between DM pairs creation from SM particles and reciprocal processes, pretty much as in the standard WIMP framework.  

\begin{figure}[!t]
\centering
\includegraphics[height=7.5cm]{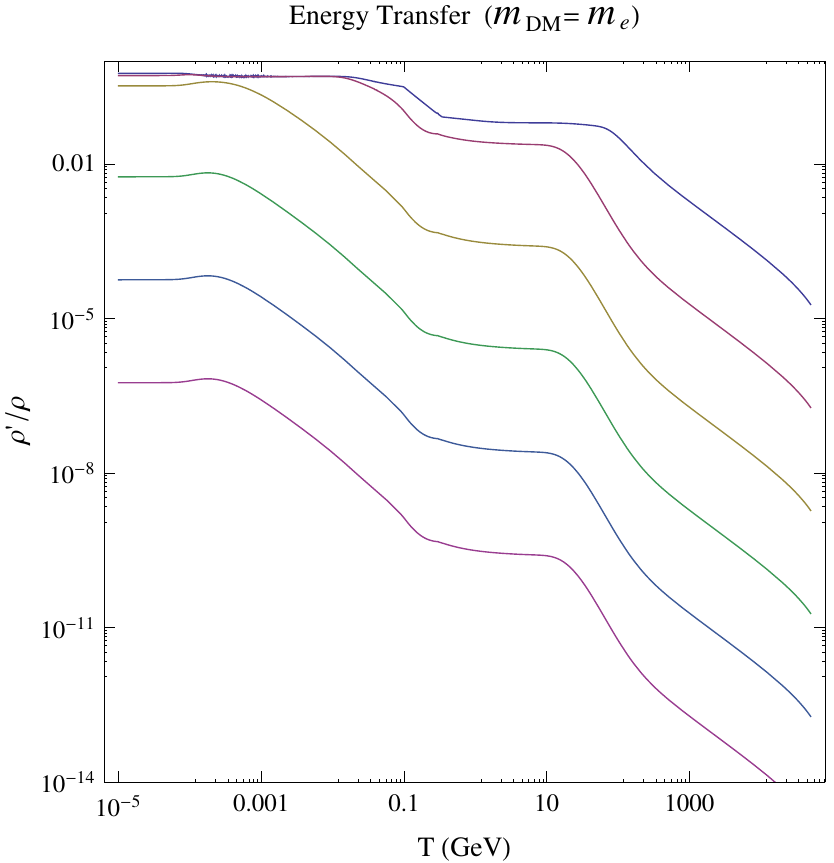}\includegraphics[height=7.5cm]{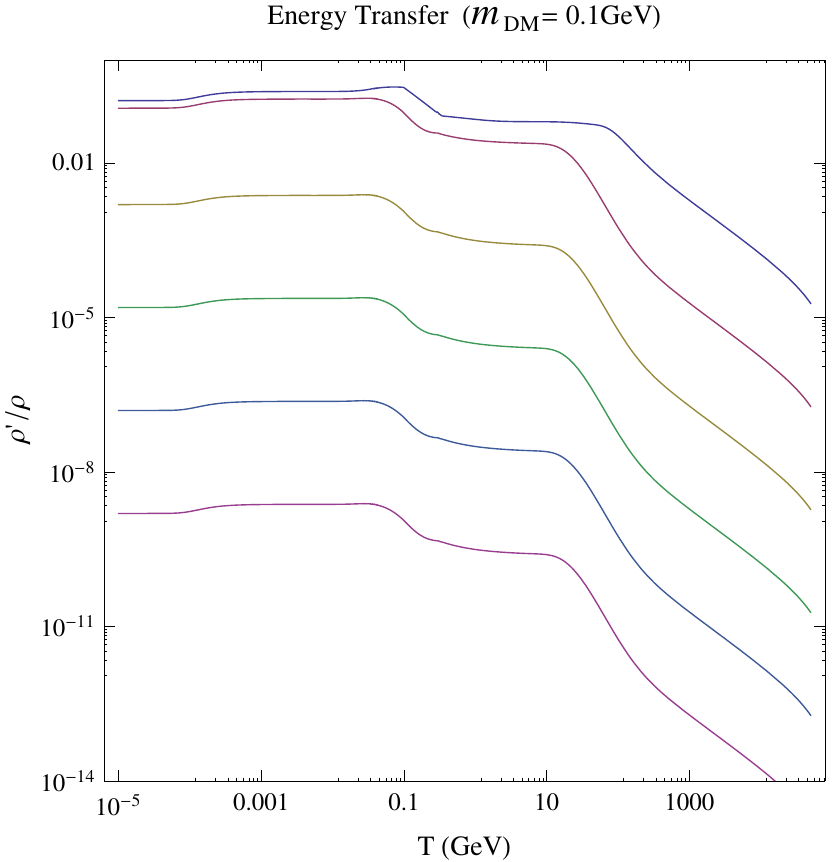}\\
\includegraphics[height=7.5cm]{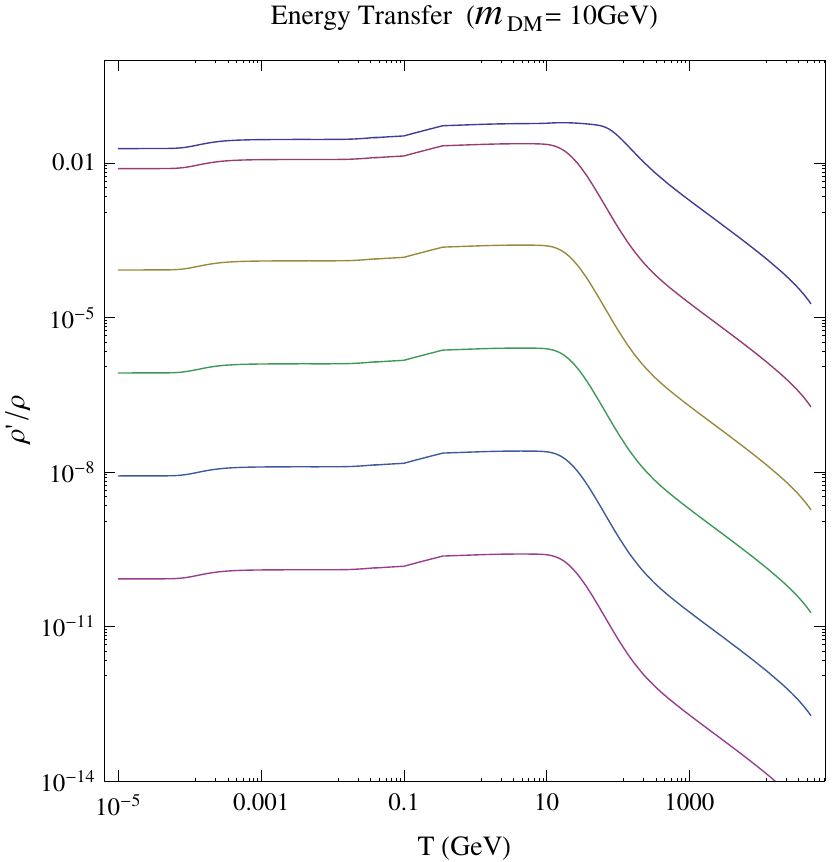}\includegraphics[height=7.5cm]{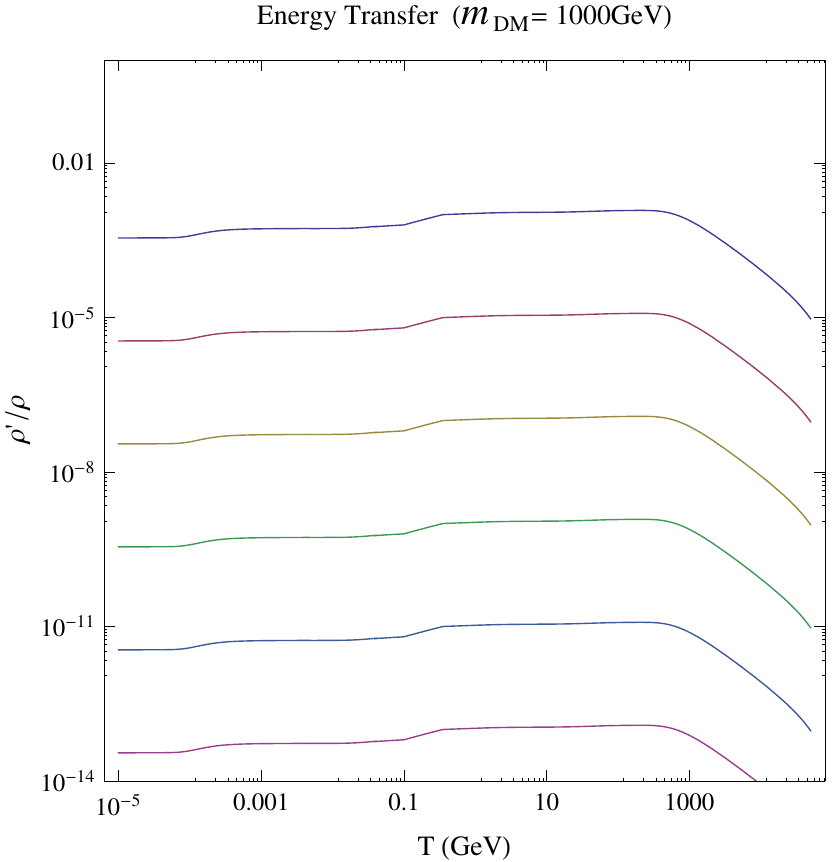}
\caption{Evolution of the ratio of the visible ($\rho$) and hidden ($\rho'$) sectors energy densities, for a range of  connector parameter, $\kappa=10^{-6,-7,-8,-9,-10,-11}$ (from up to down), and for various DM masses.}
\label{figure_GKMtempt}
\end{figure}

If, instead, the hidden sector thermalizes (in the sense that the hidden photons and the DM  reach thermal equilibrium at some temperature $T' \leq T$) the $\langle \sigma_{HS} v \rangle$ term must be taken into account. In this case one also needs to determine $Y_{eq}(T')$, and thus the evolution of the hidden sector temperature $T'$ as a function of $T$. In other words, we have to determine the energy density of the hidden sector, $\rho'$, as a function of $T$, which is given by some possible initial hidden sector energy density, plus the contribution that may be transferred from the SM sector. 
The energy density transferred to the hidden sector by a process of the form $12\rightarrow 34$ is given by the following Boltzmann equation
\begin{equation}
\frac{d\rho'}{dt}+3H(\rho'+p')=  \int \displaystyle\prod_{i=1}^{4} d^3 \bar{p}_i \cdot g_i f_1(\vec p_1) f_2(\vec p_2) |i\mathcal{M}_{1\,2 \leftrightarrow 3 \,4}|^2 (2\pi)^4 \delta^{(4)}(p_1+p_2-p_3-p_4) \Delta E_{tr} \,,
\label{Etransfereq}
\end{equation}
which, as shown in Appendix C, may be rewritten as
\begin{equation}
\frac{d(\rho'/\rho)}{dT}=-\frac{1}{H(T) T \rho}\frac{g_1 g_2}{32 \pi^4}  \int ds\cdot \sigma (s) (s-4m^2)s T K_2(\frac{\sqrt{s}}{T})\,,
\label{deltaE3}
\end{equation}
where $\rho$ is the energy density of the visible sector and $K_2$ is the usual modified Bessel function.

For $T>m_{DM}$ ($T<m_{DM}$) a numerical resolution of Eq.~(\ref{deltaE3})
gives that $\rho'/\rho$ increases like~$\sim1/T$ ($\sim$ const
respectively). This behaviour can be seen in Fig.~\ref{figure_GKMtempt} which
gives $\rho'/\rho$ as a function of the visible sector $T$, for various values
of the connector coupling and DM mass $m_{DM}$.\footnote{{Notice that the
    wiggles for  $T \gtrsim 10$ GeV seen in Fig.~\ref{figure_GKMtempt} in the
    panels corresponding to  the three lightest candidates are associated to
    resonant  energy transfer through the $Z$ resonance. For the two lightest
    candidates, this contribution has little impact on $\rho'/\rho$ (note the
    log-scale). For the  $m_{DM} = 10$ GeV candidate, the energy transfer is
    on the contrary dominated, and enhanced, by Z decay. See Appendix D and also section 6.}}

Notice that for these calculations we have assumed that the DM is in thermal equilibrium with the $\gamma'$ as, for all practical purposes, energy transfer is relevant only in this case. Indeed, if the hidden sector does not thermalize, the knowledge of $\rho'$ is unnecessary in Eq.~(\ref{generalboltzmann1}) since both the $Y_{eq}^2(T')$ source term and the small contribution of the hidden sector to the Universe expansion rate may be neglected. 
Now the condition for thermal equilibrium of the visible and hidden sector through  the connector is simply $\langle \sigma_{connect} v\rangle n_{eq} \gtrsim H$ which, taken at $T \simeq m_{DM}$, translates into roughly $\kappa \gtrsim 10^{-8,-7,-7,-5}$ for $m_{DM}=m_e,\,0.1\,\hbox{GeV},\,10\,\hbox{GeV},\,1\,\hbox{TeV}$ respectively. For larger $\kappa$, $T'$ reaches $T$ and $\rho'/\rho$ stops increasing. As can be seen in Fig.~\ref{figure_GKMtempt} it may even  decrease because, after the visible and hidden sectors have decoupled, reheating occurs in the visible sector each time a SM particle species becomes non-relativistic  ({\em i.e.} $T'/T$ decreases). We also notice that, below the critical value of the connector coupling at which the hidden sector may thermalize with the visible sector, $\rho'/\rho$ reaches a plateau at $T\sim m_{DM}$, which is given by 
$\rho'/\rho\simeq \kappa^2 \alpha^2 m_{Pl}/(m_{DM} (g^{eff}_\ast(T=m_{DM}))^{3/2})$.

%%%%%%%%%%%%%%%%%%%%%%%%%%%%%%%%%%%%%%%%%%%%%%
\section{Dark matter relic density: the {\it Mesa} phase diagram}

In the following we will first assume that the initial hidden sector energy density is negligible with respect to the energy which has been transferred from the visible sector at the temperature where the DM relic density freezes. At the end of this section we will discuss what changes when this is not the case.

Assuming a negligible initial hidden sector energy density, the relic density is determined by three parameters only: the DM mass $m_{DM}$,  the connector parameter $\kappa$, and the hidden sector coupling $\alpha'$. By construction there is no
 dependence on any other interactions or new particle masses, since in the visible sector we assume nothing but the SM. 
\begin{figure}[!htb]
\centering
\includegraphics[height=7.5cm]{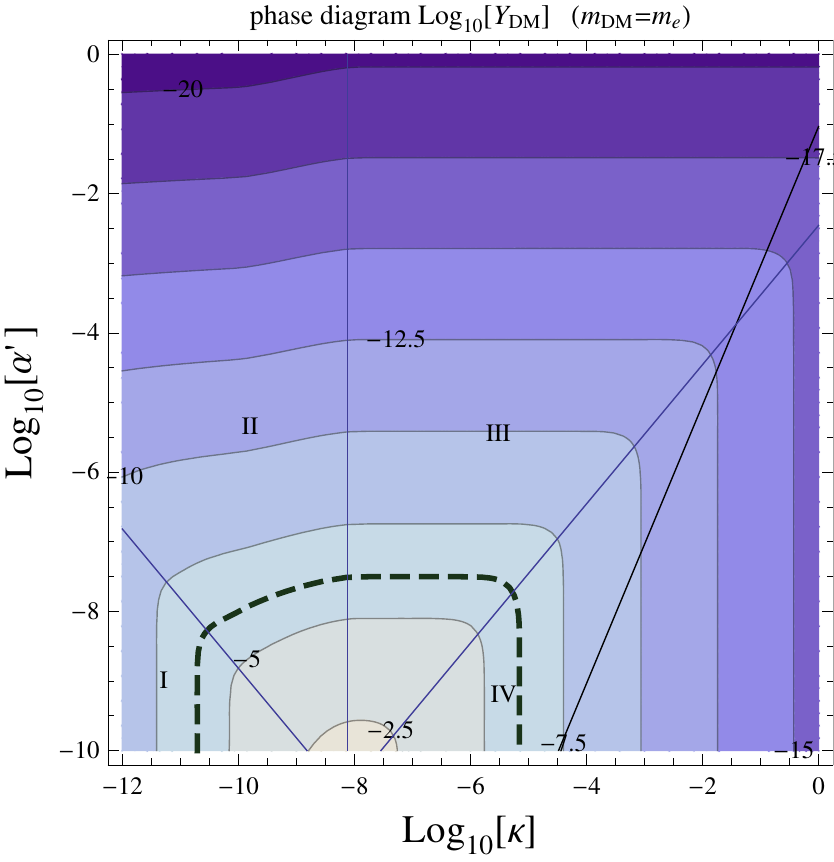}\includegraphics[height=7.5cm]{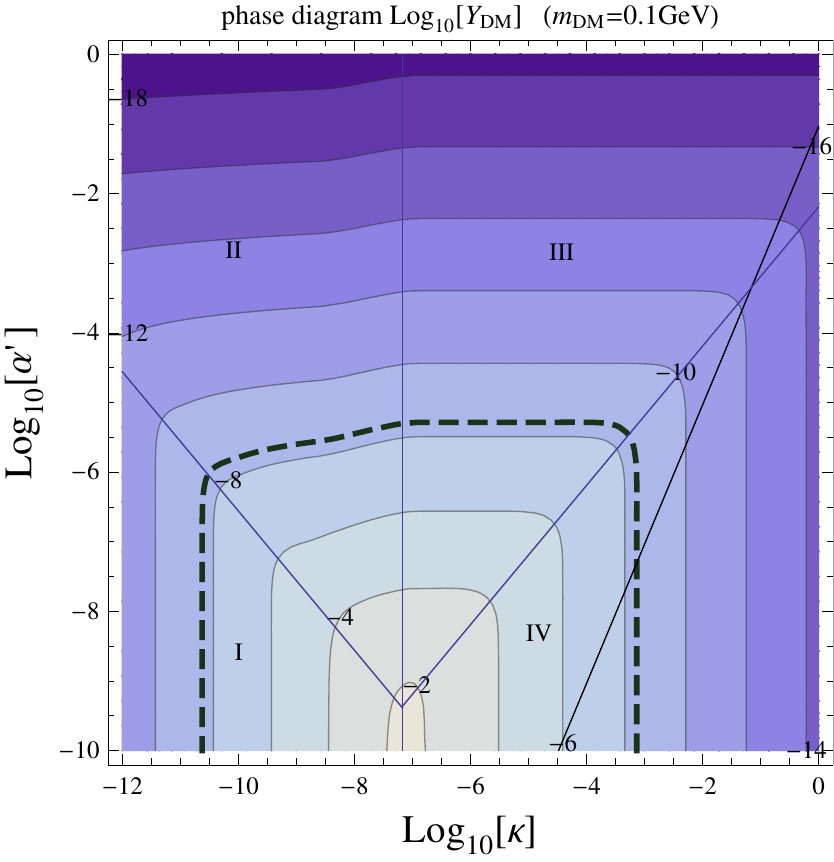}\\
\includegraphics[height=7.5cm]{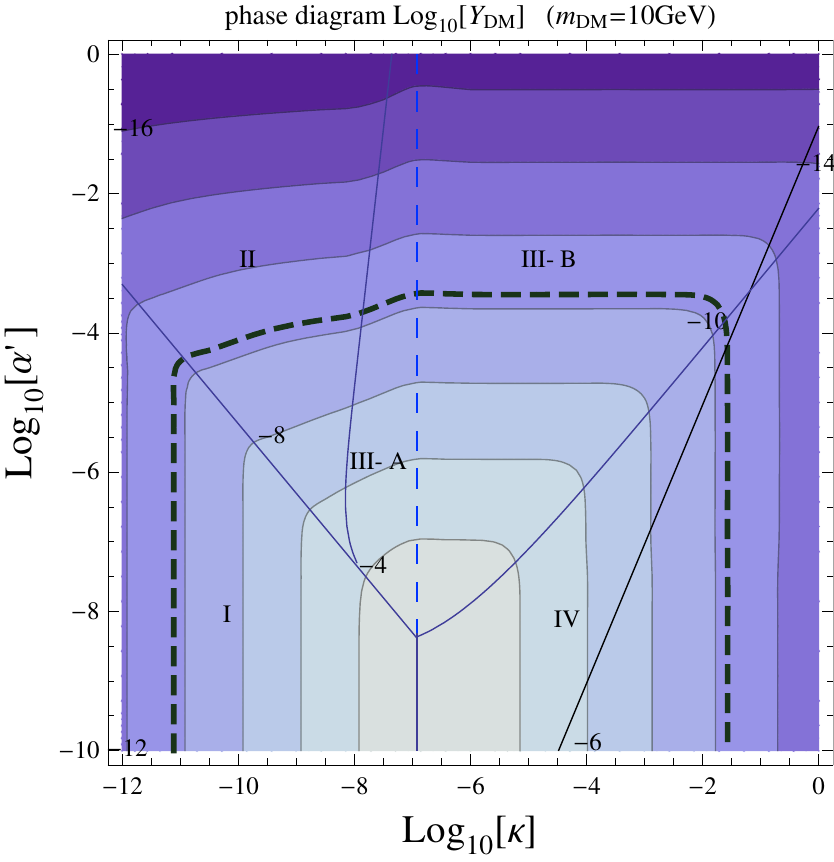}\includegraphics[height=7.5cm]{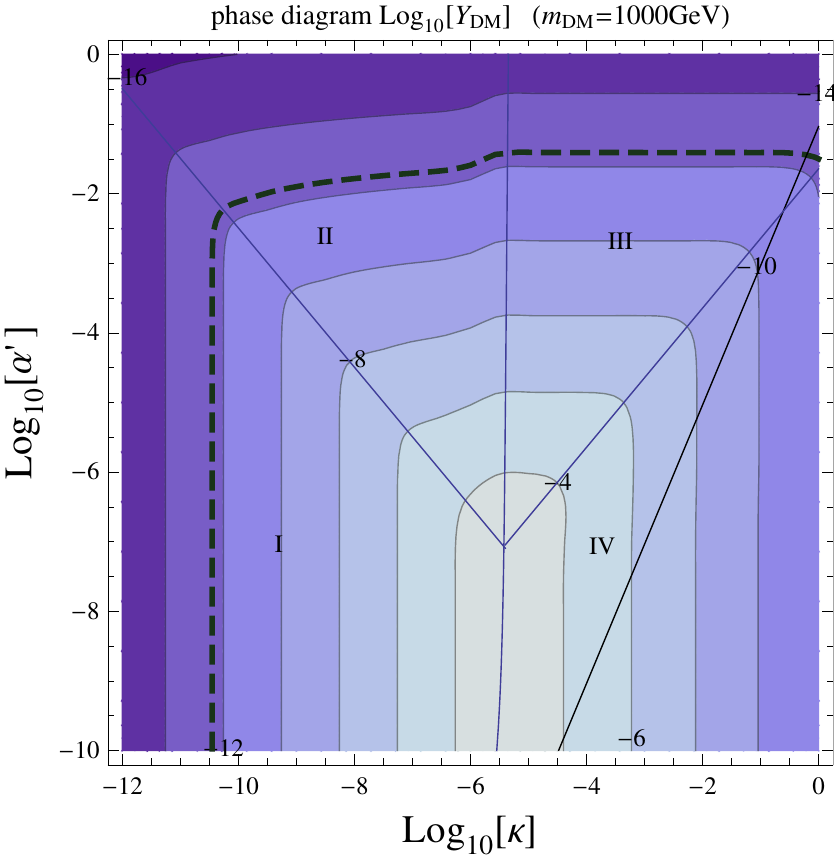}
\caption{Phase diagrams for the kinetic mixing portal: contours of $Y_{DM}$ as a function of $\kappa$ and $\alpha'$ for $m_{DM}=m_e,\,0.1\,\hbox{GeV},\,10\,\hbox{GeV},\,1\,\hbox{TeV}$.
The dashed thick line gives $\Omega_{DM}h^2=0.11$, or in other words $Y_{DM} m_{DM} =4.09\cdot10^{-10}$~GeV. We have drawn the "transition lines" delimiting the 4 phases. Phases I, II, III, IV correspond to the freeze-in, reannihilation, hidden sector freeze-out and connector freeze-out phases respectively. There are transition lines between I and II, II and III, III and IV and I and IV. The solid black line corresponds to $\epsilon={1/\sqrt{4 \pi}}$. Below this line the connector interaction is not expected to be perturbative. {Note that, for $m_{DM}=10 $GeV, the $Z$ boson leads to a ``Mesa'' phase diagram that is slightly more complex, but analogous to the one obtained in the case of the Higgs portal (see Section 6 and Appendix D).}}
\label{figure_tableGKM}
\end{figure}

Fig.3 shows the results we get for the relic abundance as a function of
$\kappa$ and $\alpha'$ and for various values of the DM mass.  
The figure has the simple characteristic shape of an isolated flat-topped hill, or what we call here the "Mesa".  The figures reveal the existence of  essentially four distinct regimes or phases, which we will explain in details in the rest of this section.\footnote{Note that in Ref.~\cite{Cheung:2010gj}  phase diagrams have been considered for the more general case where 
there is an additional feebly coupled $A$ particle in the {\em visible} sector that decays or annihilates slowly to  DM particles. In this case, the phase diagram depends, in addition to the DM mass and the connector and hidden sector interactions, also on the mass of the particle $A$ and its annihilation rate in the visible sector. The phase diagrams we have obtained, and shown in Fig.~\ref{figure_tableGKM}, are simpler and more predictive,  as the visible sector, on which they are based, depends only on known interactions and parameters.}

 The first regime (phase I) is that of freeze-in, which corresponds to no thermalization,  either through the connector or the hidden gauge interaction. Alternatively, if thermalization takes place, 
freeze-out of the hidden gauge (phase III) or connector (phase IV) interactions can occur, depending on which interaction is dominant.
A fourth possibility, which we will explain in details,
is due to the existence of an intermediate reannihilation regime (phase II). In this regime, there is a subtle inter-play between the connector and hidden gauge interactions. Like in the other regimes (freeze-in and out) we show that, in the reannihilation regime, the relic abundance is given by a simple analytical expression.

%%%%%%%%%%%%%%%%%%%%%%%%%
\subsection{Phase I: the freeze-in regime}

If both $\kappa$ and $\alpha'$ are sufficiently small, none of the interactions may thermalize the DM particle, neither with the SM sector, nor with the $\gamma'$. The relic density is therefore given by the (by now standard) freeze-in mechanism \cite{McDonald:2001vt,Hall:2009bx}: the number of DM particles is simply given by twice  the number of pair creations. This leads to the left-hand-side cliff of the Mesa shown in Fig.~\ref{figure_tableGKM}: the relic density is independent of $\alpha'$ since these interactions are negligible, and in this regime the
relic density only depends on $\kappa$.

In practice it means that the last three terms of Eq.~(\ref{generalboltzmann2})
can be neglected and the relic density is simply given by  integrating  $\gamma_{connect}$ over time, which gives the number of $e'$ particles produced per unit time per unit volume. Equivalently it is given by  the integral over temperature of $dY/dT=-\gamma_{connect}/(T\,H(T)\,s)$.
The freeze-in production is infrared dominated because, for large $T$, one has $dY/dT\sim 1/T^2$ (for a cross section which behaves like $1/s$ for large values of $s$, as in Eq.~(\ref{sigmaf})).
%-(\ref{sigmaW})).
As a result we can approximate the total number of $e'$ particles produced by the number of $e'$ produced per unit time, $\gamma_{connect}$, times the Hubble time at $T=max[m_i,m_{DM}]$, with $m_i$ the mass of the initial SM particle in the pair production process. At later times the production becomes Boltzmann suppressed because less and less SM particles may produce DM pairs. In other words, $Y$ grows monotonously as $1/T$ until it reaches a plateau where it freezes-in, with the value
\begin{equation}
\label{purefreezein}
Y=c\, \frac{\gamma_{connect}}{s H}\Big|_{T=max[m_i,m_{DM}]}\,,
\end{equation}
 with  $c$ is a coefficient of order 1 (and with, as said above, $Y_{DM}=2Y$).
Numerically we get $c_a=5.0 (4.8)$ for the $e^+ e^-\rightarrow e' e'$ channel and $m_{DM}<m_e$ (resp. $m_{DM}>m_e$).
For $\mu^+ \mu^-\rightarrow e' e'$, we have $c=9.1 (4.8)$ for $m_{DM}<m_\mu$ (resp. $m_{DM}>m_\mu$).
The $c$ coefficient depends on the $s$ dependence of the cross section considered. For $m_{DM}>>m_i$ the actual value of $m_i$ is irrelevant and the various fermions distinguish themselves only through their respective electric charge (and number of colors). The $c$ coefficients are slightly larger than unity because the production occurs mostly at a temperature that is a factor of 2-3 times smaller than $max[m_i,m_{DM}]$. This is the temperature at which Boltzmann suppression is actually effective. 
 
\begin{figure}[!t]
\centering
\includegraphics[height=6.8cm]{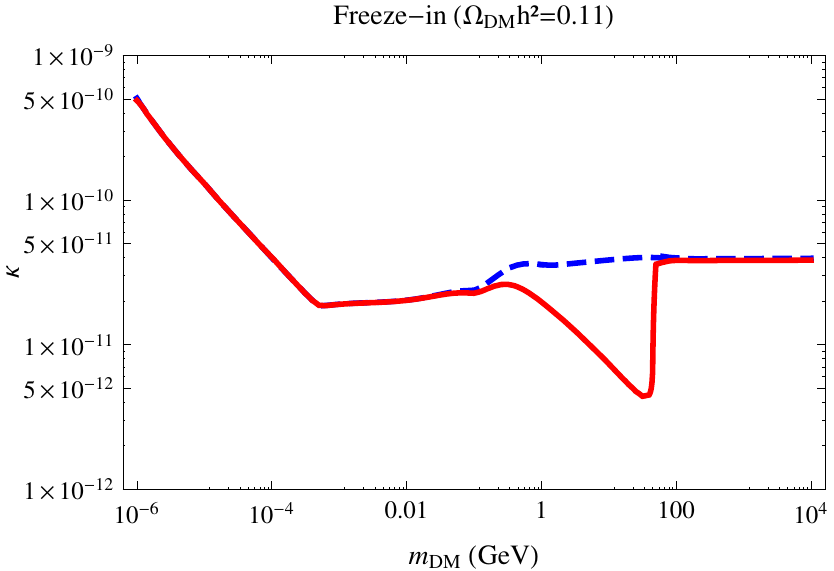}\caption{Values of $\kappa$ that give the observed relic density through freeze-in ($\alpha'$ is assumed to be negligible). The continuous line corresponds to the contribution of both the $\gamma$ and $Z$ channels, the dashed line is only for the $\gamma$.} 
\label{kappain}
\end{figure}

Provided that the production of DM is non-resonant, analytically we get that the abundance of DM particles scales as $Y \simeq \gamma_{connect}/(s H)|_{T=max[m_f,m_{DM}]}\propto \kappa^2/m_{DM}$ if $m_f<m_{DM}$, and as
$Y \propto \kappa^2/m_f$ if $m_f > m_{DM}$. 
The channels that dominate the production are those for which $m_f < m_{DM}$, since they are effective down to $T\sim m_{DM}$. 
Therefore, if the candidate is heavier than the electron, $Y$ is independent of $m_f$ and scales as $\kappa^2/m_{DM}$, and so $\Omega_{DM}$ is independent of $m_{DM}$ (modulo threshold effects each time $m_{DM}$ becomes larger than the mass of one of the SM particles).
This can be seen in Fig.~\ref{kappain}, which gives the value of $\kappa$ required to have the observed DM  relic density.
If instead $m_{DM}<m_e$, then $\Omega_{DM}$ scales like $\kappa^2 m_{DM}$, as can also be seen in Fig.~\ref{kappain}. 
Finally, for $m_{DM} \gtrsim 1$ GeV and up to $m_Z/2$, DM production is dominated by a resonance, {\em i.e.}  $Z$ decay. In this case, the DM abundance is independent of $m_{DM}$ (see Appendix D and also section 6) and scales like $\kappa^2$, so that $\Omega_{DM}$ scales like $\kappa^2 m_{DM}$. This behaviour corresponds to the dip
in Fig.~\ref{kappain}. Notice that it implies that a smaller value of $\kappa$
 is required to reach the observed DM abundance, a feature also visible
  in Fig.\ref{figure_tableGKM}.

\begin{figure}[!t]
\centering
\includegraphics[height=5.5cm]{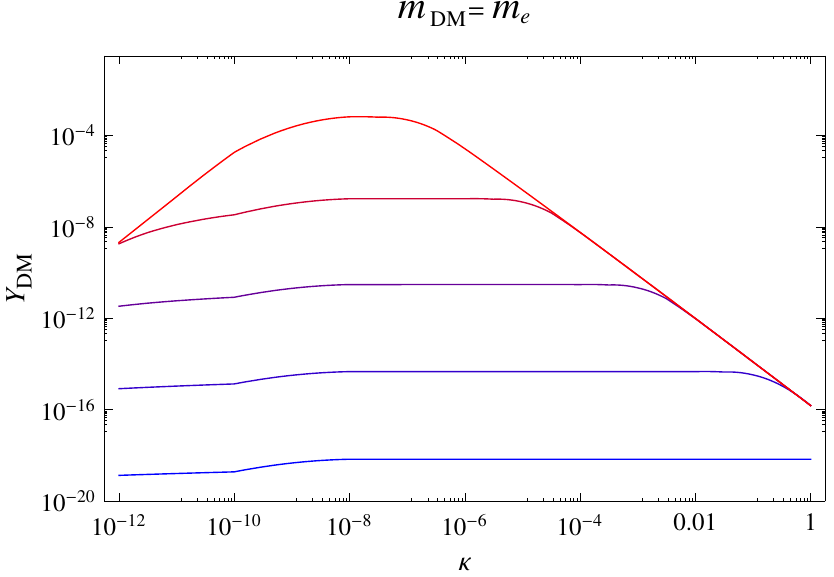}\includegraphics[height=5.5cm]{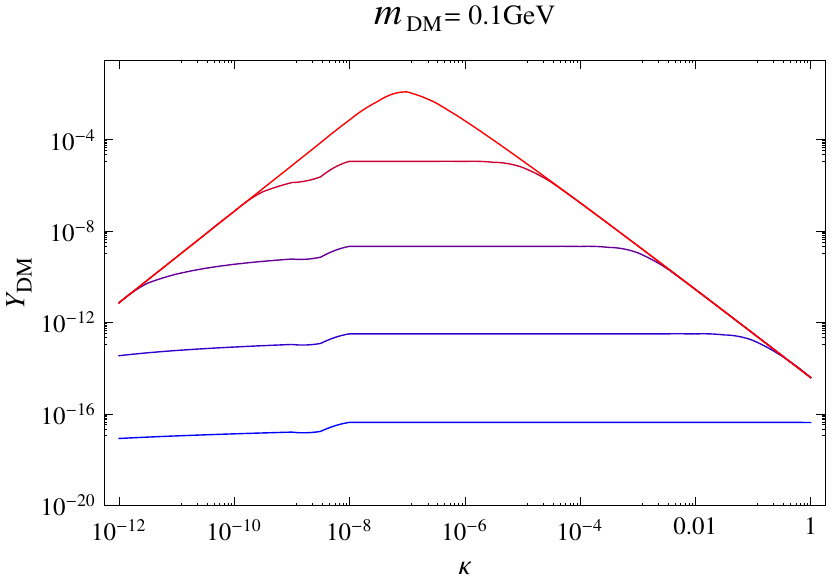}\\
\includegraphics[height=5.5cm]{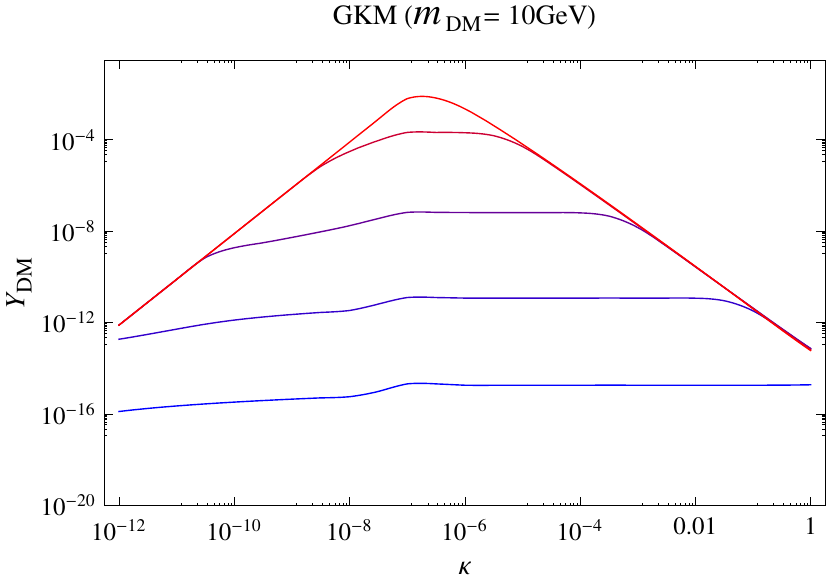}\includegraphics[height=5.5cm]{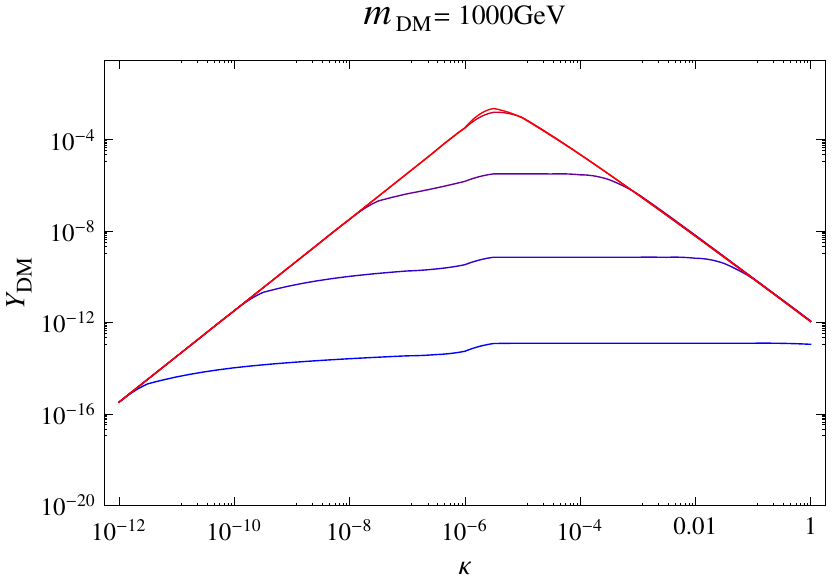}
\caption{DM relic abundance $Y_{DM}$ as a function of the connector parameter $\kappa$ for different DM masses $m_{DM}$ and values of the hidden sector interaction, $\log_{10}(\alpha'/\alpha)=1,-1,-3,-5,-7$ (bottom-up).}
\label{YDM-kappa}
\end{figure}

The various scaling properties discussed above can also be seen in Fig.~\ref{YDM-kappa}, which displays $Y$ as a function of $\kappa$ for various values of $\alpha'$ and $m_{DM}$. In this figure one clearly sees, for tiny values of $\alpha^\prime$, the characteristic volcano shape of the transition between the freeze-in and freeze-out regime, from the processes driven by $\kappa$. The top of the volcano corresponds to the point where the connector interaction thermalizes, delimiting the $\Omega_{DM}\sim \kappa^2$ freeze-in behaviour from the $\Omega_{DM}\sim 1/\langle \sigma_{connect} v \rangle\sim 1/\kappa^2$ freeze-out behaviour. In Ref.~\cite{Hall:2009bx}  similar transitions have been obtained for other types of interactions. For larger values of $\alpha'$ however, there is no more  freeze-in-to-freeze-out transition. The volcano becomes a truncated volcano and
the transition from freeze-in to freeze-out undergoes an intermediate regime of reannihilation, which we will now discuss.

%%%%%%%%%%%%%%%%%%%%%%%%%%

\subsection{Phase II:  freeze-out with a source term: the reannihilation regime}

Starting from a freeze-in situation, if one increases $\kappa$ and/or $\alpha'$, at some point, to be defined below, there are enough DM particles and the interactions in the hidden sector are fast enough for the hidden photon $\gamma'$ and the dark matter particle $e'$ to thermalize, even if 
the connector interaction remains out-of-equilibrium. Therefore one may in principle define a hidden sector temperature $T'$, with $T^\prime \leq T$. In this case, we may determine the dark matter relic density in two steps. First we estimate the hidden sector energy density, $\rho^\prime$, as a function of the visible sector temperature $T$, as explained above, and define a hidden temperature $T^\prime$, which depends only on $\kappa$ and $m_{DM}$, through $\rho^\prime \propto T^{\prime4}$. This step gives us in turn a way to define the DM equilibrium number density {\em in the hidden sector},  $Y_{eq}(T') \neq Y_{eq}(T)$. From these, in a second step, we may  integrate
the  Boltzmann equation for the DM number density, Eq.~(\ref{generalboltzmann1}), taking into account both the source connector  and annihilation in the hidden sector.

Starting from an initially insignificant abundance of hidden sector particles, the DM number density follows the freeze-in regime until the $\gamma'$ and $e'$ thermalize.
The condition for chemical equilibrium between the $\gamma'$ and $e'$ is 
\begin{equation}
\Gamma_{annih} \equiv \langle \sigma_{HS} v\rangle n_{eq}(T^\prime) > H\,,
\label{gammaannihH}
\end{equation}
whereas  kinetic equilibrium holds provided
\begin{equation}
\frac{1}{H}\frac{d\rho'/dt}{\rho'}>1\,,
\end{equation}
where $\rho'$ here stands for the energy transferred from the DM particles to the $\gamma'$.
Actually both conditions essentially coincide because the DM and $\gamma'$ thermalize for $T\gtrsim m_{DM}$ when the DM particles are still relativistic.
In terms of the cross sections, Eq.~(\ref{gammaannihH}) can be written as 
\begin{equation}
\langle \sigma_{eff} v \rangle n_{eq}(T) \sqrt{c} >H \quad \hbox{with}\quad \langle \sigma_{eff} v \rangle \equiv \sqrt{\langle \sigma_{HS} v\rangle \langle \sigma_{connect} \rangle} \,,
\end{equation}
where $c$ is a constant of order unity defined in Eq.~(\ref{purefreezein}).
This reflects the fact that $n_{eq} \langle \sigma v_{HS} \rangle$ is proportional to the number density of DM particle, which itself is proportional to $\langle \sigma_{connect} v\rangle n_{eq}^2(T)/H$. Taking this condition at $T\simeq m_{DM}$ gives the "phase transition" line between the freeze-in and reannihilation regimes. It gives a value of $\alpha'$ larger than a quantity which scales as $\propto m_{DM}/(m_{Pl} \kappa)$  (numerically we get $\kappa \alpha' > 1\cdot 10^{-15} \hbox{GeV}^{-1} \cdot m_{DM}/(g^{eff}_{*}(T=m_{DM}))^{1/4}$ with $g^{eff}_*$ the effective number of degrees of freedom of both SM and hidden sectors).
Once the hidden sector has thermalized, the DM number density is just that of equilibrium, which for $T' \gtrsim m_{DM}$ is
\begin{equation}
Y_{eq}(T')=\frac{45 \zeta(3)}{2\pi^4}\frac{g_{e'}}{g_{*s}}\,\xi^3\,,
\end{equation}
in which $\xi=T'/T < 1$, $g_{e'}=2$ is the number of degrees of freedom of $e'$ and $T'$ is defined by the equilibrium relation
\begin{equation}
\rho'=\frac{\pi^2}{30}g^{HS}_*\, T'^4.
\end{equation}
Later on, 
%once the hidden sector temperature $T'$ drops below $m_{DM}$, 
the number density $Y$ follows the equilibrium one $Y_{eq}(T^\prime)$ and does so until the latter becomes sufficiently Boltzmann suppressed.
However the Boltzmann equation at this stage is different from the one in the standard freeze-out as it contains an extra source term from the connector. This makes the reannihilation regime a bit complex, so let us explain what happens step by step. 

The reannihilation regime corresponds to a situation in which interactions within the hidden sector are fast (that is, compared to the expansion rate), while the energy transfer from the visible sector is comparatively slow. In this case, the hidden sector reaches thermal equilibrium, but at a lower temperature than that of the visible sector, $\xi=T'/T\ll 1$. Hence the particles in the visible sector are more abundant (per species) than in the hidden sector, $Y \ll Y_{eq}(T)$. Starting from the full Boltzmann equation Eq.(\ref{generalboltzmann2}), we thus have
\begin{eqnarray}
z \frac{H}{s} \frac{dY}{dz}&=&  \langle \sigma_{connect} v \rangle
(Y_{eq}^{2}(T)- Y^2)+\langle \sigma_{HS} v \rangle (Y_{eq}^{2}(T')-Y^2) \nonumber\\
&\simeq &  \langle \sigma_{connect} v \rangle
Y_{eq}^{2}(T) +\langle \sigma_{HS} v \rangle (Y_{eq}^{2}(T')-Y^2) \,,
\label{generalboltzmannreannih}
\end{eqnarray}
(here we have dropped the initial SM particle $i$ index to avoid the cluttering of symbols). 
%In other words, if the connector interaction is sufficiently large (in a sense to be make clear below) the visible sector may act as a reservoir of DM particles. 

By assumption, early on the processes $DM DM \leftrightarrow \gamma' \gamma'$ are fast, and the abundance of DM is able to track $Y_{eq}(T')$, but when $T'\lesssim m_{DM}$ the equilibrium abundance $Y_{eq}(T')$ becomes Boltzmann suppressed and so does the rate of $\gamma' \gamma' \rightarrow DM DM$. 
If the connector source term is negligible, decoupling takes place when $Y_{eq}(T^\prime)$ drops below the critical value
\begin{equation}
Y_{crit} \equiv H /\langle \sigma_{HS} v \rangle s \simeq  Y_{eq}(T^\prime_{crit}) \,,
\label{Ycrit}
\end{equation}
a condition that corresponds to 
\begin{equation}
 \langle \sigma_{HS} v \rangle n_{eq}(T^\prime)  \simeq H(T).
\label{GammaH}
\end{equation}
This is just like in the standard freeze-out mechanism. 

If, however, the connector source term is large enough, then, at some temperature $T_a > T_{crit}$, the $\gamma' \gamma' \rightarrow DM DM$ process may become sub-dominant compared to the production of DM through the connector. This occurs provided 
%its contribution to DM production becomes smaller than the one of the source term, from the $SM SM  \rightarrow DM DM$ process. This occurs when
\begin{equation}
\langle \sigma_{connect} v \rangle Y_{eq}^2(T) \gtrsim \langle \sigma_{HS}v \rangle Y_{eq}^2(T')
\label{primaryfreeze-out}
\end{equation}
for $T \lesssim T_a$.
Clearly, the connector source term may become more important than the hidden sector source term simply because  $T^\prime < T$, so that $Y_{eq}(T') \ll Y_{eq}(T)$ for $T^\prime < m_{DM}$. This is illustrated in Fig.~\ref{reannihilation-example} (in particular the right panel), where $Y_{eq}(T^\prime)$ becomes Boltzmann suppressed while $Y_{eq}(T) \sim $ const ({\em i.e.} $T > m_{DM} > T^\prime$).
 In this situation
the $\gamma' \gamma' \rightarrow DM DM$ process becomes
irrelevant around $T_a$,
{and this is true even if its rate is still larger than the Hubble rate}. Consequently, from Eq.~(\ref{generalboltzmannreannih}) we get that, for $T \lesssim T_a$, the DM number density is determined by the following simpler Boltzmann equation\footnote{A similar equation has been considered  in Ref.~\cite{Cheung:2010gj} for the case of a decay process (rather than annihilation) as a source term from the visible sector. 
}
\begin{eqnarray}
z \frac{dY}{dz} &\simeq& \frac{ \langle \sigma_{connect} v \rangle s}{H} Y_{eq}^2(T)- \frac{\langle \sigma_{HS}v \rangle s}{H} Y^2 \nonumber\\
&\equiv & \frac{ \langle \sigma_{connect} v \rangle s}{H} Y_{eq}^2(T)-\frac{Y^2}{{Y_{crit}}}\,.
\label{boltz-qse}
\end{eqnarray}

Now, down to $T\simeq T_a$,  $Y \simeq Y_{eq}(T')$ holds, as the rate for  DM pair creation by hidden photons is faster than the Hubble rate (and the rate of production through the connector). 
%hidden sector interaction is in thermal equilibrium, 
%$Y \simeq Y_{eq}(T')$ holds, so}
%that
Moreover, at $T \simeq T_a$, first $Y >Y_{crit}$ holds, and second, using Eq.({\ref{primaryfreeze-out}), one has
\begin{equation}
\frac{\langle \sigma_{connect} v \rangle s}{H} Y_{eq}^2(T) \simeq {\langle \sigma_{HS} v \rangle s \over H}Y^2 \simeq {Y^2\over Y_{crit}} > Y_{crit}\,
\end{equation}
so that both terms on the RHS of the Boltzmann equation ({\ref{boltz-qse}) are relevant at $T \simeq T_a$.\footnote{That is to say, each term gives in the Boltzmann equation a contribution   to $(z/Y) (dY/dz)$  that is larger than 1.} Hence below $T_a$ the tracking solution is given by $Y= Y_{QSE}$, 
\begin{equation}
{Y^2_{QSE}} \equiv Y_{crit} \, \frac{ \langle \sigma_{connect} v \rangle s}{H} Y_{eq}^2(T) = \frac{ \langle \sigma_{connect} v \rangle}{\langle \sigma_{HS} v \rangle} Y_{eq}^2(T)\,,
\label{QSE}
\end{equation}
that is dubbed the Quasi Static Equilibrium (QSE) abundance \cite{Cheung:2010gj} to emphasize its intermediate character, see  Fig.~\ref{reannihilation-example},
\begin{equation}
Y_{eq}(T^\prime)  < Y_{QSE} < Y_{eq}(T)\,.
\end{equation}

The abundance of DM tracks the QSE  until a temperature $T\equiv T_f$ at which $Y_{QSE}$ drops below the equilibrium condition critical value of Eq.~(\ref{Ycrit}). Indeed at $T=T_f$ the condition $Y_{QSE} \simeq Y_{crit}$ is equivalent to the conditions $\langle \sigma_{connect} v \rangle s Y_{eq}^2(T)/H \simeq Y$ and  $Y\simeq Y_{crit}$, which shows that both terms decouple at the same temperature $T_f$. From this point on the abundance of DM  is frozen.
\begin{figure}[!t]
\centering
\includegraphics[height=8cm]{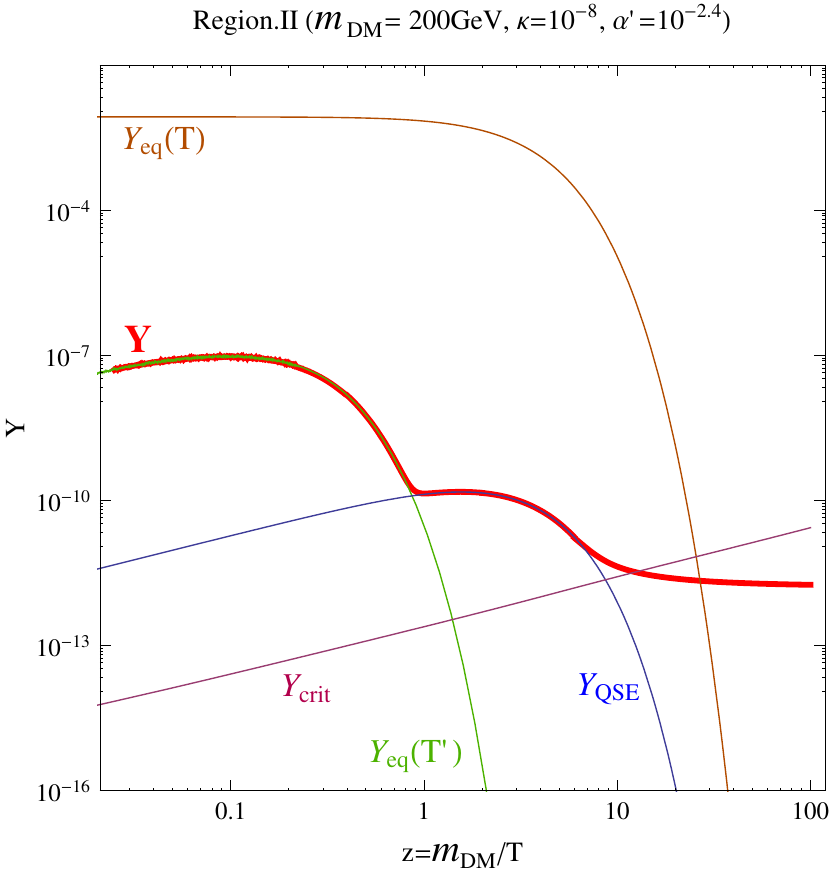}\includegraphics[height=8cm]{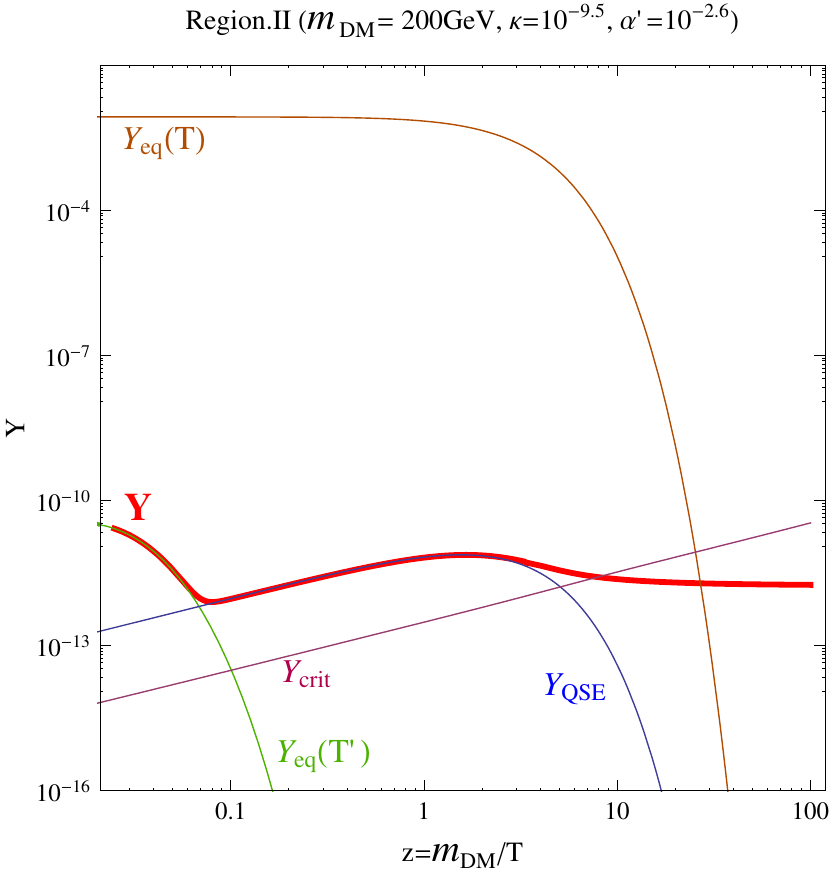}
\caption{Examples of evolution of the DM number density $Y$ as a function of $z\equiv m_{DM}/T$, in the reannihilation phase (red line). Also shown are $Y_{QSE}$, $Y_{crit}$, $Y_{eq}(T')$ and $Y_{eq}(T)$. The left panel  is obtained for $m_{DM}=200$~GeV, $\kappa=10^{-8}$ and $\alpha'=10^{-2.4}$.  It gives $T'/T\simeq 0.067$ at $T=T_a$ and
$T'/T\simeq 0.072$ at $T=T_f$. The right panel is obtained with $m_{DM}=200$~GeV, $\kappa=10^{-9.5}$ and $\alpha'=10^{-2.6}$. Both cases lead to a relic density that is in agreement with observations.}
\label{reannihilation-example}
\end{figure}

In other words, and to summarize, the  condition that has to be satisfied to stay in the reannihilation regime is that Eq.~(\ref{primaryfreeze-out}) holds 
before the freeze-out of interactions in the hidden sector,  Eq.~(\ref{GammaH}).
If it is not the case, we have a standard freeze-out scenario. Otherwise, as illustrated in Fig.~\ref{reannihilation-example}, the abundance of DM begins to follow the equilibrium density of the hidden sector $Y_{eq}(T^\prime)$ until
it intercepts at a temperature $T=T_a$ the QSE abundance, which becomes the new tracking solution. Equilibrium between the connector source term and annihilation into hidden photon freezes at $T = T_f$, and the final DM abundance is given by the value of the QSE abundance when it intercepts the critical density of Eq.(\ref{Ycrit}),
\begin{equation}
Y(T_f)=Y_{crit}(T_f)=Y_{QSE}(T_f) \,.
\label{YDMreannihil}
\end{equation}

From the Boltzmann equation (\ref{QSE}) we may get an analytical estimate of the final relic density. 
First using Eqs.~(\ref{Ycrit}) and (\ref{QSE}), we get from Eq.~(\ref{YDMreannihil})
 \begin{equation}
Y^2(T_f)=\frac{\langle \sigma_{connect} v \rangle}{ \langle \sigma_{HS} v \rangle} Y^2_{eq}(T_f) \,.
\label{Yanalinterm}
\end{equation} 
that, using Eq.~(\ref{Ycrit}) once more in Eq.~(\ref{Yanalinterm}), gives
\begin{equation}
n_{eq}(T_f) \sqrt{\langle \sigma_{connect} v \rangle  \langle \sigma_{HS} v \rangle}=H(T_f)\,.
\label{freeze-outreacond}
\end{equation}
This condition is like the one for standard freeze-out, but with an effective cross section given by 
$\langle \sigma_{eff} v \rangle\equiv \sqrt{\langle \sigma_{connect} v \rangle  \langle \sigma_{HS} v \rangle}$.\footnote{Equivalently, for $Y \sim Y_{QSE}$ we may rewrite the Boltzmann equation~(\ref{boltz-qse}) as
\begin{equation}
z {H\over s}{dY\over dz} \approx 2 \sqrt{\langle \sigma_{connect} v \rangle \langle \sigma_{HS} v \rangle} Y_{eq}\left(Y_{QSE} - Y\right ) \,,
\end{equation}
which shows that freeze-out takes place at a temperature $T= T_f$ such that  $H\sim  \sqrt{\langle \sigma_{connect} v \rangle \langle \sigma_{HS} v \rangle} n_{eq}(T)$ and $Y = Y_{QSE}(T_f)$.
}
Solving for $T_f$ in the usual way we get 
\begin{eqnarray}
x_f&=&\log [ 0.038\frac{g_{e'}}{\sqrt{g^{eff}_*}}m_{Pl} m_{DM} \langle \sigma_{eff} v \rangle c (c+2) ]\nonumber\\
&&-\frac{1}{2} \log[\log[0.038\frac{g_{e'}}{\sqrt{g^{eff}_*}} m_{Pl} m_{DM} \langle \sigma_{eff} v \rangle c(c+2)]]
\label{xfreannihil}
\end{eqnarray}
and
\begin{equation}
Y_{eq}(T_f)=\frac{3.79 x_f \sqrt{g^{eff}_*}}{g_{\star s}m_{Pl} m_{DM} \langle \sigma_{eff} v \rangle}  \,,
\end{equation}
with $x_f\equiv m_{DM}/T_f$ and $c$ a numerical constant of order unity.  Using Eq.~(\ref{Yanalinterm})
we get finally
\begin{equation}
Y(T_f) \equiv Y_{QSE}(T_f) =   \frac{3.79 x_f \sqrt{g^{eff}_*}}{g_{\star s}m_{Pl} m_{DM}\langle \sigma_{HS} v \rangle}  .
\label{Yreanfinal}
\end{equation}
For the standard choice of numerical constant $c(c+2)=1$, this expression approximates the exact result within a factor less than  2-3 in the reannihilation regime of Fig.~\ref{figure_tableGKM}. Better accuracy may be obtained by adopting sligthly smaller values, for instance $c=0.1$ for $m_{DM}=1$~TeV. Note that $Y$ depends on $\langle \sigma_{connect} v \rangle$ only through the logarithmic dependence of $x_f$.

To sum up the final relic density obtained in the reannihilation scenario is inversely proportional to the hidden sector cross section, Eq.~(\ref{Yreanfinal}). This is on one hand the standard expectation. On the other hand its decoupling temperature, $x_f$, is determined by an effective cross section $\langle \sigma_{eff} v \rangle$ which is the geometric mean of both connector and hidden sector cross sections. In practice since the relic density depends linearly on $x_f$, this can change the result by up to one order of magnitude. For example with a hidden sector cross section {of the order of the one that is typically needed in the standard} freeze-out mechanism, $x_f$ can be as low as $\sim$ 2-3 instead of the usual $\sim 20$. In this circumstance, to get the observed relic density, the hidden sector cross section has to be one order of magnitude smaller than in the usual freeze-out scenario. Note also that $x_f$ cannot be smaller than 
 $\sim$ 2-3 because the source term, and therefore $Y_{QSE}$, gets Boltzmann suppressed when $T\lesssim m_{DM}$. Also it cannot be larger than the usual $\sim 20$ value because, {using the observed relic density in Eq.~(\ref{Yreanfinal}), one gets anyway a value of $\langle \sigma_{HS}v\rangle$ which, plugged in Eq.~(\ref{xfreannihil}),}
 cannot give a larger value of $x_f$ (as $\sigma_{connect}< \sigma_{HS}$ in this regime). As for $x'_f\equiv m_{DM}/T'_f$ it cannot be smaller than $\sim$ 2- 3 too because $Y_{eq}(T')$ decouples when $T'\lesssim m_{DM}$, but it could be orders of magnitude larger than unity (if $T'/T<<1$).
 In all cases DM freezes when non-relativistic because the DM particles left mainly consist of particles which have been pair created at $T\sim T_f$ with $T_f\lesssim m_{DM}$ ({the particles produced early on having annihilated to $\gamma'$ since a while) and because anyway, when they freeze, they are still in kinetic equilibrium with the $\gamma'$ and this down to a temperature $T'$ which is below $T'_f$.} 

From the results above, the structure of the phase diagram in the reannihilation regime can be easily understood. For a fixed value of the connector $\kappa$, as $\alpha'$ increases, $\Omega_{DM}$ obviously decreases because, as usual,  $Y(T_f) \propto 1/\langle \sigma_{HS}v \rangle$ decreases. For a fixed value of $\alpha'$ the dependence on $\kappa$ on the other hand is milder because it enters only logarithmically, through $x_f$, Eq.(\ref{xfreannihil}). In the phase diagram, this explains the near flatness of the part of the Mesa corresponding to the regime II in Fig.~\ref{figure_tableGKM}.

We note also that, in practice, if the DM production is dominated by processes where the mediator mass is irrelevant (as applies obviously for $m_{DM}> m_Z/2$, or for $m_{DM}\lesssim 1$~GeV where the $\gamma$ mediated processes dominate), 
%and except along the transition regime between the freeze-in and reannihilation regimes, 
we did not find any case for which, without thermalization of the connector, the DM particle would freeze-out without experiencing a period of reannihilation, {\em i.e.}~where $Y$ follows the $QSE$ distribution for a period of time.
This can be understood in the following way. Starting from $Y=0$, if the hidden sector thermalizes, {\em i.e.}~if $Y$ goes over $Y_{crit}$, $Y$ increases quickly up to a point at which $Y=Y_{eq}(T') >Y_{QSE}>Y_{crit}$.\footnote{This can be anticipated from Eq.~(\ref{QSE}), which at the thermalization point gives $Y_{QSE}^2 \simeq Y_{crit} Y$, {\em i.e.} $Y_{QSE}=Y_{crit}$ (since, before thermalization, $Y\simeq \langle \sigma_{connect} v \rangle s Y_{eq}^2(T)/H$), and the fact that just after thermalization ($Y> Y_{crit}$) $Y_{QSE}$ still increases essentially as the geometric mean of $Y$ and $Y_{crit}$.} Later on this hierarchy is not modified down to $T'\lesssim m_{DM}$ because the three abundances
have similar dependence in $T$.\footnote{$\langle \sigma_{HS} v \rangle$ is essentially constant during this period because $T'$ never go much above $m_{DM}$. Therefore from Eq.~(\ref{Ycrit}) we get $Y_{crit}\sim 1/T$ and from Eq.~(\ref{QSE}) we get $Y_{QSE}\sim 1/T$. As for $Y_{eq}(T')$, using the $\rho'/\rho\sim 1/T$ relation above, it scales as $Y_{eq}(T')=n_{eq}(T')/s\sim T'^3/T^3\sim 1/T^{3/4}$. These scaling properties can be seen in Fig.~\ref{reannihilation-example}.} At $T'\lesssim m_{DM}$, $Y$ becomes rapidly Boltzmann suppressed while $Y_{QSE}$ (which is Boltzmann suppressed only at $T\lesssim m_{DM}$) and $Y_{crit}$ do not change their behaviour. As a result $Y$ intercepts $Y_{QSE}$ before it intercepts $Y_{crit}$ and reannihilation occurs before freeze-out.
Only on the border line between freeze-in and reannihilation {does DM just reach} the critical value before immediately decoupling (without knowing either a sizable period of reannihilation or a sizable period of thermal equilibrium).

The range of parameters for which the reannihilation regime actually occurs is quite large, see Fig.~\ref{figure_tableGKM}. 
In particular, values of parameters which could lead to a direct detection rate compatible with, for instance, the
results of the DAMA or CoGeNT experiments (see Section 4) turn out to lie in the reannihilation phase. 
%As an example the values considered in the first panel of Fig.~\ref{reannihilation-example} are consistent with the rate observed by the CoGeNT experiment (see Section 4 for details). 
Note also that the period of reannihilation can cover several orders of magnitude in $T$.

%%%%%%%%%%%%%%%%%%%%%%%%%%%%%%%%%%%%%%%%%%%%%%%
\subsection{Phases III \& IV: Freeze-out  with hidden gauge or connector interactions}

Phase III of the Mesa in Fig.~\ref{figure_tableGKM} corresponds to the situation in which both the hidden sector and connector interactions are fast enough to thermalize 
but in such a way that the hidden sector interaction annihilation rate remains dominant in the Boltzmann equation. In other words both sectors share the same temperature, $T=T'$, but once they have thermalized, the connector not only does not lead to any further energy transfer between both sectors, but also  play no role in the freeze-out of DM, which is thus determined only by $\alpha'$. This explains the shape of the phase diagram for large values of $\alpha^\prime$  (top of the Mesa, where the relic density is independent of $\kappa$ for fixed $\alpha^\prime$).
In this case, the relic density simply scales as $\Omega_{DM}\propto 1/\langle \sigma_{annih.} v \rangle \propto m_{DM}^2/\alpha'^2$ as can be seen in Fig.~\ref{figure_tableGKM} and \ref{YDM-kappa}.
Also, in this case there is no reannihilation processes since the source term can be neglected as soon as $T\simeq T'$. 

The "phase transition" line between the reannihilation and hidden sector freeze-out regimes corresponds to the thermalization condition $\langle \sigma_{connect} v\rangle n_{eq}(T)/H|_{T\simeq m_{DM}}>1$, considered at the end of Section 2. This condition applies also for the freeze-in to connector freeze-out regime phase transition, relevant for very small values of $\alpha'$. Numerically we get 
$ \kappa \gtrsim 8\cdot 10^{-7}\cdot m_{DM}^{1/2} / (g^{eff}_*(T=m_{DM}))^{3/8}$.

Finally, if the connector parameter is so large that it does all the job of thermalizing both sectors and also dominates the DM freeze-out,  then, obviously, the hidden sector interaction becomes irrelevant, which explains the vertical behaviour of the phase IV of the Mesa in Fig.~\ref{figure_tableGKM} with a relic density which decreases as the connector increases, as usual through standard freeze-out, $\Omega_{DM}\propto 1/\langle \sigma_{annih.} v \rangle \propto m_{DM}^2/\kappa^2$ (as can also be seen in Figs.~\ref{figure_tableGKM} and \ref{YDM-kappa}).  In this case  $\alpha'$ plays little role,  apart from creating a thermal population of $\gamma^\prime$. The phase transition line between both freeze-out regimes simply stems from the condition $\langle \sigma_{connect} v \rangle > \langle \sigma_{HS}v\rangle$. 
 Numerically we get $\kappa> 4.6 \cdot 10^{2} \cdot \alpha' /\left (g^{SM}_*(T=m_{DM})\right)^{1/2}$.

Note that, for fixed values of $\alpha'$ and increasing values of $\kappa$, the reannihilation phase goes necessarily towards the hidden sector interaction freeze-out phase, rather than directly towards the connector freeze-out phase. This stems from the fact that in the reannihilation phase the hidden sector interaction is already in thermal equilibrium whereas the connector one is not. The border between the reannihilation and hidden sector freeze-out phase corresponds to the stage where the connector is just large enough to begin to thermalize, whereas the hidden sector is already in deep equilibrium. Therefore in this case the connector decouples before the hidden sector interaction does.

Note also that all phase transition lines meet at a single point. This is due to the fact that at the point where the 
II-III and I-IV line (where $\Gamma_{connect}=H$) meets the I-II line (where $\Gamma_{HS}=H$), one has necessarily $\Gamma_{connect}=\Gamma_{HS}$ which is the condition the III-IV line fulfills. At the meeting point both cross sections decouple as soon as they thermalize.

%%%%%%%%%%%%%%%%%%%%%%%%%%%%%%%%%%%%%%%%%%%%
\subsection{What if the primordial hidden sector energy density does not vanish?}

As can be seen from the above discussion, to assume an initial population of DM particles in the hidden sector, and 
to neglect the effect of any possible connector ({\em i.e.}~considering $T'/T$ as a constant),
as has been done in previous works (see Refs.~\cite{Ackerman:2008gi,Feng:2009mn,Feng:2008mu}), is justified only if the connector is very tiny, actually even smaller than in the pure freeze-in scenario.
If this is the case 
one is left with an ordinary freeze-out scenario in the hidden sector, induced by the hidden sector annihilation, but still mediated by a Hubble expansion rate that receives contributions from the abundance of both the visible and hidden sectors particle species. The relic density depends on the parameters
 $m_{e'}$, $\alpha'$ and  $\xi \equiv T'/T$. 
In the non-relativistic limit, the s-wave, dominant part of the cross section of Eq.~(\ref{sigma1}) (reversing the in and out particles in this process) gives $\langle \sigma_{annih} v\rangle \simeq \pi \alpha'^2/ m^2_{DM}$.
 For the freeze-out temperature $T_f$ we obtain:
 \begin{eqnarray}
x_f&=&\xi \cdot ln[0.038 \cdot\xi^{5/2}\langle\sigma_{annih}v\rangle m_{Pl}m_{DM}
\frac{g_{e'}}{\sqrt{g^{eff}_{*}}} c(c+2)]\nonumber\\
&&-\xi \frac{1}{2}\cdot ln\{\xi \cdot ln[0.038 \cdot \xi^{5/2}\langle\sigma_{annih}v\rangle m_{Pl}m_{DM} 
\frac{g_{e'}}{\sqrt{g^{eff}_{*}}}c(c+2)]\} \,,
\label{xf}
\end{eqnarray}
which gives
\begin{equation}
\Omega_{DM} h^2=2 \frac{1.07\times 10^9 x_f \cdot GeV^{-1}}{({g_{*s}}/\sqrt{{g^{eff}_{*}}})m_{Pl}\langle\sigma_{annih}v \rangle }.
\label{omegaeprime}
\end{equation}
where here too a factor 2 comes from the fact that DM is composed of both a particle and its antiparticle. $c$ is a numerical constant of order unity.
We get that the values $c=0.22,0.35,0.4$ give the correct abundance for $\xi=0.01,0.1,1$ respectively.
\begin{figure}[t]
\begin{center}
\vglue -.1cm
\includegraphics[width=8.80cm]{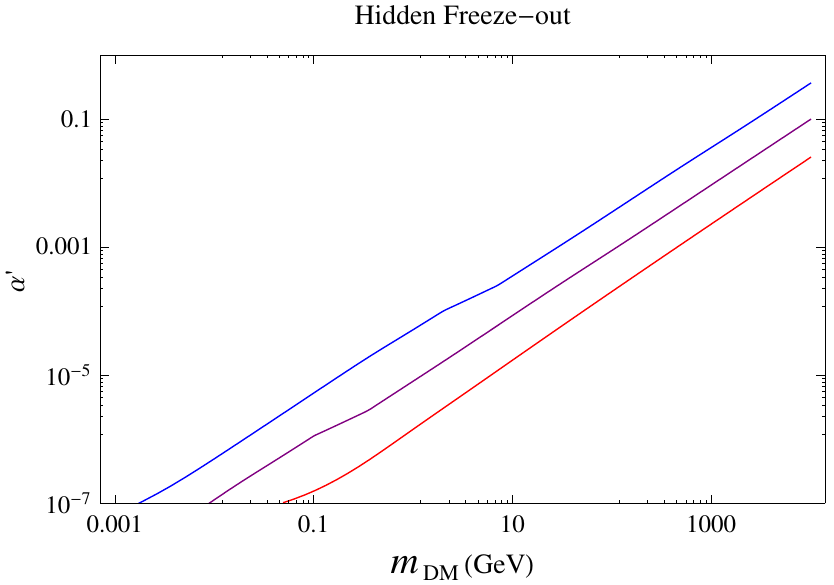}
\end{center}
\vglue -.8cm
\caption{Values of $\alpha'$ required to get the WMAP relic density as a function of $m_{DM}$ assuming  no connector between the hidden sector and the SM sector, for different values of the temperature ratio $\xi\equiv T'/T=0.01, 0.1, 1$ (bottom-up).}
 \label{fermion}
 \vspace{-3mm}
\end{figure}

Here we have  assumed that the present value of the ratio of temperatures, or better $\xi^3$, is small, so that $g_{*s}$(today)$\approx 3.91$ (which implicitly means that the Hubble expansion rate is dominated  by the visible sector particle species). In Fig.~\ref{fermion} we show the values of $m_{DM}$ and
  $\alpha^\prime$ which, from a numerical integration of the Boltzmann equation, lead to a relic density of $\Omega_{DM} h^2=0.11$. 
This plot can be understood in the following way.  
Given the Boltzmann suppression of the DM equilibrium number density around freeze-out,  $n_{DM}(T')\propto (T')^{3/2}e^{-m_{DM}/T'}$, 
the value of $m_{DM}/T'_f=x_f/\xi$ does not change much with $\xi$, even though the smaller $\xi$ is, the larger is the Hubble parameter  at $T'\simeq T'_f$ (when $\Gamma_{annih}\simeq H$): the dependence is only logarithmic. In other words $x_f=\xi x'_f$ changes linearly in $\xi$, up to a logarithmic factor, see Eq.~(\ref{xf}). But if one considers a smaller $\xi$ the value of the Hubble constant (dominated by the visible sector, {\em i.e.}~$H\sim$ const $\cdot T_f^2$) at freeze-out changes as $\xi^{-2}$ since $H(T_f)\propto T_f^2\propto T'^2_f/\xi^2$. This implies that the
number of DM particles left at freeze-out, $n_{DM}(T'_f)$, increases as $\xi^{-2}$  (since $H(T_f)=\Gamma_{annih}(T'_f)\simeq \langle \sigma_{annih} v \rangle \,n_{DM}(T'_f)$). However the entropy at freeze-out (also dominated by the visible sector, $s\sim$ const $\cdot T^3$) increases similarly as $\xi^{-3}$. Therefore $\Omega_{DM}=m_{DM} \,n_{DM}^{Today}/\rho_{Today}\simeq m_{DM} (n_{DM}(T'_f)/s(T'_f)).(s_{Today}/\rho_{Today})$ is proportional to $\xi$, in the same way as $x_f$, see Eq.~(\ref{omegaeprime}). For smaller $\xi$ this must be compensated by taking a smaller $\sigma_{annih}$, as shown in  Fig.~\ref{fermion}, so that freeze-out occurs earlier. Note that these results agree
with the one of Ref.~\cite{Feng:2009mn} up to the power 5/2 instead of 3/2 in Eq.~(\ref{xf}) (which is freeze-out prescription dependent and has a moderate numerical effect).

The condition that determines which DM source dominates the relic density today, the primordial one or the connector induced one,
reduces essentially to a condition on energy densities.
For instance, taking values of parameters leading to the observed relic density in Fig.~\ref{figure_GKMtempt}, an initial hidden sector population has essentially no influence if the initial $\rho'/\rho|_{init.}$ is smaller than the value of $\rho'/\rho|_{T_f}$ which, starting from {\em  zero} initial abundance, one obtains at the temperature where the abundance freezes (in or out). Conversely, if this condition is not satisfied, the connector can be essentially neglected for the determination of the relic density.
Therefore to know which contribution dominates, one can essentially compare the initial $\rho'/\rho$ with the value one obtains from Fig.~\ref{figure_GKMtempt}. It is {\em a priori} unlikely that both contributions would be of same magnitude.

%%%%%%%%%%%%%%%%%%%%%%%%%%%%%%%%%%%%%%%%%%%%
\section{Testing freeze-in with direct detection experiments?}

An interaction between dark matter and quarks mediated by a massless gauge boson is interesting for DM direct detection in many respects. To begin with, it allows to give an observable signal even for very tiny value of the coupling because the cross section has a collinear infrared divergence leading to a direct detection elastic cross section proportional to $1/E_r^2$, {\em i.e.} which is enhanced at the low nuclear recoil energies $E_r$. More mundane dark matter particles, which are supposed to interact with a nucleon through a massive particle (or at least heavier than the recoil energies which are typically in the few keV range, see {\em e.g.} \cite{arXiv:1006.3318} in the case of massive $Z'$), have a constant cross section on quarks. This is also a feature which is absent in other feebly interacting scenarios. A corollary of the  energy dependence of the elastic cross section is that the recoil energy spectrum is distinct from that of  generic DM candidates. In particular, this  behaviour favours experiments with a low recoil energy threshold, a feature which has been invoked as a way of reconciling for instance the DAMA/LIBRA measurements with the exclusion limits by other experiments in the framework of the mirror model for dark matter \cite{Foot:2008nw,Foot:2010hu}. Notice that the velocity distribution of dark matter in the context of mirror dark matter is distinct from that usually assumed for WIMPs. Here we are conservative and make the standard choice  of a Gaussian distribution with mean velocity that we set to $v_0 =220$ km/s at the position of the Sun, truncated at an escape velocity $v_{esc}= 550$
km/s (see for instance \cite{Jungman:1995df}).

\begin{figure}[!htb]
\centering
\includegraphics[height=9.0cm]{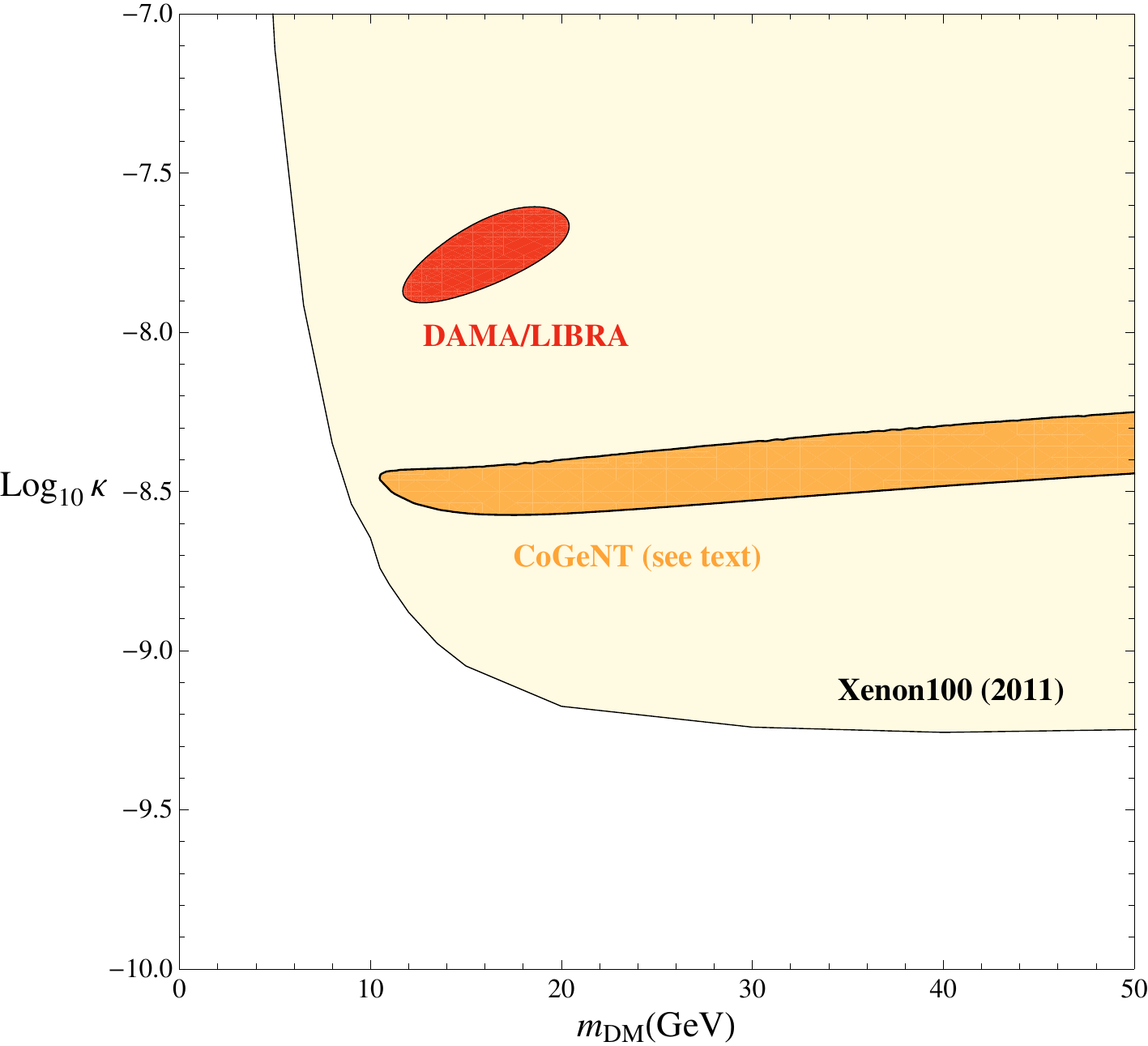}
\caption{Regions in the plane $\log_{10}(\kappa)-m_{DM}$ compatible with DAMA modulation data (red/dark grey blob) and CoGeNT total rate (orange/light grey finger)  at 99 \% C.L., together with the exclusion limit at 90 \% C.L. from Xenon100 (see text for details).}
\label{DAMA_CoGeNT}
\end{figure}

In this section we consider a few benchmark experimental results and confront the model with the data. The main conclusion is that present and future experiments may probe an interesting part of the parameter space of the model. In particular we show that Xenon1T will probe all the regimes of the model, including the freeze-in, a feature that is specific to the present scenario. Also, as already alluded to in the previous section, the reannihilation regime corresponds to candidates that may explain the DAMA \cite{Bernabei:2008yi} or CoGeNT data (see \cite{Aalseth:2010vx,Aalseth:2011wp} and also the discussion below), but these solutions are already strongly disfavoured by the current exclusion limit set by Xenon100 \cite{Aprile:2011hi}.  
The main results are summarized in Figs.~\ref{DAMA_CoGeNT} and \ref{Xenon1T}.

The spin-independent (SI) elastic cross section between dark matter and a nucleus of mass number $A$ and atomic number $Z$ is, for kinetic mixing, given by 
\begin{equation}
\frac{d\sigma}{dE_r}=\frac{1}{E_r^2 v^2}\frac{2 \pi \kappa^2 Z^2 \alpha^2}{m_A}  F_A^2(q r_A) \,,
\label{sigmaSI}
\end{equation}
with $F_A$ the Helm nucleus form factor
\begin{equation}
F_A(q r_A)=3\frac{j_1(qr_A)}{q r_A} e^{-(qs)^2/2} \,,
\end{equation}
$q=(2 m_A E_r)^{1/2}$ is the momentum transferred, $r_A=1.14 A^{1/4}$~fm, $s=0.9$~fm and 
$j_1$ is the $n=1$ spherical Bessel function of the first kind \cite{Foot:2010hu}.
The cross section ${d\sigma}/{dE_r}$ may be cast in a form which is reminiscent of the SI cross section used by direct detection experiments 
\begin{equation}
\frac{d\sigma}{d E_r}= \sigma^{n}_{eff} \frac{m_A}{2 v^2 \mu_{nDM}^2}\frac{[f_pZ +f_n(A-Z)]^2}{f_p^2} F_A^2(E_r) \,,
\end{equation}
with 
\begin{equation}
\sigma^n_{eff}=\frac{\mu_{nDM}^2}{m_A^2 E_r^2} 2 \pi \kappa^2  \alpha^2\frac{Z^2 f_p^2}{[f_p Z + f_n(A-Z)]^2} \,,
\end{equation}
and where $\mu_{nDM}$ is the nucleon/DM reduced mass and $f_{p,n}$ are the coupling strengths of $p,n$ to the mediator.
However this is of little practical use because, unlike in most models with a heavy mediator,  $\sigma^n_{eff}$ depends on $E_r$. Hence we rather present the results in the $\kappa-m_{DM}$ plane, with $\alpha'$ fixed by the requirement $\Omega_{DM} h^2=0.11$. 

\begin{figure}[!t]
\centering
\includegraphics[height=9.0cm]{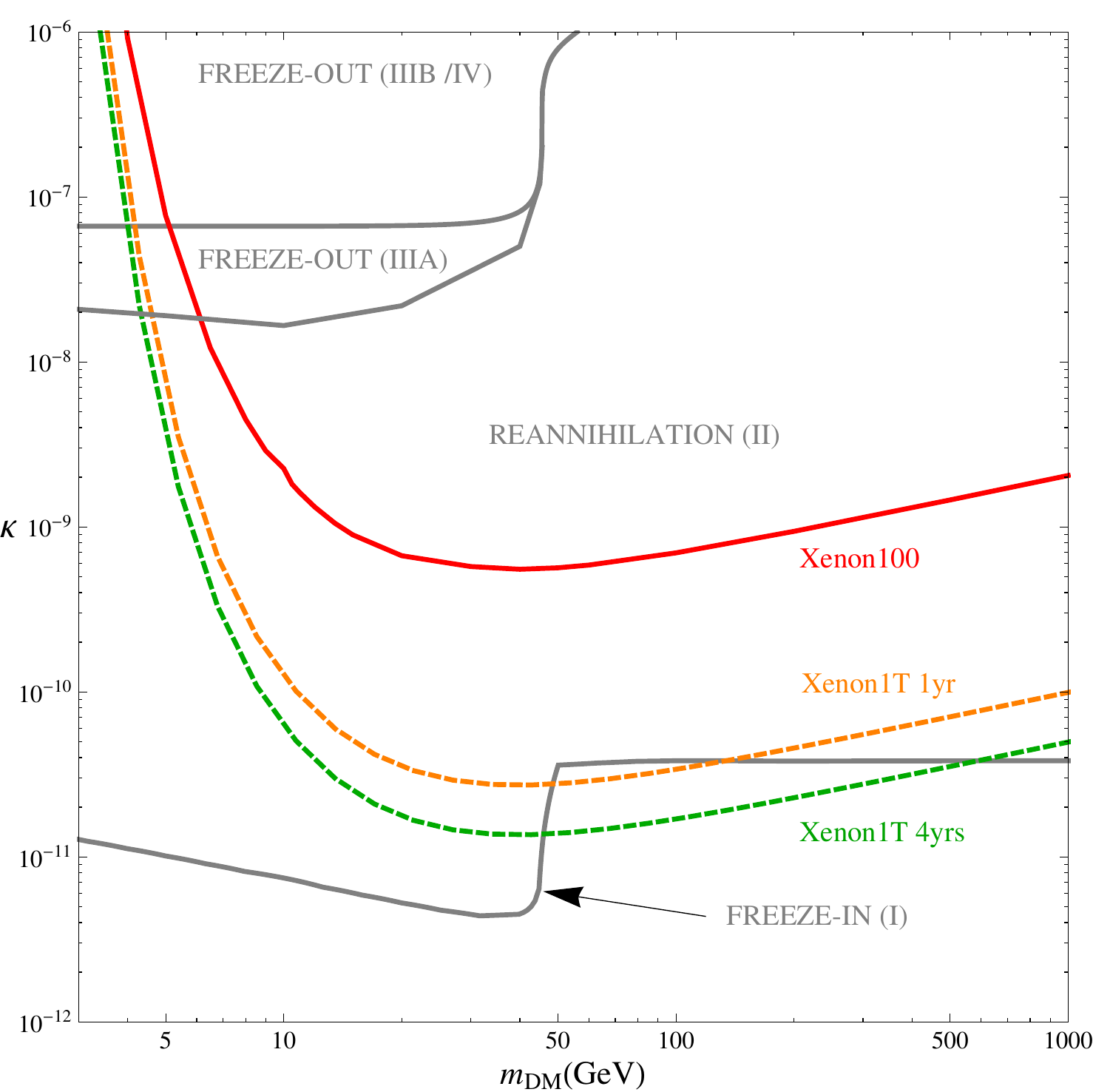}
\caption{Exclusion limits at 90 \% C.L. from the current Xenon100 data and forecast for Xenon1T for one year (dashed, orange) and 4 years (dashed, green) exposures. The light grey  line at the bottom corresponds to the pure freeze-in regime. The dip is due to the effect of the $Z$ resonance on freeze-in.
 The upper, light grey lines delimit the region where reannihilation (below it) and hidden sector freeze-out (above it) regimes are possible (imposing the DM relic density constraint). There is a further division in the regime of hidden sector freeze-out (III), depending on whether the connector thermalizes (IIIB) or not (IIIA), see Appendix D.}
\label{Xenon1T}
\end{figure}

These are given in Figs.~\ref{DAMA_CoGeNT} and \ref{Xenon1T}. In the first figure we show, for the sake of illustration, a zoom in the region of parameters compatible with the DAMA/LIBRA modulation at 99 \% C.L.\footnote{To determine the DAMA/LIBRA region, we have used the detector resolution given 
in Ref.~\cite{Fairbairn:2008gz}. Also we have assumed no channelling, and the most standard choice of quenching parameters (concretely we have taken $q_{I} = 0.09$ an $q_{Na} = 0.3$). The contour is given by a $\chi^2$ with $n=10$ degrees of freedom (best fit point corresponding to $\kappa = 1.7 \cdot 10^{-8}$ and $m_{DM}=16$~GeV, with $\chi^2 = 5.2$).}. The best fit is for $\kappa \approx 2\cdot 10^{-8}$ and $m_{DM} \sim 15$ GeV. A glance at Fig.~\ref{figure_tableGKM} reveals that the candidates consistent with both DAMA and WMAP are in the reannihilation regime II (actually close to the freeze-out regime IIIA). As Fig.~\ref{DAMA_CoGeNT} shows, they are also excluded by Xenon100.\footnote{To derive our exclusion limit we have used a simple fit to the mean scintillation efficiency extrapolated to low recoil energies, as given in Fig. 1 of \cite{Aprile:2011hi}. Concretely we have set $L_{eff}= 0$ for $E_{nr} \leq 1$ keV, $L_{eff}=-0.154 + 0.22 \ln E_{nr}$ for 2 keV $< E_{nr} \leq 3$ keV and $L_{eff} = 0.053 + 0.0353 \ln E_{nr}$ for $E_{nr} > 3$ keV but below 100 keV. As in \cite{Aprile:2011hi} we have taken into account the Poisson fluctuation in the number of photo-electrons (PE) events and have set the threshold at $3$ PE. To set the limit we have used the Monte-Carlo code for the Optimal Gap Method, provided by Yellin and described in details in \cite{Yellin:2002xd}.} For further illustration, we also show candidates corresponding to a recent, albeit preliminary, reassessment of the CoGeNT unmodulated data, as apparently a dominant part of the signal may at the end be attributed to background \cite{CollarTaup2011}.\footnote{See the talk of CoGeNT at Taup2011. A substantial fraction of the original (unmodulated) CoGeNT signal \cite{Aalseth:2010vx,Aalseth:2011wp} is now being attributed to background, so that the signal is much reduced, and also less peaked at small recoil energies.
% (see figure on page 20 of the talk). 
The residual signal is no more excluded by the CDMS low-energy analysis \cite{Ahmed:2010wy}, but it also moves away from the preferred DAMA region, making it more difficult to reconcile both experimental results, which is anyway not our purpose in the present work. Moreover, all the CoGeNT (tentatively reevaluated data) are excluded by Xenon100.} As our purpose here is not to explain/reconcile the current direct detection experiments, we do not refer to the recent CRESST and modulated CoGeNT data.

The bottom-line of this section is that currently, for $m_{DM}>
\hbox{few}$~GeV, both freeze-out regimes are excluded, whereas the
reannihilation regime is strongly constrained by Xenon100. The freeze-in
regime is allowed for any DM mass. As for the future, we refer to the Xenon-1T
experiment. Assuming a {one ton-one year exposure, and no  event,
the orange dashed curve of Fig.~\ref{Xenon1T} gives the sensitivity reach of Xenon1T in the
$\kappa-m_{DM}$ plane\footnote{For Xenon1T, we have applied the same
  prescriptions as for Xenon100 regarding scintillation efficiency, the
  Poisson statistics for low PE and a threshold at 3 PE.}. Interestingly, this
experiment may probe a large fraction of the reannihilation regime. 
Furthermore, it may also be possible to test the freeze-in regime. For the sake of illustration, we draw the exclusion limit that would correspond to an exposure of 1 year (orange dashed line) and 4 years respectively (green dashed line). In the first case, it could be possible to probe the freeze-in mechanism for a DM mass range between about 45 GeV and about 100 GeV. With 4 years, the range extend to a candidate with a mass of about 500 GeV. 

Positive signals with a recoil energy spectrum that is in accordance with a $1/E_r^2$ scattering cross section on nuclei would allow to distinguish this model from a more standard DM candidate, which usually predicts a constant cross-section.
{In principle, it should be possible to distinguish this model from mirror models, which {\em a priori} display the same $1/E_r^2$ dependence (see for instance \cite{McDermott:2011hx}), but which have a distinct velocity distribution and, in general, a multi-component halo of dark matter with particles masses in the few GeV range.

%%%%%%%%%%%%%%%%%%%%%%%%%%%%%%%%%%%%%%%%%%%%
\section{Compatibility with cosmological constraints}

Various cosmological constraints on DM particles interacting with massless gauge bosons have been considered in the literature \cite{Ackerman:2008gi,Feng:2009mn,Feng:2008mu,McDermott:2010pa,Berezhiani:2000gw,Berezhiani:2008gi,Jaeckel:2010ni}. In this section we review the constraints that are the most relevant for our purpose, referring to the original literature for more detailed discussions if necessary, and extend some of the constraints to the case of a massive hidden photon\footnote{Notice that if we break the $U(1)'$ symmetry (either spontaneously or through the Stueckelberg mechanism), we lift the degeneracy between the visible and the hidden photons. In this section, we work directly in the basis of mass eigenstates.}. We show in particular how the stringent galactic ellipticity constraint gets considerably relaxed when one considers a slightly massive, rather than a massless, $\gamma'$.
\bigskip

\noindent\underline{Primordial nucleosynthesis constraints:} To begin with, let us dispose right away of the most basic constraint, based on Big Bang nucleosynthesis (BBN) and the increase of the expansion rate of the Universe in presence of additional relativistic degrees of freedom around $T \sim 1$ MeV (see for instance \cite{Berezhiani:2008gi}). This is usually phrased in terms of limits on the number of neutrino families \cite{Nakamura:2010zzi,Cyburt:2004yc} which, at 95 \% C.L., we may take to be $\Delta N_\nu \leq 1.4$. In our simple model, there are only two extra degrees of freedom around the time of BBN (expect for the marginal case -{\em i.e.} in our work-  of a light DM candidate, $m_{DM} \leq m_e$), corresponding to the polarization states of the hidden photon, so that 
\begin{equation}
\frac{\rho^\prime}{\rho} = \frac{2}{10.75}\, \xi^4 \equiv {0.16}\, \Delta N_\nu
\end{equation}
where, we recall, $\xi = T^\prime/T$. BBN nucleosynthesis gives then $\xi \leq 1.05$.
If the DM is lighter than 1 MeV, the bound is a bit stronger $\xi \leq 0.8$
but still irrelevant for our purpose. A CMB constraint on the
  number of neutrino families $N_\nu \leq 4$ leads to a similar conclusion \cite{Berezhiani:2000gw}. The Planck
  satellite experiment is expected to narrow down the range to $\Delta N_\nu \leq 0.2$.
 
\bigskip
\noindent\underline{Galactic dynamics constraints:}
As one may expect, the most important  constraints on our scenario rest on the (milli-)charged character of the DM and the associated long range interaction. 
In particular, this interaction may affect the way dark matter clusters in the universe, with potential consequences for the formation of clusters of galaxies, including the dynamics of the so-called Bullet Cluster, and modifications of the dark matter halo profile in galaxies. It turns out that the latter effect gives the strongest constraint on the interactions between DM particles, and so on $\alpha^\prime$ \cite{Feng:2009mn}.\footnote{The CMB puts very mild constraints on our model. For all practical purposes our candidates behave as cold dark matter and are essentially decoupled at the time of matter-radiation equality (this is not the same for the so-called Mirror Dark Matter, see for instance \cite{Berezhiani:2008gi}. }  A long range force may affect  the shape of dark matter halo in galaxies and clusters of galaxies. In particular they may erase the observed/inferred tri-axial elliptic shape of halo of elliptic galaxies. They may also lead to the formation  of a cored (a central region with almost constant DM density) instead of a cuspy profile. The ellipticity of elliptic galaxies gives an upper bound on $\alpha'$ which is quite stringent \cite{Feng:2009mn}. In this reference, the bound is estimated to be given approximately by $\alpha' \lesssim 10^{-7}\,(m_{DM}/\mbox{\rm GeV})^{3/2}$ (modulo an extra logarithmic dependence on $m_{DM}$ and $\alpha^\prime$). Comparing this constraint with the values of $\alpha^\prime$ reported in  Figs.~\ref{figure_tableGKM} and \ref{figure_GKMdd} (see also Fig.~\ref{ellipticity}), we observe that the freeze-in regime, as well as the connector freeze-out regime is always allowed, since they work with any values of $\alpha'$ smaller than a given value which depends on the DM mass. {But the reannihilation and freeze-out (see Figs.~3, 7 and Fig.1 of Ref.~\cite{Feng:2009mn}) in the hidden sector regimes both require a fairly massive DM candidate, estimated to be $m_{DM}\gtrsim 1$~TeV and $m_{DM}\gtrsim 10$~TeV respectively.}
We notice however that the ellipticity constraint derived in \cite{Feng:2009mn} is tentative and that more work could change the limits on $\alpha^\prime$ and $m_{DM}$. For instance, if we relax the ellipticity bound on $\alpha'$ by a factor of 3, reannihilation and freeze-out in the hidden sector are viable regimes for $m_{DM}\gtrsim 100$~GeV and $m_{DM}\gtrsim 1$~TeV respectively. 

\begin{figure}[!htb]
\centering
\includegraphics[height=9cm]{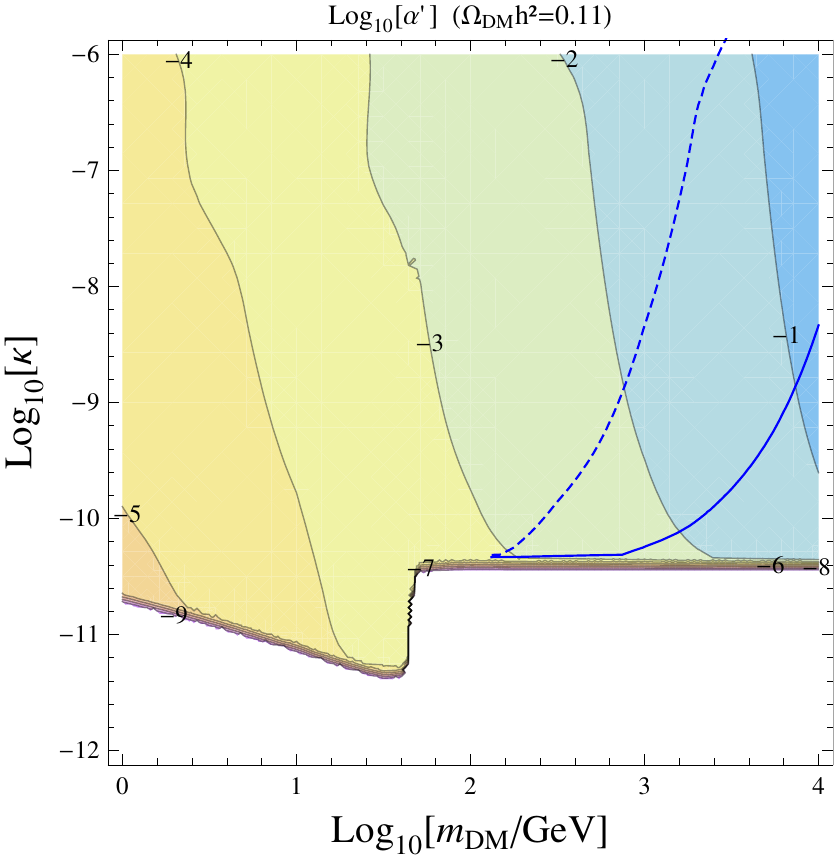}
\caption{Upper limit on $\kappa$ obtained from the ellipticity bound (solid line) on $\alpha'$ imposing that $\Omega_{DM}h^2=0.11$, as a function of $m_{DM}$. The dashed line gives the bound we get if we relax the ellipticity bound on $\alpha'$ by a factor 3. The lower line gives the value of $\kappa$ required along the freeze-in regime. This regime, ellipticity constraint included, allows any value of $m_{DM}$ since it works with any arbitrary small value of $\alpha'$. The contour lines give the value of $\alpha'$ one gets imposing the relic density constraint, in accordance with Fig.~\ref{figure_tableGKM}.}
\label{figure_GKMdd}
\end{figure}

Similarly the ellipticity of  DM halos may put bounds directly on the coupling between DM and ordinary matter, {\em i.e.}~on $\kappa$. From Fig.~1 in \cite{McDermott:2010pa}, we see that the constraint is mild, $\kappa \lesssim 10^{-2}$ for $m_{DM}~\sim~10$ GeV. Other effects~\cite{McDermott:2010pa}, in particular the requirement of decoupling at recombination are more stringent, $\kappa \lesssim 10^{-6}$ again around $m_{DM} \sim 10$ GeV. However, these constraints are weaker than the exclusion limits set by direct detection experiments (\cite{McDermott:2010pa} for CDMS-Si and  Fig.~\ref{Xenon1T} of the present manuscript for Xenon100).

\bigskip
\noindent\underline{Possible depletion of DM in the Galaxy:}
 Direct detection constraints rest  on the assumption that the energy density of DM in the vicinity of the Sun is $\rho_{DM} \sim 0.3$ GeV/cm$^{-3}$. 
This might not be the case for milli-charged particles, as has been emphasized in \cite{Chuzhoy:2008zy} and further discussed in \cite{McDermott:2010pa}  (see also \cite{SanchezSalcedo:2010ev}), as magnetic fields in the Galaxy  may lead to a depletion of the DM abundance at the position of the Sun \cite{Chuzhoy:2008zy}. To be sure, this effect does not {\em per se} exclude a candidate. Rather it puts limits on its potential relevance for DM search through direct and, implicitly, indirect detection experiments.  Three effects are {\em a priori} relevant here. First the large scale  magnetic field of the Galaxy may prevent the DM particles penetrating the Galaxy plane if their gyromagnetic radius is smaller than the height of the disc. Concretely, for roughly
\begin{equation}
\kappa \gtrsim 6 \cdot 10^{-13} \left({m_{DM}\over {\rm GeV}} \right) \left({ v_{DM}\over 300 \mbox{\rm km/s}}\right)  \left({ 5 \mu G\over B}\right) \,,
\end{equation}
a charged DM particle from the halo can not penetrate the Galactic disk \cite{McDermott:2010pa}. Here $B$ is the mean, large-scale magnetic field, and $v_{DM}$ the velocity of DM. {\em A priori} this may be counterbalanced by diffusion of DM by the small-scale, turbulent component of the Galactic field. The diffusion time scales like
\begin{equation}
\label{tau_diff}
\tau_{\rm diff} \approx 10^{18} \kappa \left({H_d\over 100 \mbox{\rm pc}}\right)^2 \left({\mbox{\rm GeV}\over m_{DM}}\right)\left({300 \mbox{\rm km/s}\over v_{DM}}\right)^2 \left({B \over 5 \mu G}\right) \; \mbox{\rm yrs} \,.
\end{equation}
with $H_d$ the height of the Galactic disk. If we compare this with the propagation time scale is $H_d/v_{DM} \sim 4\cdot 10^5$ yrs, we get that diffusion is effective compared to magnetic shielding only if 
$\kappa \lesssim 4 \cdot 10^{-13}$, a very small value which puts the model beyond the reach of direct detection experiment. One alternative is to assume that  DM is present {\em ab initio} within the Galactic disk. However, this component of DM  may be expelled from the Galactic disk by Fermi acceleration in supernovae shocks \cite{Chuzhoy:2008zy,McDermott:2010pa}. The time scale for DM acceleration by supernovae shock waves is estimated to be of the order of $\tau_{\rm acc} \sim 10^7$ years \cite{McDermott:2010pa}. Expulsion is efficient if either the time scale for energy loss, $\tau_{\rm e.l.}$, or the diffusion time scale $\tau_{\rm diff}$, is large compared to the acceleration time. The former gives $\kappa \lesssim 3.4\cdot  10^{-4} (m_{DM}/\mbox{\rm GeV})^{1/2}$, while the latter requires $\kappa \gtrsim 10^{-11} (m_{DM}/~$GeV), using the fiducial values of the parameters in Eq.(\ref{tau_diff}). Hence, at the end of the day, we must assume that
\begin{equation}
\kappa \lesssim 10^{-11} \left({m_{DM}\over \mbox{\rm GeV}}\right) \,,
\end{equation}
 is required in order to match the expected local energy density, $\rho_{DM} \sim 0.3$ GeV/cm$^3$. A glance at Fig.~\ref{figure_GKMdd} reveals that this condition only concerns candidates with $m_{DM}\lesssim~2$~GeV. 

\bigskip
\noindent\underline{Summary of cosmological constraints for a massless hidden photon:}
Taken literally  {the constraints discussed above} point towards freeze-in as the only possible {testable} ({\em i.e} possibly no screening by magnetic fields) or {viable} (compatible with, say, the ellipticity bound) scenario, except for values of $m_{DM}$ larger than a few hundreds GeV, for which  the reannihilation and hidden sector freeze-out regime scenarios are also possible.
This also excludes the possibility that the model would be responsible for the DAMA or CoGeNT data (regardless of the exclusion set by Xenon100). These require  $m_{DM}$ in the few GeV range, and a $\kappa \sim 10^{-8}$, values that can lead to the observed relic density only for $\alpha'\sim10^{-4}$, which is much larger than what is allowed by the ellipticity bound, $\alpha'< 8 \cdot 10^{-6}$. In other words, taking into account the ellipticity bound, such candidates would have a relic density about two order of magnitudes larger than what is observed.

\bigskip
\noindent\underline{Ellipticity bound and depletion for a massive hidden photon:}
Note that the bounds of Ref.~\cite{Ackerman:2008gi,Feng:2009mn,Feng:2008mu} apply exclusively for a massless mediator.  
One simple way to relax the ellipticity bound and/or the impact of magnetic fields, and so expand the scope of our scenario, is to assume that the $U(1)'$ group is spontaneously broken in such a way  that the hidden photon becomes a light $\gamma'$.\footnote{The fact that the $U(1)'$ gauge group is spontaneously broken does not affect the stability of the lightest fermion of the hidden sector. But still, even if broken, it forbids a possible Yukawa coupling to a SM lepton doublet and a Higgs doublet, which otherwise would make the DM particle unstable.} 
To assume that the $\gamma'$ is massive does not affect either the relic density analysis above as long as $m_{\gamma'}< m_{DM}$, or the direct detection analysis above, as long as $m_{\gamma'}\lesssim 1$~MeV. But it does relax largely the cosmological constraints just discussed.
To show that, we adapt the analysis of \cite{Feng:2009mn} to the case of a massive hidden 
photon.\footnote{From the cross section for the scattering  between DM particles interacting through a massive photon,
$$
{d\sigma\over d\Omega} = {4 \alpha^{\prime 2} m_{DM}^2\over \left( 4 m^2_{DM}  v^2 \sin^2(\theta/2) + m_{\gamma^\prime}^2\right)^2},
$$
 the rate for energy loss in the Born approximation is given by 
$$
\dot E = 
{4 \sqrt{\pi}\alpha^{\prime 2} \rho_{DM} \over  m_{DM}^2 v_0^3} \int dv v e^{-v^2/v_0^2} 
\left[
\log\left(4 m_{DM}^2 v^2\sin^2(\theta/2) + m_{\gamma^\prime}^2\right) 
+ {m_{\gamma^\prime}^2\over  4 m_{DM}^2 v^2\sin^2(\theta/2) +m_{\gamma^\prime}^2}
\right]_{\theta= \theta_{min}}^{\theta =\pi} \,,
$$
where  the minimum scattering angle $\theta_{min}$ is set by Debye screening in the hidden plasma \cite{Feng:2009mn}.
The integral over DM velocity $v$ is peaked at $v=v_0$, so that for a massless $\gamma^\prime$, this reduces to the result of \cite{Feng:2009mn}, with the familiar divergence from Rutherford scattering, $\dot E \propto - \log(\sin^2(\theta_{min}/2)$. The divergence is cut-off for  a massive $\gamma^\prime$, $\dot E \propto - \log(m_{\gamma^\prime}/v m_{DM})$ but the dependence is essentially logarithmic for small $m_{\gamma^\prime}$, so the impact is moderate unless $m_{\gamma^\prime} \gtrsim $ few keV (see Fig.~\ref{ellipticity}). The turnover behaviour of the ellipticity bound is related to the change of sign of the log at for $m_{\gamma^\prime} \sim m_{DM} v_0$. 
}
Our results are shown in Fig.~\ref{ellipticity}.\footnote{{One caveat is that our calculations are done in the Born approximation. Born approximation is only reliable for ``weak coupling'', {\em ie} $\alpha^\prime m_{\gamma^\prime}/m_{DM} v^2 \lesssim 1$ (see \cite{Feng:2009hw}), a condition which does not hold over the whole parameter range we consider. In particular, for $m_{\gamma^\prime} = 1$ MeV, our calculations are reliable for $m_{DM} \gtrsim 1 $GeV, while for $m_{\gamma^\prime} = 10$ MeV, weak coupling approximation is only valid for a more restricted range, $10$ GeV $\lesssim m_{DM} \lesssim 300 $ GeV. Although the general trend of giving a mass to the hidden photon should be clear from the figure, one should keep in mind that, in the strong coupling regime, the scattering amplitude may possibly get a resonant enhancement, leading for some parameters to  more stringent constraints than those shown in the figure.}} As an example we may infer that, for $m_{DM} \sim 10$ GeV, taking $m_{\gamma^\prime} \sim $ 1 MeV allows to take $\alpha^\prime \sim 10^{-4}$, which puts such a  candidate in the reannihilation regime for the production of the DM abundance (see Fig.~\ref{figure_tableGKM}). 
Note that the red curve in Fig.~\ref{ellipticity} corresponds to a massless hidden photon and has to be compared with Fig.~1 of \cite{Feng:2009mn}. The DM candidates below the red curve are excluded. Our bound is compatible, but slightly milder than in \cite{Feng:2009mn}. This, we believe, is  because we have taken an average DM energy density $\rho_{DM} \sim 1 $ GeV/cm$^3$ and velocity dispersion $v_0 = 300$ km/s, rather than profiles for these parameters as in \cite{Feng:2009mn}. However this approximation serves our purpose which is to illustrate how the constraint from ellipticity changes for a massive hidden photon. 
Notice also that the constraint from interaction with the magnetic fields in the galaxy and supernovae shocks quickly drops if the $\gamma'$ is massive.
This stems from the fact that the correlation length $m_{\gamma^\prime}^{-1} \lesssim 1$ MeV$^{-1}$ of the $\gamma^\prime$ is much smaller than the typical particle separation length, ${\cal O}(1$~cm). 

To show how the mass of the $\gamma'$ affects the direct detection analysis above,
let us consider for the massive case the differential cross section for scattering on a nuclei, substituting in Eq.(\ref{sigmaSI}) \footnote{See also \cite{Fornengo:2011sz,Schwetz:2011xm} for recent work on long range forces in direct detection.}
\begin{equation}
\left.{d\sigma\over d E_R}\right\vert_{m_{\gamma^\prime}=0} \propto {1\over E_R^2} \longrightarrow {d\sigma\over d E_R} \propto {1\over (E_R + m_{\gamma^\prime}^2/2 m_A)^2} \,.
\end{equation}
For $A=$Xe and $E_R$ in the few keV range, the mass of the hidden photon becomes significant only for $m_{\gamma^\prime} \sim \sqrt{m_{Xe} E_R} \gtrsim 50$ MeV. From Fig.~\ref{Xenon1T}, we see that a candidate with $m_{DM} \sim 10$ GeV and $\kappa \sim 10^{-9}$ is below the Xenon100 exclusion limit, but within reach of Xenon1T. Hence, light mediators open the possibility to also probe the reannihilation regime of the model, while being consistent with laboratory and astrophysical constraints.

\begin{figure}[!t]
\centering
\includegraphics[height=9cm]{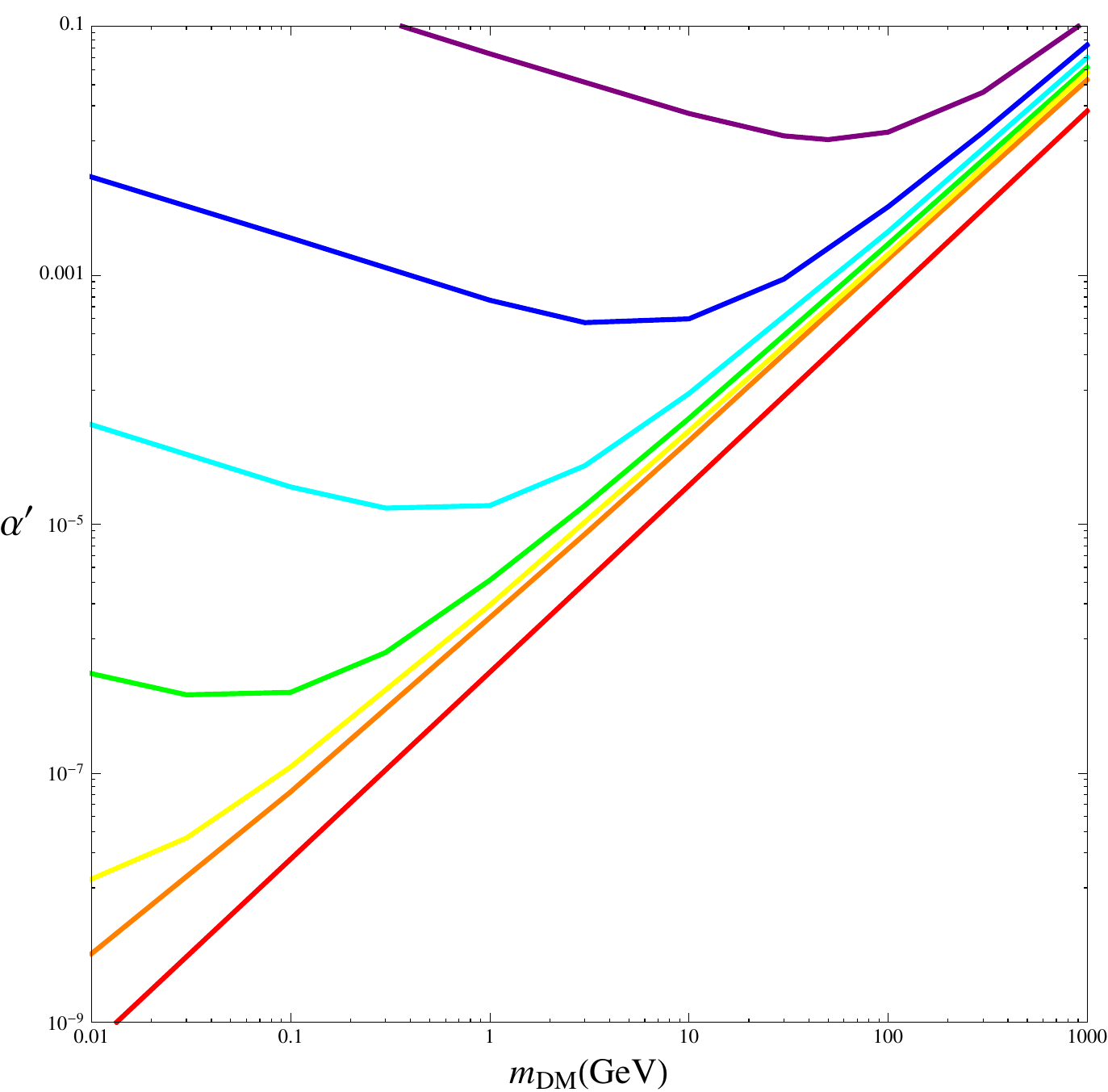}
\caption{Ellipticity bounds (see text) in the $\alpha^\prime-m_{DM}$ plane, for interactions through a massless hidden photon (lower line), and $m_{\gamma^\prime}=1$ keV, 10 keV, 100 keV, 1 MeV, 10 MeV, 100 MeV (bottom-up).}
\label{ellipticity}
\end{figure}

\bigskip
\noindent\underline{Other constraints for massive hidden photons:}
There are numerous astrophysical, cosmological and laboratory constraints on massive hidden photons and their coupling to ordinary matter. Here we may refer to Fig.~10 of Ref.~\cite{Redondo:2010dp} (see also \cite{Jaeckel:2010ni}) where constraints on the mixing parameter (our $\epsilon$, their $\chi$) and hidden photon mass are summarized.\footnote{In the same reference \cite{Redondo:2010dp} one may find in Fig.9 various constraints on milli-charged particles summarized. There $Q\equiv \kappa$.} The range of interest for our discussion corresponds to $m_{\gamma^\prime} \gtrsim 100$ keV. Below this value, there are strong constraints on the mixing parameter from constraints on energy loss in stars. For 1 MeV $ \lesssim m_{\gamma^\prime} \lesssim $ 100 MeV, there is a constraint on $\epsilon \lesssim 10^{-8}$ from the duration of the SN1987A neutrino burst, which is consistent with the values of $\kappa~\approx~\epsilon \sqrt{\alpha^\prime/\alpha} \sim 10^{-9}$ and $\alpha^\prime \sim 10^{-4}$ required in  the reannihilation regime. Above $m_{\gamma^\prime} \sim $ 100 MeV, the constraints are much milder, but, in our opinion, we may not really speak of a light $\gamma^\prime$ anymore.\footnote{Lighter hidden photons $m_{\gamma^\prime} \lesssim 1$ eV are also less constrained but they have little impact on the ellipticity bound.} 
If the hidden photon is in the MeV range, one should worry about constraints from nucleosynthesis, possibly ruining the production of light elements \cite{Pospelov:2007mp}. For the sake of illustration, let us consider again the reannihilation scenario for $m_{DM} \sim $ 10 GeV, and so $\epsilon \sim 10^{-8}$. To begin with, we may be confident that the Universe is radiation dominated at the time of BBN, as the energy density in the DM and in the hidden photons are comparable around the MeV scale. Furthermore, $T^\prime \ll  T$ (for instance $T^\prime \approx T/10 $ for $m_{DM} \sim 10$ GeV and $\alpha^\prime \sim 10^{-4}$), so that a massive photon in the MeV mass range becomes non-relativistic way before the epoch of BBN. Then, for $m_{\gamma^\prime}$ in the MeV range, the only relevant decay is into $e^+-e^-$ pairs, with a decay rate $\Gamma \sim \alpha \epsilon^2 m_{\gamma^\prime}$ (see for instance \cite{Batell:2009yf}), the lifetime of a hidden photon is ${\cal O}(10^{-3} s)$, and so takes place well before nucleosynthesis and thus is harmless \cite{Pospelov:2007mp}. One last potentially interesting consequence is for indirect detection. While annihilation in visible sector degrees of freedom is suppressed by the mixing angle in the case of massless hidden photons, in the massive case, DM-antiDM pairs may annihilate in a pair of on-shell massive hidden photons, which subsequently decay into electron/positron pairs with energy $m_{DM}/2$  (the secluded scenario \cite{Pospelov:2007mp}). In the reannihilation and annihilation in the hidden sector regimes, the annihilation cross section is canonical $\sigma v \sim 10^{-26}$ cm$^3 \cdot s^{-1}$.  For heavy DM candidates, a substantial boost is necessary to lead to observable consequences, but for lighter ones, $m_{DM} \sim $ few GeV, there may be strong constraints, for instance from synchrotron radiation (see for instance \cite{Boehm:2010kg} and \cite{Fornengo:2011iq}). 

\bigskip\noindent\underline{Hidden atoms:}
Another easy way to evade the cosmological and astrophysical bounds for
a massless hidden photon is to consider that there are more than one
species of particle (for instance two, say $e_{1,2}^\prime$) in the hidden sector, which may combine into neutral states after freeze-out. Both particles would be stable because QED conserves flavour. Positronium-like bound states would still annihilate, but the bound state $e'_{1,2}-\bar{e}'_{2,1}$ ones would still be stable. If the binding energy is in the keV range, direct detection would probe the constituents $e_{1,2}$ particles, but we have checked that the value of $\alpha^\prime$ required to have recombination of $e_{1,2}$ in the early universe is much larger than the ones required to have the right relic abundance, so we do not envisage this option as a viable scenario within our (minimal) framework.

 %%%%%%%%%%%%%%%%%%%%%%%%%%%%%%%%%%%%%%%%%%%%%%%%%%

\section{Creating a DM hidden sector through the Higgs portal}

Besides the kinetic mixing portal, there is {another way} to connect two sectors charged under different gauge groups: through the Higgs portal. Recently it has been considered in many different contexts.
It is interesting to see how, in the same way as above, a DM hidden sector could be generated from the SM sector through this portal. {In particular, we may wonder whether there is some universality in the structure we have seen emerging in the previous sections (the Mesa diagram)? This will turn out to be the case, even though in details there are some new phenomena related to the fact that the mediator, here the Higgs, may be heavier than the DM candidate}. 

To this end we consider the simple case where the DM consists of
 a scalar charged under an extra $U(1)'$, in which case it can couple
 to the SM through the Higgs portal interaction, with coupling $\lambda_m$,
\begin{equation}
{\cal L} \owns D_\mu^\prime \phi^\dagger D^{\prime\mu}\phi -\lambda_m \phi \phi^\dagger H^\dagger H -\mu^2_{\phi} \phi \phi^\dagger -\lambda_{\phi}({\phi} \phi^\dagger)^2- \mu^2 H^\dagger H -\lambda (H^\dagger H)^2 \,,
\label{LHiggs}
\end{equation}
with $H=(H^+,({h}+v)/\sqrt{2})$. 
We assume that the scalar has no {\it vev}
so that the $U(1)'$ gauge symmetry is unbroken and 
the scalar
 is stable.
Neglecting the possibility of a
kinetic mixing portal,\footnote{{In this scenario, we could consider the possibility to have simultaneously the kinetic mixing and the Higgs portals. However, in practice, unless the parameters of both portals are finely adjusted, only one portal should be relevant in a given regime. As our purpose here is more to confront a massless (or essentially massless) mediator to a heavy one, we do not consider further this possibility. }} the processes which can create the $\phi$ from the SM sector are
\begin{eqnarray}
\sum_i \Gamma(h \rightarrow \phi_i^2)&=&\sum_i \frac{\lambda_m^2v^2}{8E_h\pi}\sqrt{1-\frac{4m^2_{\phi_i}}{m_h^2}}\,,\\
\sum_i \sigma(W^+ W^- \rightarrow \phi_i^2)&=&\sum_i \frac{\lambda_m^2}{72\pi}\sqrt{\frac{s-m^2_{\phi_i}}{s-m^2_W}}\frac{s^2-4m_W^2s+12m_W^4}{s(s-m_h^2)^2}\,,\\
\sum_i \sigma(ZZ \rightarrow \phi_i^2)&=&\sum_i \frac{\lambda_m^2}{72\pi}\sqrt{\frac{s-m^2_{\phi_i}}{s-m^2_Z}}\frac{s^2-4m_Z^2s+12m_Z^4}{s(s-m_h^2)^2}\,,\\
\sum_i \sigma(f\bar{f} \rightarrow \phi_i^2)&=&\sum_i \frac{\lambda_m^2}{16\pi}\frac{m_f^2\sqrt{(s-4m_{\phi_i})(s-4m_f^2)}}{s(s-m_h^2)^2}\,,\\
\sum_i \sigma(h h \rightarrow \phi_i^2)&=&\sum_i \frac{\lambda_m^2}{8\pi}\frac{1}{s}\sqrt{\frac{s-4m_{\phi_i}^2}{s-4m_h^2}}(1+\frac{3m_h^2}{s-m_h^2})^2\,.
\end{eqnarray}
where $i=1,...,n$ denotes the number of degrees of freedom of $\phi$.
In the following we will consider the Boltzmann equation for the {charged
scalar} of Eq.~(\ref{LHiggs}), $n=2$. Since the Boltzmann equation
is the same for all real components of $\phi$, the DM relic density for
another value of $n$ can be obtained by multiplying the DM relic density by
$n/2$ (up to a small logarithmic dependence in the number of degrees of freedom).

\begin{figure}[!htb]
\centering
\includegraphics[height=6.6cm]{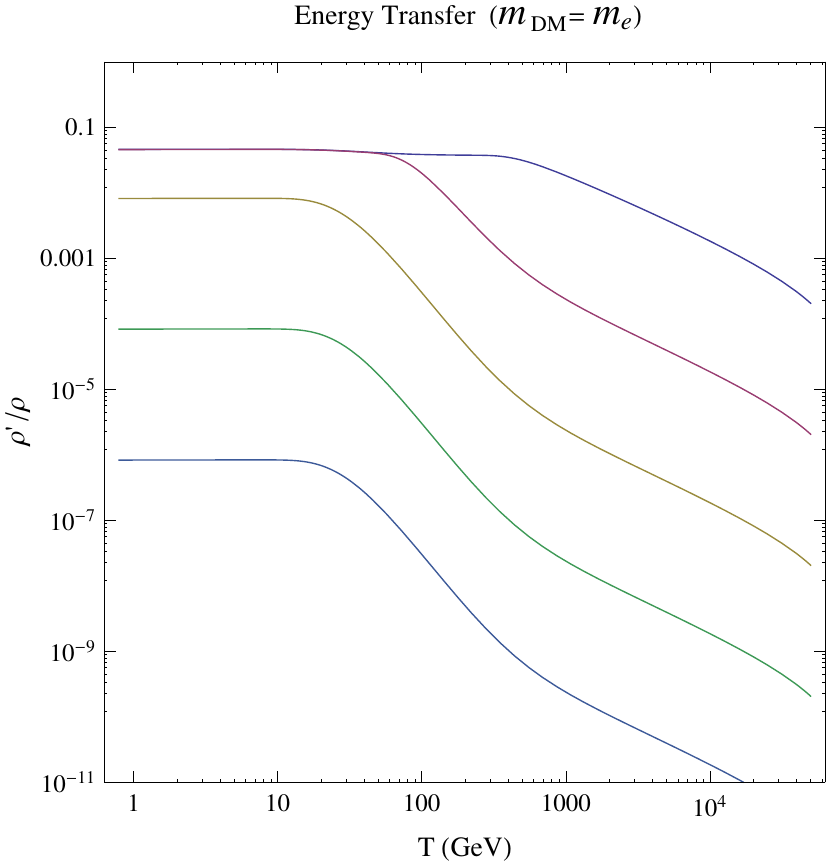}\includegraphics[height=6.6cm]{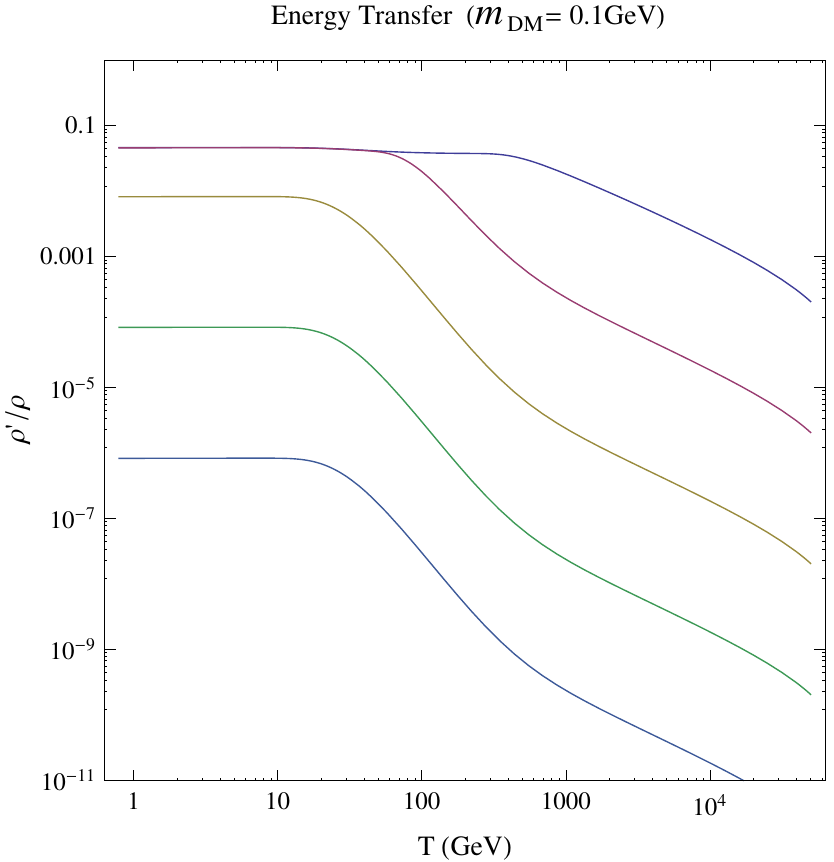}\\
\includegraphics[height=6.6cm]{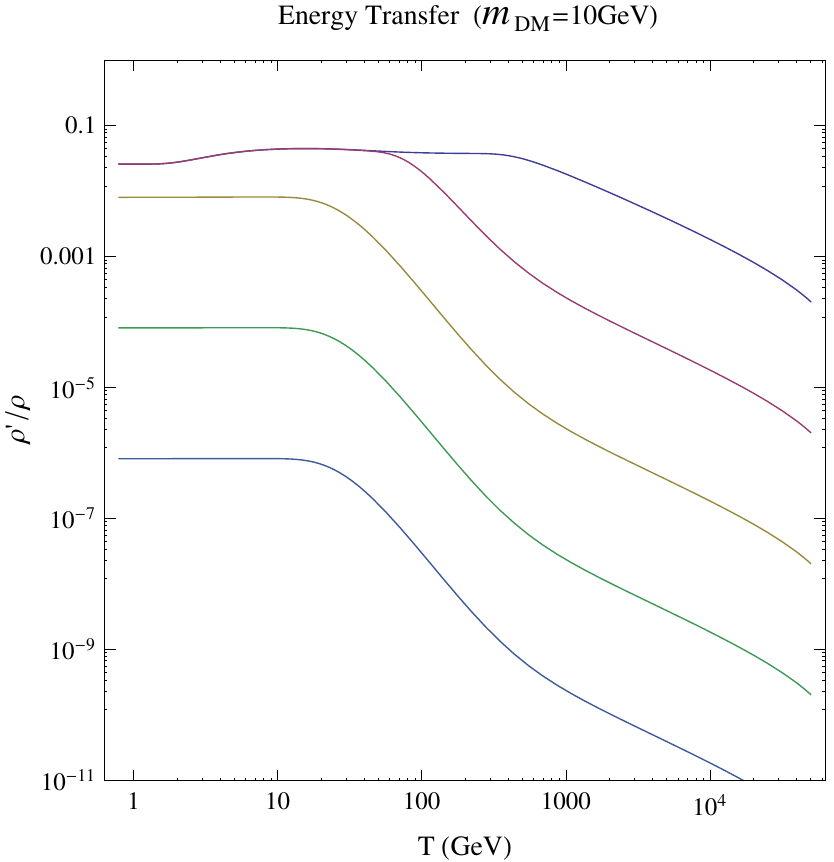}\includegraphics[height=6.6cm]{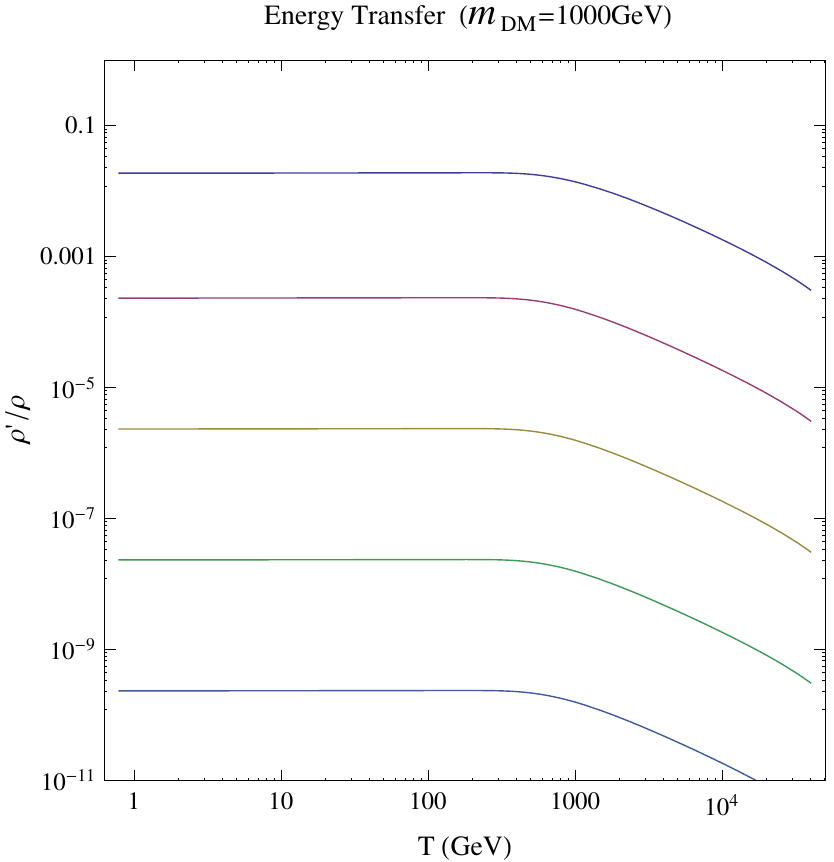}
\caption{Evolution of the ratio of the visible and hidden sectors energy densities, for a range of  connector parameter, $\lambda_m=10^{-6,-7,-8,-9,-10}$ (from up to down), and for various DM masses.}
\label{HP}
\end{figure}

The main differences between the kinetic mixing and Higgs portal cases come
from the fact that in the later case the mediator is quite massive,
$m_h>114.4$~GeV.\footnote{In the following for all numerical results we take the value $m_h=120$~GeV.} The consequences are two-folds, (a) DM can be also created by the decay of the mediator and (b) the various production channels are suppressed at small temperatures, either Boltzmann suppressed if the Higgs boson is real in the process, or by the mass of the Higgs boson at the fourth power if the Higgs boson is virtual. As we will see this implies important differences between 
the kinetic and Higgs portals but, still, in both cases 
there is a characteristic "Mesa" shape phase diagram.

\begin{figure}[!t]
\centering
\includegraphics[height=7.2cm]{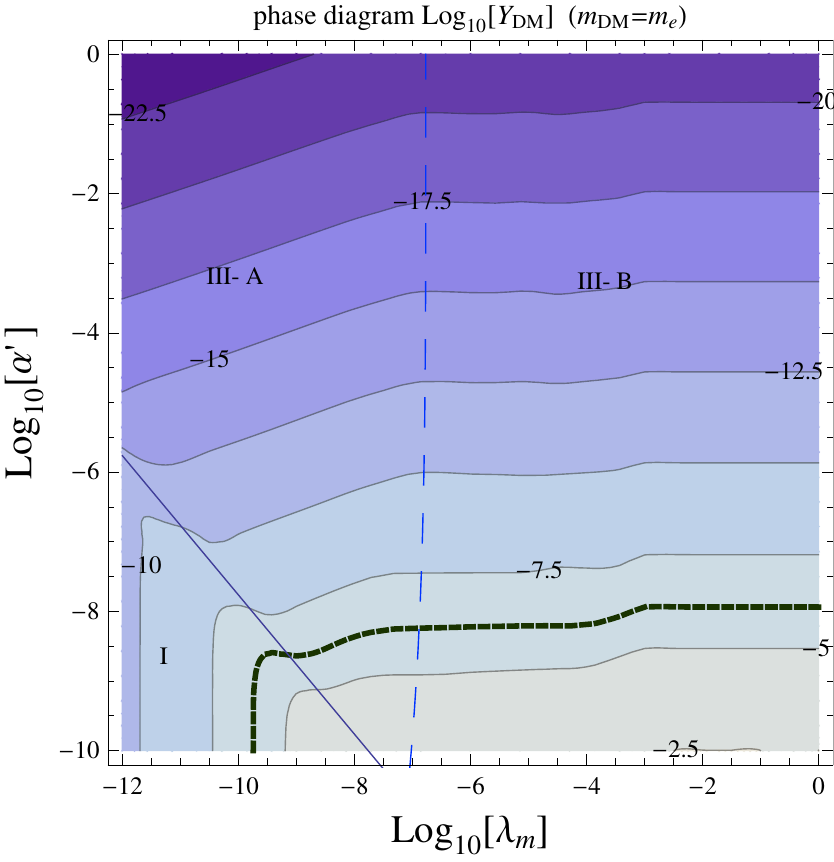}\includegraphics[height=7.2cm]{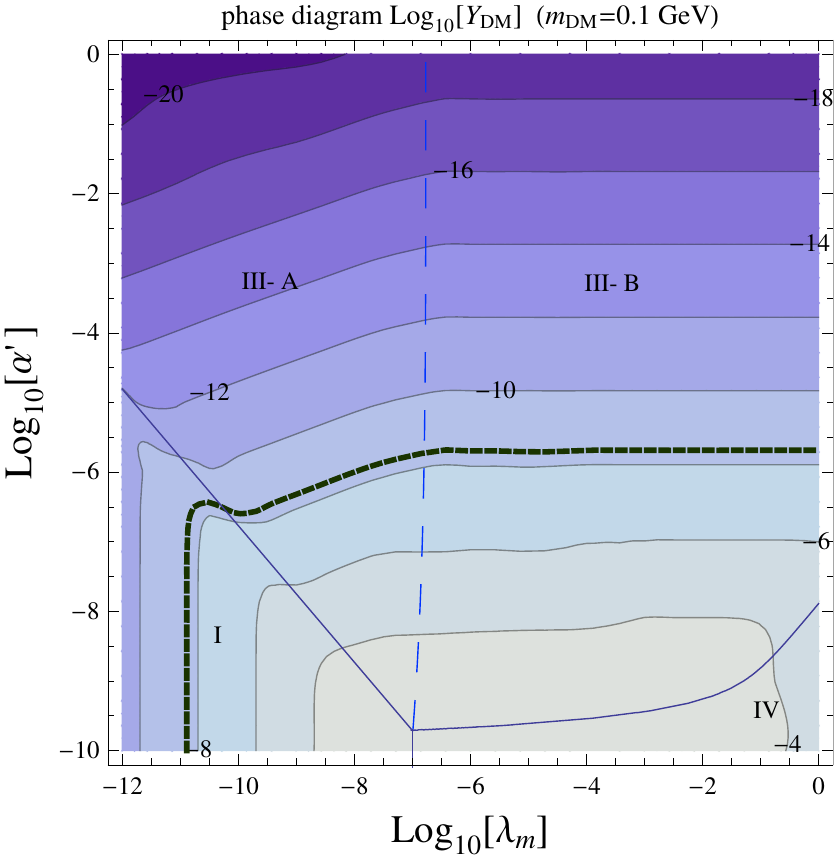}\\
\includegraphics[height=7.2cm]{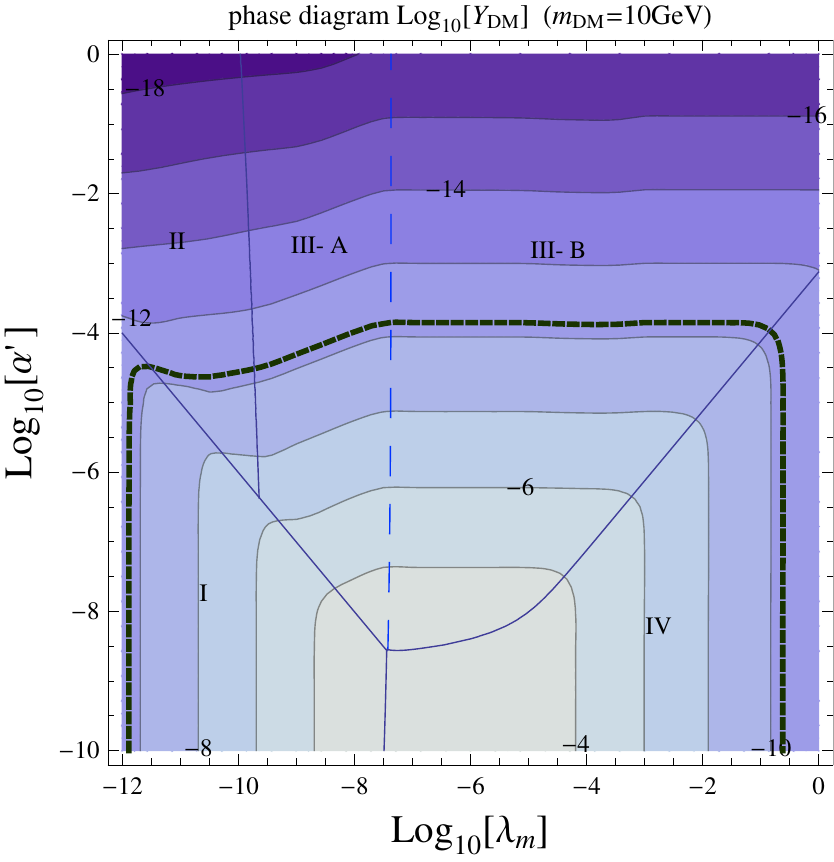}\includegraphics[height=7.2cm]{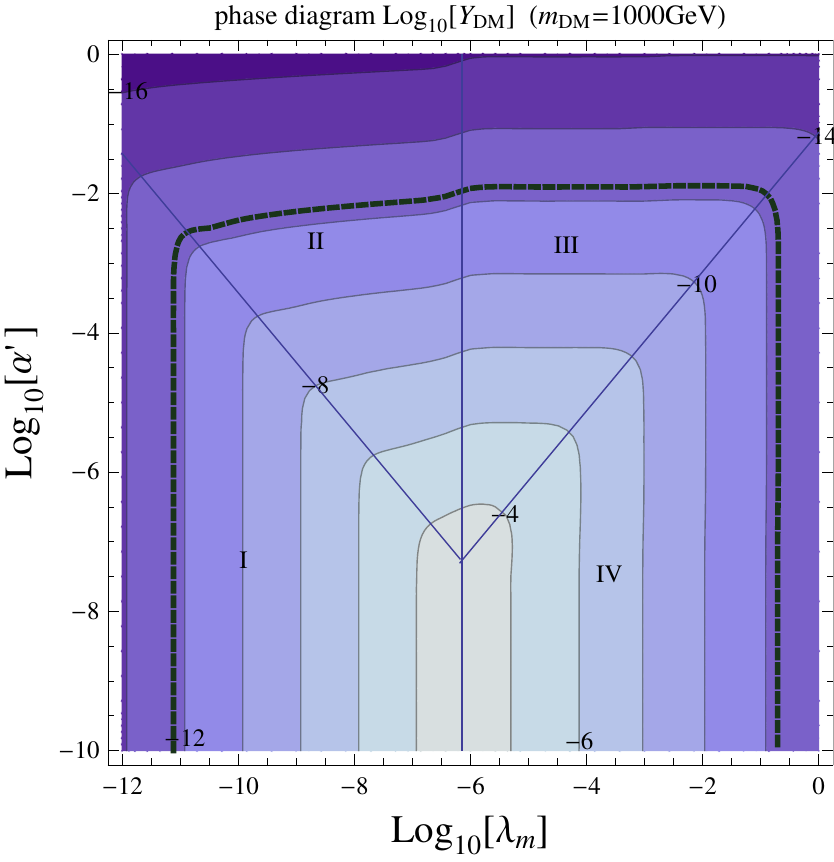}
\caption{Higgs portal phase diagrams for
  $m_{DM}=m_e,\,0.1~\hbox{GeV},\,10~\hbox{GeV},\,1~\hbox{TeV}$ (same label
  definitions than in Fig.~\ref{figure_tableGKM}, and the dashed
  blue line stands for the transition between IIIA and IIIB, as explained in
  the text).}
\label{HPphase}
\end{figure}

In the same way as for the kinetic mixing portal, for all phases but for freeze-in, it is required
to determine $\rho'/\rho$ as a function of $T$. For $m_{DM}>m_h/2$ the DM is exclusively produced by scattering  processes and, since it is produced in pairs, Eq.~(\ref{deltaE3}) applies. In this case $\rho'/\rho$ scales  in the same way as for the kinetic mixing, {\em i.e.}~as $\sim 1/T$ ($\sim$ const) for $T>m_{DM}$ ($T<m_{DM}$). This can be seen in Fig.~\ref{HP} which
gives
$\rho'/\rho$ 
as a function of $T$ for various values of $m_{DM}$ and $\lambda_m$.
For $m_{DM}<m_h/2$, on top of the energy transferred through the various scattering processes, given by Eq.~(\ref{deltaE3}), there is also a contribution from the Higgs boson decay, which turns out to dominate. The energy transferred by a decay is given by
\begin{equation}
\frac{d\rho'}{dt}+4H \rho'=\frac{g_i}{(2\pi)^3}\int f(p) E_p\Gamma_h(E_p) d^3p=\frac{g_i}{2\pi^2}\Gamma_h m_h^3TK_2(\frac{m_h}{T}) \,,
\end{equation}
or in terms of $\rho'/\rho$ and $T$
\begin{equation}
\frac{d(\rho'/\rho)}{dT}=-\frac{1}{H(T) T \rho}\frac{g_i}{2\pi^2}\Gamma_h m_h^3TK_2(\frac{m_h}{T}) \,,
\end{equation}
where $\Gamma_h(E_p)$ and $\Gamma_h$ stand for the decay width to DM pairs of a Higgs boson of energy $E_p$ and at rest, respectively.
This gives $\rho'/\rho\sim 1/T^{3}$ for $T\gtrsim m_h$. For $T\lesssim m_h$ ({\em i.e.}~typically for $T\lesssim m_h/6$), 
the DM pair creation, and therefore the increase of
$\rho'/\rho$, basically stop
as the Higgs
mediated 
channels become 
very suppressed.
 This is distinct
 from the kinetic mixing case {where, because the mediator is massless, there} is no such cut-off. These properties can be seen in Fig.~\ref{HP}.\footnote{Note that at $T>>m_h$ the energy transfer is nevertheless dominated by the scattering terms because, as shown above, they give $\rho'/\rho\sim 1/T$. This explains the change of behaviour around $T\sim 1$~TeV observed in Fig.~\ref{HP}.} 
The thermalization condition for the connector, $\Gamma_{connect}>H$, which gives $T'/T\simeq 1$, taken at $T=m_{DM}$, translates into $\lambda_m> 5\cdot 10^{-8}$ for $m_{DM}$ below $\sim m_h/2$ and $\lambda_m> 6.3 \cdot 10^{-7}$ for $m_{DM}=1 \hbox{TeV}$. If $\lambda_m$ is small enough for the connector not to thermalize, $\rho'/\rho$ reaches a plateau at $T\sim \hbox{Max}[m_{DM},m_h/6]$ with value, $\rho'/\rho\simeq  7\cdot 10^{-6}\lambda_m^2 m_{Pl}/m_{DM}$ ($\rho' / \rho \simeq 10^{-6} \lambda_m^2 m_{Pl} / m_{DM}$) for $m_{DM}<m_h/2$ ($m_{DM}>m_h$). In the intermediate $m_h$ regime a rough estimate is $\rho'/ \rho \simeq10^{-6} \lambda_m^2 m_{Pl} / m_W$.

From the knowledge of $(\rho'/\rho)(T)$, one can integrate the DM number density Boltzmann equation, which takes the same form as the kinetic mixing one, Eq.~(\ref{generalboltzmann1}),  
but since the decay term dominates for $m_{DM}<m_h/2$ it is useful to write it down explicitly, subtracting it from the scattering contribution 
\begin{equation}
s z H \frac{dY}{dz}=\gamma^{D}_{connect} \Big(1-\frac{Y^2}{Y_{eq}^{2}(T)}\Big)+
 \sum_i \gamma^{i(sub)}_{connect}
\Big(1-\frac{Y^2}{Y_{eq}^{2}(T)}\Big)+\gamma_{HS}   \Big(1-\frac{Y^2}{Y_{eq}^{2}(T')}\Big) \,,
\label{generalboltzmann1HP}
\end{equation}
where
 $Y=n_\phi/s=n_{\phi^*}/s$ stands for the number density of $\phi$ or $\phi^*$ particles ($Y_{DM}=2Y$), with  $\gamma^{D}_{connect}=n^h_{eq}  \Gamma_{h}(T)$ the decay reaction density of the Higgs boson  to $\phi \phi^*$, and
$$\Gamma_{h}^i(T)\equiv \Gamma(h\rightarrow \phi \phi^*) \frac{K_1(m_h/T)}{K_2(m_h/T)}.$$
The phase diagram, 
Fig.~\ref{HPphase}, turns out to have {the same  characteristic "Mesa" shape as}
 for the kinetic mixing portal,
despite
 the fact that the mediator is massive.
It also displays 4 phases: freeze-in, reannihilation, freeze-out in the hidden sector and connector interaction freeze-out. 
Similarly, as for the kinetic mixing case, one gets a truncated volcano shape in Fig.~\ref{YDM-HP}, where  $Y$ is displayed as a function of $\lambda_m$ for various values of $\alpha'$ and $m_{DM}$. There are nevertheless important differences one observes by looking closer at the way these diagrams are obtained.

\begin{figure}[!t]
\centering
\includegraphics[height=5.5cm]{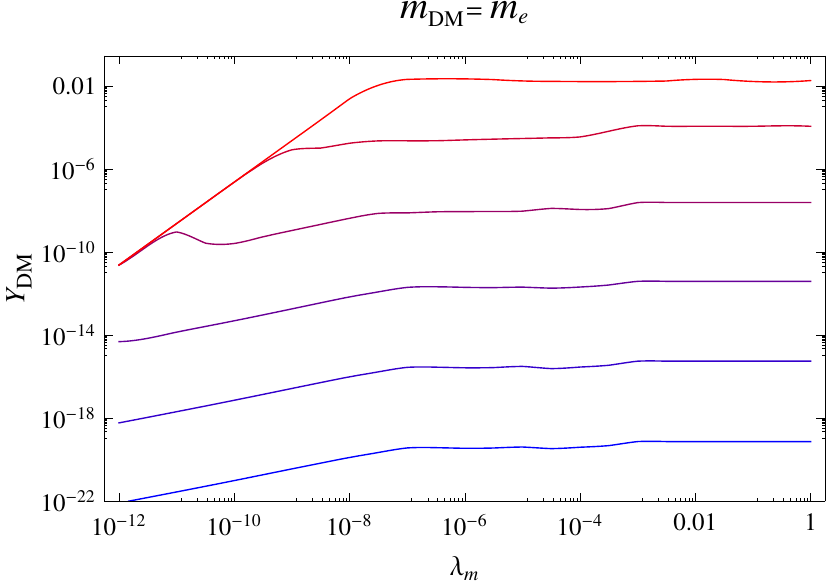}\includegraphics[height=5.5cm]{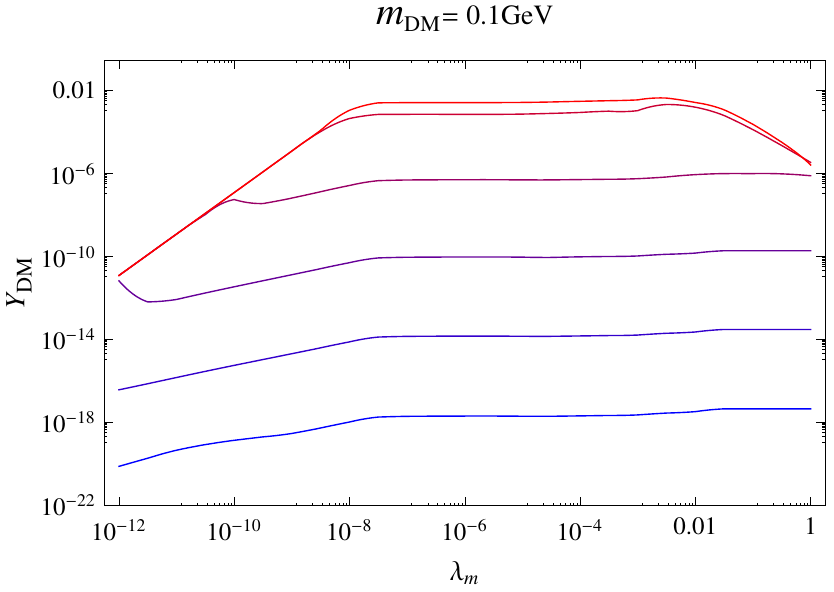}\\
\includegraphics[height=5.5cm]{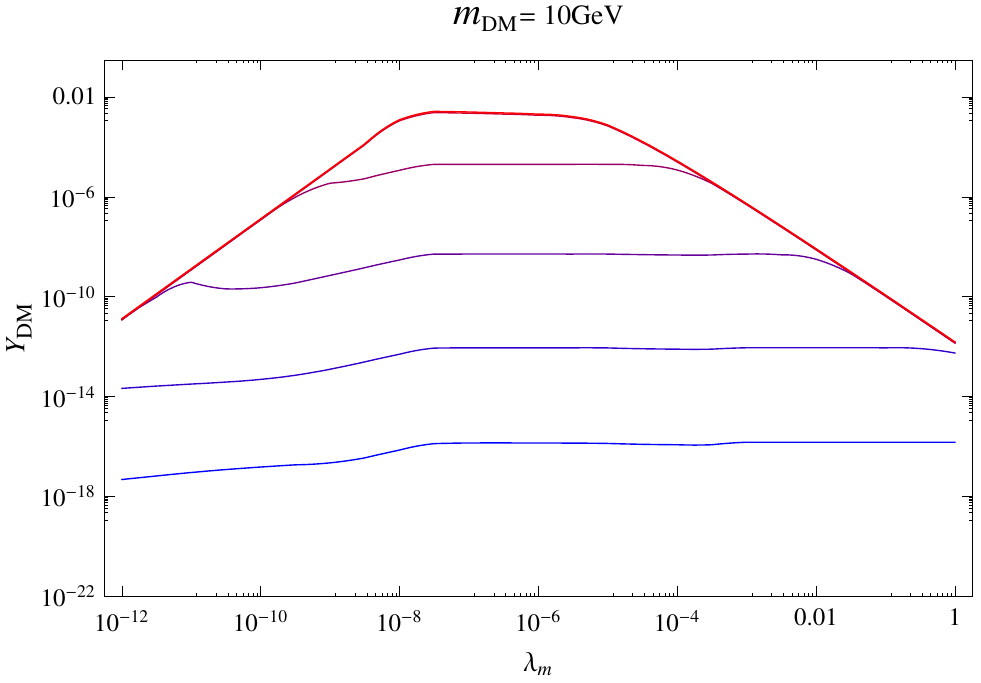}\includegraphics[height=5.5cm]{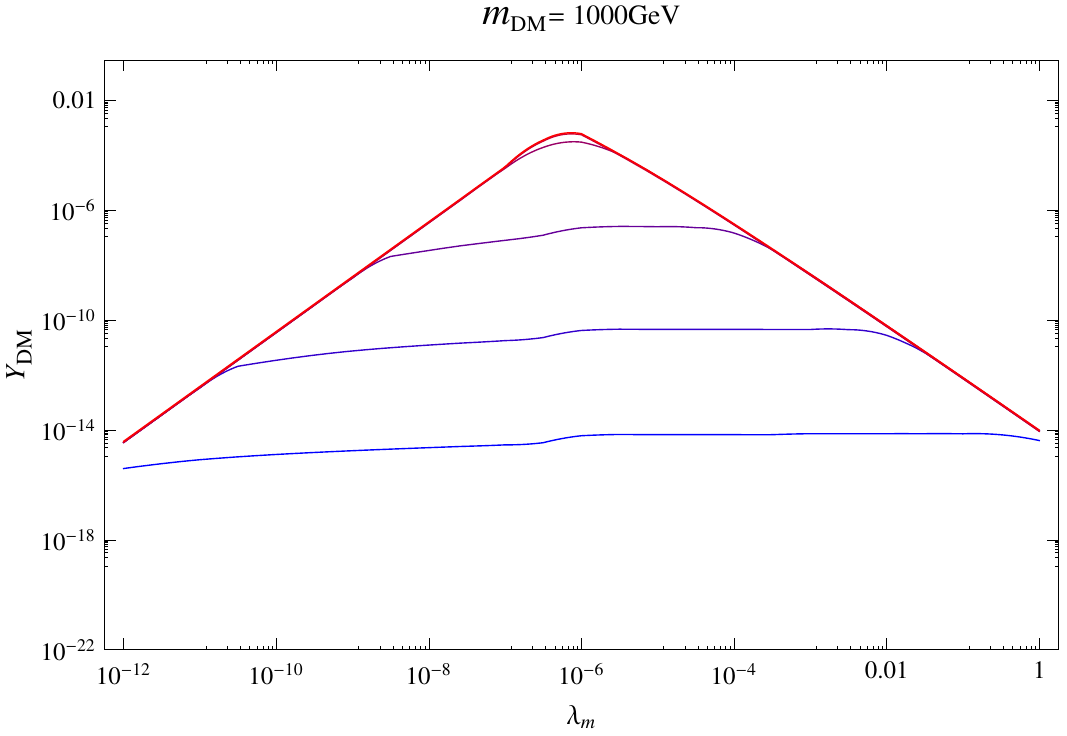}
\caption{DM relic abundance $Y_{DM}$ as a function of the connector parameter $\lambda_m$ for different DM masses $m_{DM}$ and values of the hidden sector interaction, $\log_{10}(\alpha'/\alpha)=1,-1,-3,-5,-7,-9$, bottom-up (the last two lines are the same for $m_{DM}=10$~GeV, as well as for $m_{DM}=1$~TeV).}
\label{YDM-HP}
\end{figure}

%%%%%%%%%%%%%%%%%%%%%%%%%%%%%%%%%%%%%%%%%%%%%%
\subsection{Phase I: the freeze-in regime}

The freeze-in production of a pair of scalar DM particles through the Higgs portal has been already considered in details in Ref.~\cite{Frigerio:2011in} for $m_{\phi}=m_{DM}<< m_h$, see also Ref.~\cite{Yaguna:2011qn}. In this case the decay process dominates the DM production because it involves 
a smaller number of
 couplings and is not more Boltzmann suppressed than these scattering processes
(unlike in the freeze-out case).  The scatterings are responsible for less than 1$\%$ of the total DM production.
Since the decay production is infrared dominated (until the number of Higgs particles becomes Boltzmann suppressed) one finds that the number of DM particles produced is simply the number of Higgs bosons times the decay rate times the Hubble time,\footnote{Notice that, in the freeze-in regime, the production of DM is indeed proportional to the decay rate of the Higgs boson into DM particles, and not to the branching ratio, because the Higgs boson number density is the equilibrium one, no matter the Higgs decays to SM particles are dominant. What is relevant is the competition between DM production from a equilibrium number density of Higgs boson and the expansion rate.} times a  constant $c$ of order unity
\begin{equation}
Y=c \,\frac{n_h^{eq} \Gamma(h\rightarrow DM DM)}{s H}\Big|_{T=m_h} \,.
\label{YFIHP}
\end{equation}
$c$ is independent of the decay, $c=3\pi/(2 K_2(1))\simeq 2.9$. Here too $c$ is bigger than unity because the maximum value of the production rate $n_h^{eq} \Gamma(h\rightarrow DM DM )$ occurs at $T\sim m_h/3.5$ rather than $T\sim m_h$, see Fig.~2 of Ref.~\cite{Frigerio:2011in}.
Note that the DM abundance $Y$ generated by the decay is independent of $m_{DM}$ (modulo channel threshold effects) since 
its production stops at a temperature within the range $m_h\gtrsim T>m_{DM}$, and because the Higgs decay width is independent of $m_{DM}$ (to lowest order in $m_{DM}^2/m_h^2$).
Therefore, taking into account the dependence $\Gamma(h\rightarrow DM DM)\propto \lambda_m^2v^2/m_h$, 
the DM relic density scales as $m_\phi\lambda_m^2/m_h^3$ and, the larger the Higgs boson mass is, the larger $m_\phi \lambda_m^2$ has to be to reproduce the observed relic density.
This parametric dependence can be seen
in Fig.~\ref{HP-varyingmass-FI}, which gives
the value of $\lambda_m$ necessary to get the observed relic density as a
function of $m_{DM}$ (see also Fig.~\ref{YDM-HP}). Note that in this case the DM particles
are produced when they are relativistic, since $m_h/2>m_{DM}$, but as well-known, as long as $m_{DM}$ is above the $\sim keV$ scale, this is still compatible with the constraints on {the fact that DM should be cold.}

\begin{figure}[!t]
\centering
\includegraphics[height=7.2cm]{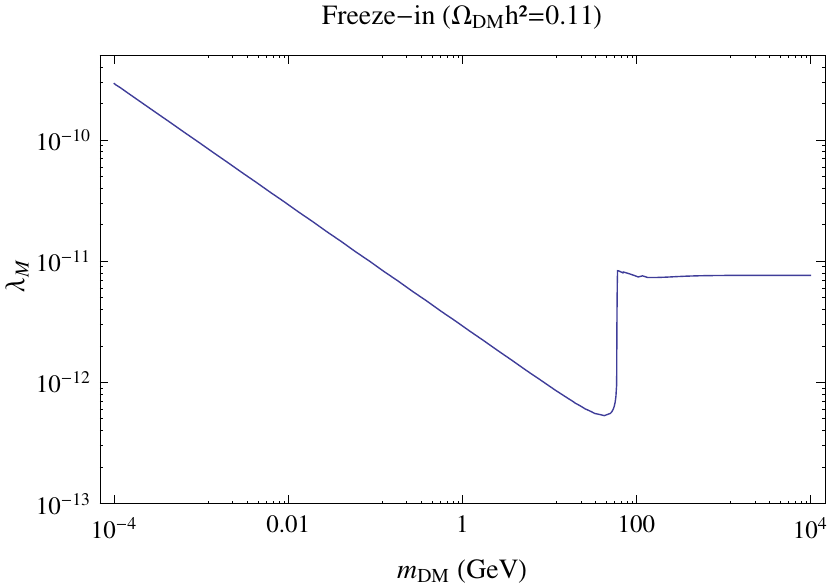}
\caption{Higgs portal parameter required to get the observed DM relic density through freeze-in ($\alpha'=0$).}
\label{HP-varyingmass-FI}
\end{figure}

In the opposite regime, $m_{DM}> m_h/2$, the Higgs boson decay production ceases to contribute
 and the scattering processes {become the relevant ones},
 producing DM down to $T\sim m_{DM}$. This gives in a similar way
\begin{equation}
Y=c \,\frac{\gamma_{connect}}{sH}\Big|_{T=m_{DM}}
\end{equation}
with $c$ of order unity.
For the $WW,ZZ,hh$ channels we get $c=2.5,2.5,2.4$ respectively.
This leads to a relic density, $\Omega_{DM}\propto \lambda_m^2$, independent of $m_{DM}$. As a result, as can be seen in Fig.~\ref{HP-varyingmass-FI}, the value of $\lambda_m$ which gives the observed relic density,
\begin{equation}
\lambda_m \simeq 7.6 \cdot 10^{-12},
\end{equation}
is also independent of $m_{DM}$ in this regime.
The $WW$ process is  dominant but is not the only
relevant one. For example for $m_{DM}=1$~TeV the $WW,ZZ,hh$ and $tt$ processes contribute respectively to
  $49\%,25\%,25\%$ and $0.5\%$ of the relic density.

%%%%%%%%%%%%%%%%%%%%%%%%%%%%%%%%%%%%%%%%%%
\subsection{Phase II, III and IV: the reannihilation and freeze-out regimes}

In the case of the kinetic mixing we have seen that, if the hidden sector thermalizes, but not the connector, the DM abundance freezes 
according to the reannihilation regime,
where
 both the source term and the hidden sector thermalization process are active down to the freezing temperature. In this case, as explained above, $Y$ first follows the thermal density $Y_{eq}(T')$, then follows the quasi static number density $Y_{QSE}$ until it intercepts $Y_{crit}$ where it freezes.
$Y$ follows $Y_{QSE}$ because $Y_{eq}(T')$ intercepts necessarily $Y_{QSE}$
 before it intercepts $Y_{crit}$ ($Y_{eq}(T')$ gets suppressed at $T'\lesssim
 m_{DM}$, that is to say before $Y_{QSE}$, since the later ({\em i.e.}~the
 connector) is suppressed  at $T\lesssim m_{DM}$).

For the Higgs portal, if $m_{DM}>m_h/2$, the situation is the same. For example for $m_{DM}=1$~TeV in Fig.~\ref{HPphase} we get a phase diagram which has the very {same structure as} for the kinetic mixing interaction.
However if $m_{DM}<{m_h/2}$ the situation is different.
In this case, instead of being suppressed at $T=m_{DM}$, as for the massless mediator case, the connector interaction (and hence $Y_{QSE}$) gets cut-off at $T\lesssim m_h/6$. 
As a result, unless $T'/T$ is small, $Y_{QSE}$  gets always suppressed before $Y_{eq}(T')$, since the later gets suppressed only at $T'\lesssim m_{DM}$.
In other words 
the system passes through a period with $m_{DM} < T^\prime < T\lesssim m_h/6$, where the connector has already decoupled but not the hidden sector interaction. An example of evolution of $Y$ where this holds
is shown in the left panel  of Fig.~\ref{HPreannih}.
There one sees that, as the hidden sector thermalizes,  $Y_{eq}(T')$ gets larger than $Y_{QSE}$ (as in the massless case) and then stays so, {\em i.e.}~$Y$ follows $Y_{eq}(T')$ until it intercepts $Y_{crit}$ without ever following $Y_{QSE}$. In this case there is no period of reannihilation (where $Y$ follows $Y_{QSE}$), even though $T'/T<1$, but  a hidden sector interaction freezes-out. This freeze-out is standard in the sense that it occurs when the connector has already decoupled ({\em i.e.}~when $T'/T$ is already constant). But still, the final relic abundance depends on the size of the connector because, prior to freeze-out, the larger the connector is, the larger $T'/T$ is, and so the larger is $\Omega_{DM}$. This regime has no equivalent in the massless case. It is a regime which comes between the reannihilation regime and the hidden sector freeze-out regime with thermalization of the connector. In the later case, the relic density is independent of the size of the connector (since $T'/T\sim1$).

\begin{figure}[!t]
\centering
\includegraphics[height=8cm]{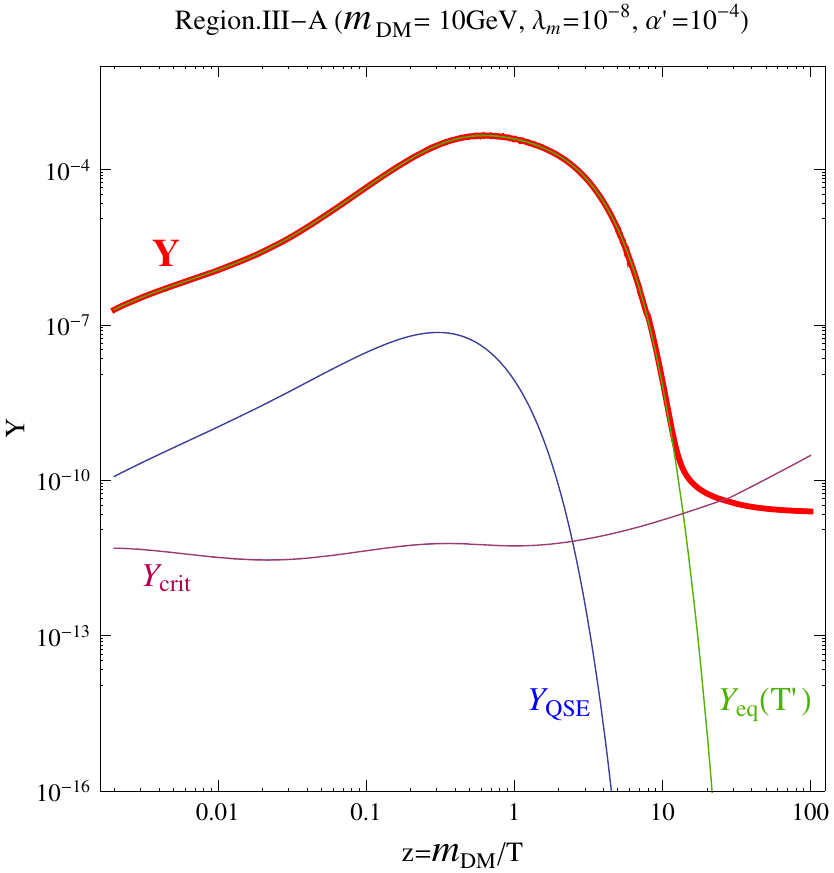}\includegraphics[height=8cm]{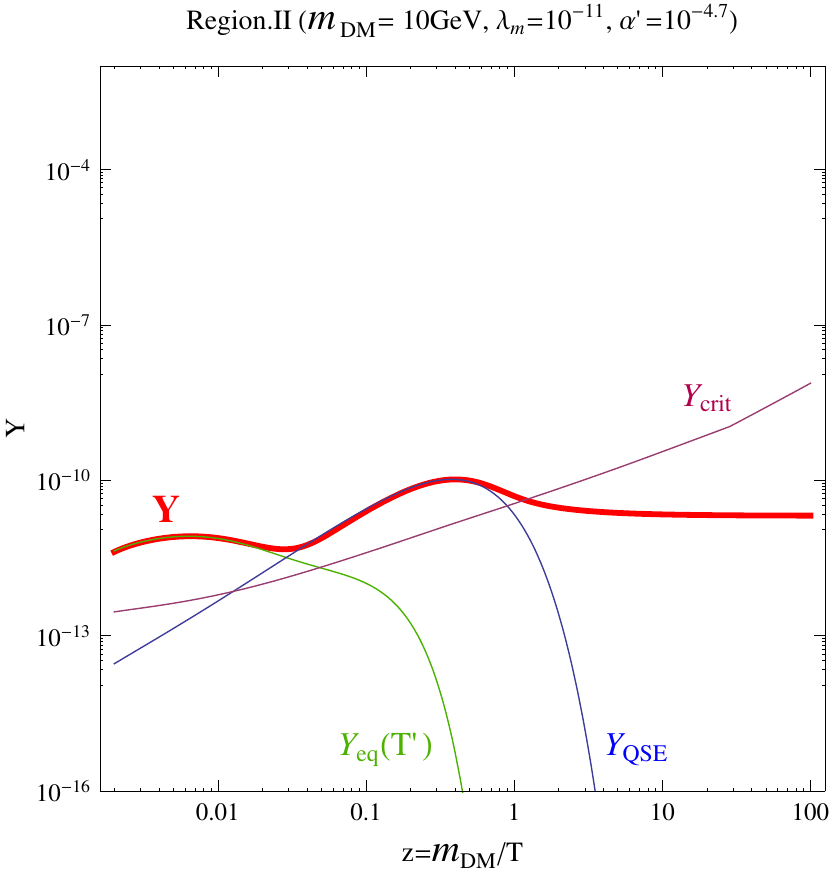}
\caption{Examples of evolution of the DM number density $Y$ as a function of $z\equiv m_{DM}/T$. The first example corresponds to a case of hidden sector interaction freeze-out without thermalization of the connector (region IIIA), obtained with $m_{DM}=10$~GeV, $\lambda_m=10^{-8}$ and $\alpha'=10^{-4}$. The second example corresponds to a reannihilation case (region II), obtained with $m_{DM}=10$~GeV, $\lambda_m=10^{-11}$ and $\alpha'=10^{-4.7}$. 
Also shown are $Y_{QSE}$ (blue), $Y_{crit}$ (purple) and $Y_{eq}(T')$ (green).}
\label{HPreannih}
\end{figure}

Practically it is clear that without thermalization of the connector, the lighter is $m_{DM}$, the easier the system can evolve through a situation where  
$m_{DM} < T^\prime < T\lesssim m_h/6$, hence the larger is the region of parameter space where this freeze-out phase occurs, and conversely the smaller is the reannihilation region. For instance for $m_{DM}$ below $\sim1$~GeV there is basically no more relevant reannihilation phase region, {\em i.e.}~the reannihilation occurs with values of $\lambda_m$ too tiny to give a large enough relic density. Similarly for $m_{DM}\gtrsim m_h/6$, there is, as with the kinetic mixing interaction, no sizable region of parameter space where the hidden sector interaction would freeze-out without thermalization of the connector. As for the intermediate DM mass values, they allow both regions. An example of reannihilation relic density evolution for an intermediate mass value,  $m_{DM}=10$~GeV, is given on the right panel of Fig.\ref{HPreannih}.
 Let us now discuss each regime with equations, for the case $m_{DM} \lesssim m_h/2$.

\bigskip
\noindent
\underline{Reannihilation}: region II in Fig.~\ref{HPphase} for the $m_{DM}=10$~GeV panel. 
Starting from a freeze-in regime, 
and increasing $\lambda_m$, the hidden sector 
 gets sufficiently populated to thermalize.\footnote{Unless $\alpha'$ is so small that it never thermalizes even if $T'/T\sim 1$.} For values of parameters where thermalization is hardly obtained, eventually $T'/T$ will still be sufficiently
small for $Y_{eq}(T')$ to be suppressed before $Y_{QSE}$. As a result, after having followed $Y_{eq}(T')$, $Y$ follows the $Y_{QSE}$ line, until freeze-out, see right panel of Fig.~(\ref{HPreannih}).
In this case reannihilation proceeds as in section 3.2 above, except that one has to replace everywhere the $\langle \sigma_{connect} v \rangle$ scattering term by the decay term ($\gamma_{connect}\equiv \langle \sigma_{connect} v \rangle n^2_{eq}(T) \rightarrow \Gamma_h(T) n_{eq}^h$). 
In particular $Y_{crit}(T)=H/\langle \sigma_{HS} v \rangle s$ is unchanged but $Y_{QSE}$ for a decay is given by 
\begin{equation}
Y^2_{QSE}(T)=Y_{crit} \Gamma_h(T) \frac{1}{H} Y_{eq}(T) =\frac{\Gamma_h(T)}{ \langle \sigma_{HS}v\rangle s^2}   n^h_{eq}(T) \,.  
\label{YQSEdecay}
\end{equation}
At freeze-out, where $Y=Y_{QSE}(T_f)=Y_{crit}(T_f)$
one gets therefore, in the same way as Eqs.~(\ref{Yanalinterm})-(\ref{freeze-outreacond}),
\begin{equation}
\sqrt{n^h_{eq}(T_f) \Gamma_h(T_f) 
 \langle \sigma_{HS} v \rangle}=H(T_f)\,.
\label{freeze-outreaHPcond}
\end{equation}
This equation can also be found in Ref.~\cite{Cheung:2010gj}.
Taking into account the fact that $T_f\lesssim m_h$ (when the connector source term gets suppressed), {\em i.e.}~$n^h_{eq}(T)=(\frac{m_h T}{2\pi})^{3/2}exp[-m_h/T]$ and $\Gamma_h(T)\simeq \Gamma_h$, from Eq.~(\ref{freeze-outreaHPcond}), we get
\begin{eqnarray}
x_f&\simeq&\frac{m_{DM}}{m_h}\log[0.0229c(c+2)\frac{ \langle \sigma_{HS}v\rangle
    M^2_{pl} }{g_*^{eff}}\frac{m_h^{3/2}\Gamma}{m_{DM}^{5/2}}]\notag\\
&&+\frac{5}{2}\frac{m_{DM}}{m_h}\log
[\log[0.0229c(c+2)\frac{ \langle \sigma_{HS}v\rangle
    M^2_{pl} }{g_*^{eff}}\frac{m_h^{3/2}\Gamma}{m_{DM}^{5/2}}]],
\end{eqnarray}
From this value of $x_f$ one gets therefore $Y(T_f)$ by substituting it in
\begin{equation}
Y(T_f)=Y_{crit}(T_f)=\frac{H}{\langle \sigma_{HS} v \rangle s}\Big|_{T=T_f}=\frac{5}{3} \frac{45}{2 \pi^2} \frac{\sqrt{g_*}}{g_{*s}}\frac{x_f}{\langle \sigma_{HS} v\rangle m_{DM}m_{Pl}}
\end{equation}
$c$ is a numerical constant of order unity. One gets the numerically exact result for $c$ between 0.4 (for large values of $\alpha'$) and 1 (for small values of $\alpha'$), related to the fact that $Y$ freezes-out in reality a little bit later after that $Y_{QSE}=Y_{crit}$). In this regime we find that $x_f$ varies between $\sim1$ and $\sim 5$. 
As with the kinetic mixing interaction in the reannihilation regime, the final relic density is inversely proportional to $\langle \sigma_{HS} v\rangle$ but $x_f$ depends on both interactions, so that it is smaller than the usual $\sim 20$ value.

The transition line between the freeze-in and reannihilation regimes is given by the condition
\begin{equation}
\frac{\Gamma_{HS}}{H}\Big|_{T=m_{DM}} \simeq 1
\end{equation}
which using Eq.~(\ref{YFIHP}) can be rewritten as
\begin{equation}
\frac{\sqrt{\langle   \sigma_{HS} v   \rangle}}{H}\Big|_{T=m_{DM}}\sqrt{ \Gamma_h(T) n^h_{eq}(T)}|_{T \simeq \tilde T} \Big(\frac{m_{DM}}{\tilde{T}}\Big)^{3/2}   \simeq 1
\label{HPcondI-II}
\end{equation}
with $\tilde{T}\equiv \hbox{Max}[m_{DM},\sim m_h/6]$.\footnote{One can also mention that for intermediate DM mass values,
 at the transition between freeze-in and reannihilation, the relic density
 undergoes a "kick", see Fig.~\ref{HPphase}, a feature which is not found in
 the kinetic mixing case. It can be shown that it is due to the fact that it
 corresponds to a transition from a situation where DM is relativistic (in the
 freeze-in regime, with a kinetic energy of order $\sim m_h/2$)  to a
 situation where DM, as soon as it has thermalize, looses a lot of kinetic
 energy (down to a kinetic energy of order $T'$), leading to a enhanced hidden
 sector annihilation cross section, since $\langle \sigma_{HS} v\rangle \propto 1/s$, hence a reduced DM abundance.}
 
Note finally that here too, the DM particles are produced relativistically from the decay of the Higgs boson, but, unlike in the freeze-in regime, quickly they become non-relativistic from the fact that they are in kinetic equilibrium with the $\gamma'$ down to a $T'$ temperature below $m_{DM}$.

\bigskip
\noindent
\underline{Hidden sector freeze-out without thermalization of the connector}:  region IIIA in Fig.~\ref{HPphase}.
Starting from a reannihilation or freeze-in regime situation, see Fig.~\ref{HPphase}, if one further increases $\lambda_m$, at some point $T'$ gets sufficiently high for the hidden sector scattering not to be much Boltzmann suppressed when the connector  $T\sim m_h/6$ decouples, even though $T'/T<1$. 
In this case the hidden sector freezes-out when the connector has already decoupled {for a while}. Therefore to get the freeze-out temperature as well as the relic density, one just needs to calculate the final value of $\rho'/\rho$ which is reached at $T\sim m_h$, see Fig.~\ref{HP}, {and is approximately} given by
\begin{equation}
\frac{\rho'}{\rho} \approx \frac{\Gamma_h(T) \cdot m_h n^h_{EQ}(T) }{H
  \rho}|_{T\sim \frac{m_h}{3}}
\end{equation}
which gives $\xi=T'/T=(g_*^{SM}/g_*^{HS})^{1/4}(\rho'/\rho)^{1/4}$, and to plug this value of $\xi$ in Eqs.~(\ref{xf})-(\ref{omegaeprime}).
This gives a relic density which scales as $m_{DM}^2 \lambda_m^{1/2}/\alpha'^2$.
The separation line between the reannihilation regime and this regime is given by the condition
\begin{equation}
Y_{QSE}=Y_{eq}(T')
\end{equation}
at freeze-out where $Y_{eq}(T')=Y_{crit}$. Numerically it corresponds to $\lambda_m\sim 3\cdot 10^{-8} (m_{DM}/m_h)^2$ (neglecting a logarithmic dependence on $\alpha'$). This condition, together with the condition of Eq.~(\ref{HPcondI-II}), leaves relatively little parameter space for the reannihilation regime. For example for $m_{DM}=10$~GeV, reannihilation occurs within $2\cdot 10^{-12} \lesssim \lambda_m \lesssim 2 \cdot 10^{-10}$ (imposing the relic density condition). Practically this means that apart for the region $1\,\hbox{GeV} \lesssim m_{DM} \lesssim m_h/3$, the reannihilation regime is not relevant to give a large enough relic density and this regime fills all the parameter space where the hidden sector interaction thermalizes but not the connector.
Note also that the value of $x_f$ we get in this case lies between $\sim 20$ (for $T'/T$ close to 1) to $\sim 5$ (for lower values of $T'/T$).

\bigskip
\noindent
\underline{Hidden sector freeze-out with thermalization of the connector}: region IIIB in Fig.~\ref{HPphase}.
If one increases further $\lambda_m$, at some point, the connector interaction thermalizes, when $\Gamma_{connect}=H$ at $T\gtrsim Max[m_{DM}, \sim m_h]$, and one enters in the standard hidden sector interaction freeze-out regime. In this case
$\Omega_{DM}$ scales as $\sim1/\langle \sigma_{annih} v \rangle\propto m_{DM}^2/\alpha'^2$ and $x_f\sim 20$.
It is independent of $\lambda_m$ since $T'/T\simeq 1$, at least in first approximation.
In second approximation one can mention nevertheless that the relic abundance might be not totally independent of the size of the connector for $m_{DM}<<m_h$. This stems from the fact that, as the communication between both sectors stops at $T\sim m_h/6$, the hidden sector temperature cannot track the visible sector one when $T$ gets reheated due to the decoupling of a SM species (but this is a rather small effect, as can be seen in Fig.~\ref{HPphase} for $m_{DM}=m_e$ and $\lambda_m\sim 10^{-3}$).

\bigskip
\noindent
\underline{Connector freeze-out}: region IV in Fig.~\ref{HPphase}.
Finally if the connector not only thermalizes but also leads to a reaction rate larger than the one of the hidden sector, then one enters the standard connector freeze-out regime.
For a relic density of order of the observed one it goes through the usual freeze-out of the annihilation
process with $\Omega_{DM}\sim 1/\langle \sigma_{connect} v \rangle$. In particular, for $m_{DM}<m_h/2$, as $x_f$ is of order $20$ the Higgs decay-inverse decay process can be totally neglected at freeze-out since it is obviously more Boltzmann suppressed than the scattering process.
This regime has been abundantly studied in the literature for the Higgs portal, for the case of a real scalar (n=1)~\cite{McDonald:1993ex,Andreas:2008xy,Andreas:2010dz}, as well as for $n=2$ \cite{Barger:2010yn}. 
For $m_{DM}\lesssim m_h/2$ it gives a relic density which scales as $m_W^2/\lambda_m^2$ (for $m_h\sim 120$~GeV) whereas for $m_{DM}<< m_h$ it scales as $m_h^4/(m_f^2\lambda_m^2)$ with $f$ the closest in mass fermion (with $m_f\lesssim m_{DM}$) (and for $m_{DM}>> m_h$ it scales as $m_{DM}^2/\lambda_m^2$).

Note nevertheless that,  for larger values of $\Omega_{DM}$, freeze-out in this regime occurs for smaller values of $x_f$, in which case the Higgs boson decay/inverse decay processes dominate the freeze-out process.
This explains why along this regime the transition line between region III and IV, corresponding to the condition $\Gamma_{connect}=\Gamma_{HS}$, bends towards smaller values of $\lambda_m$, see Fig.~\ref{HPphase} for the $m_{DM}=10$~GeV example.

Note also that for $\alpha'\simeq 0$, due to this bending behaviour, starting from $\lambda_m\simeq 0$ and increasing $\lambda_m$, one lays first in the freeze-in regime, then in the connector freeze-out regime dominated by the decay ($x_f\simeq 1$), then in the connector freeze-out regime dominated by the scattering processes. As a result in Fig.~\ref{YDM-HP}  the volcano is not perfectly conic as for the kinetic mixing case but remains truncated even for negligible values of $\alpha'$. This behaviour is nevertheless irrelevant for cases which give the observed relic density. 

In the same vein, note also that, as for the kinetic mixing case, all I-II-III-IV phase transition lines meet at a single point.
The meeting point corresponds to the situation where the hidden sector interaction, as well as the connector interaction, decouples as soon as they thermalize. Within the mass range we consider, $m_e<m_{DM}<10$~TeV, this situation always gives a too large relic density.

Finally let us mention  that for $m_{DM}\sim m_h/2$ the III to IV transition line can be significantly modified because in this case the connector scattering process, which determines the DM freeze-out, is resonantly enhanced.

\subsection{Testing the Higgs portal phase diagram?}

To test the freeze-in scenario, as well as the reannihilation scenario, from direct detection is much more difficult for the Higgs portal than for kinetic mixing, because the former does not lead to a $1/E_r^2$ collinear enhancement of the direct detection cross section. {For instance, the sensitivity reach from non-observation of a 100 GeV scalar at Xenon100 (Xenon1T) is  
$\lambda_m \lesssim 10^{-2}$ (resp. $2\cdot 10^{-3}$), which is several orders of magnitudes larger than the characteristic couplings required by the freeze-in or reannihilation scenarios. This upper bound nevertheless constraint the Higgs portal interaction freeze-out regime which requires typically 
$\lambda_m \sim 10^{-2}$ ($\lambda_m \sim 10^{-1}$) for $m_{DM} \sim 100$ GeV (resp. $10$ GeV)  (for $m_h = 120$ GeV, see e.g. Ref.~\cite{Andreas:2008xy,Andreas:2010dz}), as well as part of the hidden sector freeze-out regime.}
Similarly the freeze-in and reannihilation regimes cannot be tested either from collider experiments but possibilities of tests at colliders for large value of $\lambda_m$ do exist, in particular from the invisible decay width of the Higgs boson, see Ref.~\cite{astro-ph/0003350,Andreas:2010dz}. In other words even for DM masses above the GeV scale the phase diagram is much {less constrained} experimentally than in the kinetic mixing case. Only the connector freeze-out regime and perhaps a part of the hidden sector interaction freeze-out regime could be tested within a not too far future.

%%%%%%%%%%%%%%%%%%%%%%%%%%%%%%%%%%%%%%%%%%

%\newpage
\section{Summary} 

We have considered the simple possibility that, along the thermal history of the Universe, DM could have been created out of SM particles through a portal connecting the visible sector (SM) to a dark sector. This framework is particularly predictive as it is induced by the SM particles, that is to say particles whose properties (mass and interactions) and thermal number density are known. As a result, the energy transferred from the visible sector to the hidden sector and the DM relic density produced, depend, besides (obviously) the DM mass, only on the strength of the connector interaction and of the interaction(s) which may thermalize the hidden sector.

Whether we consider the gauge kinetic mixing portal, which involves a massless (or massive) mediator, or the Higgs portal, whose mediator is the Higgs doublet, we have found that there are 4 distinct ways of producing DM, leading to a phase diagram with the characteristic shape of  a ``Mesa". In two of the phases, the DM relic density depends exclusively on the connector interaction, whereas in the two others it depends essentially on the strength of the hidden sector interaction  (provided the connector interaction is within a certain range).

It is useful to summarize the parametric dependence of the relic density for the various phases, which holds for a connector lighter than the DM particle (both in the case of kinetic mixing and Higgs portals):
\begin{eqnarray}
\hbox{I}:&&\hspace{-4mm}\hbox{Freeze-in:}\quad Y\sim \frac{(n^{SM}_{eq})^2 \langle \sigma_{connect} v \rangle}{sH}\Big|_{T=Max[m_{SM},m_{DM}]} \quad\hbox{with}\quad x_f\sim \frac{m_{DM}}{\hbox{Max}[m_{SM},m_{DM}]}\,\,\,\, \\
\hbox{II}:&&\hspace{-4mm}\hbox{Reannihilation:}\quad Y\sim \frac{x_f}{\langle \sigma_{HS} v \rangle \tilde{m}^2}\quad\hbox{with}\quad x_f\sim \log  \langle  \sigma_{eff} v \rangle \tilde{m}^2\\
\hbox{III}:&&\hspace{-4mm}\hbox{Hidden sector freezeout:}\quad Y\sim \frac{x_f}{ \langle \sigma_{HS} v \rangle \tilde{m}^2}\quad\hbox{with}\quad x_f\sim \log  \langle  \sigma_{HS} v \rangle \tilde{m}^2 \\
\hbox{IV}:&&\hspace{-4mm}\hbox{Connector freezeout:}\quad Y\sim \frac{x_f}{ \langle \sigma_{connect} v \rangle \tilde{m}^2}\quad\hbox{with}\quad x_f\sim \log \langle  \sigma_{connect} v \rangle \tilde{m}^2\quad\quad\quad 
\end{eqnarray} 
with $\tilde{m}^2=m_{DM} m_{Pl}$.
If none of the interactions thermalize, the freeze-in regime holds. If the hidden sector thermalizes but the connector does not, reannihilation occurs. In this regime both interactions freeze at the same time after going through a period of quasi static equilibrium evolution.
We have shown that in this case the relic abundance is inversely proportional to the $\langle \sigma_{HS} v \rangle$ hidden sector cross section, but that the freeze-out temperature $x_f$ is determined by an effective cross section which is the geometric mean of both the hidden sector and connector cross sections $\langle \sigma_{eff} v \rangle=\sqrt{\langle \sigma_{HS} v\rangle \langle \sigma_{connect}v\rangle}$.
If instead, both interactions thermalize one gets an ordinary hidden sector or connector interaction freeze-out regime, depending on which of the two interactions is the fastest. Finally if the hidden sector interaction is so small that it never thermalizes, even if the connector thermalizes, one also lies in the connector freeze-out phase.
We have also shown that there are no other possible regimes and that, for a non negligible hidden sector interaction, the four regimes necessarily follow each other in this order. 

In the opposite case where the mediator has a mass larger than the DM, as applies for the Higgs portal for $m_{DM}\lesssim m_h/2$,\footnote{Or for the kinetic mixing portal with $m_{DM}<m_{\gamma'}/2$, a possibility we didn't consider here but which would give a same pattern.} one gets the same pattern except that there is an additional decay contribution and that the connector interaction is cut-off at the mediator mass. As a result the decay dominates both the freeze-in and reannihilation regimes, and one has also the possibility that the hidden sector interaction freezes out without thermalization of the connector. Correspondingly one has:
\begin{eqnarray}
\hbox{I}:&&\hspace{-5mm}\hbox{Freeze-in:}\quad Y\sim \frac{n^{h}_{eq} \Gamma_h}{sH}\Big|_{T=m_h} \quad \hbox{with}\quad x_f\sim \frac{m_{DM}}{m_h}\\
\hbox{II}:&&\hspace{-5mm}\hbox{Reannihilation:}\quad Y\sim \frac{x_f}{\langle \sigma_{HS} v\rangle \tilde{m}^2}\quad\hbox{with}\quad x_f\sim \log [ \langle  \sigma_{HS} v \rangle \Gamma_h \frac{m_{Pl}^2 m_h^{3/2}}{m_{DM}^{5/2}}  ]\\
\hbox{IIIA}:&&\hspace{-5mm}\hbox{HS freeze-out (no connector thermal.):}\quad Y\sim \frac{x_f}{ \langle \sigma_{HS} v \rangle \tilde{m}^2}\quad\hbox{with}\quad x_f\sim \xi \log   \langle  \sigma_{HS} v \rangle \tilde{m}^2\,\,\,\,\,\\
\hbox{IIIB}:&&\hspace{-5mm}\hbox{HS freeze-out (connector thermal.):}\quad Y\sim \frac{x_f}{ \langle \sigma_{HS} v \rangle \tilde{m}^2}\quad\hbox{with}\quad x_f\sim \log  \langle  \sigma_{HS} v \rangle \tilde{m}^2\\
\hbox{IV}:&&\hspace{-5mm}\hbox{Connector freeze-out:}\quad Y\sim \frac{x_f}{ \langle \sigma_{connect} v \rangle \tilde{m}^2}\quad\hbox{with}\quad x_f\sim \log  \langle  \sigma_{connect} v \rangle \tilde{m}^2
\end{eqnarray} 
with $\Gamma_h$ the Higgs boson decay width to DM particles, and $\xi=T'/T=(g_*^{SM}/g_*^{HS})^{1/4}(\rho'/\rho)^{1/4}$.
The reannihilation, as well as the hidden sector freeze-out regime (without connector thermalization) give a relic density that depends mainly on the size of one interaction, but also logarithmically on the other one, unlike the other regimes for which the relic density depends only on the strength of one interaction. For a heavy mediator, the reannihilation regime is relevant only for a relatively narrow intermediate DM mass region below the mediator mass, $1~\hbox{GeV} \lesssim m_{DM} \lesssim m_h/3$. 
Therefore, apart for this region, 
if the hidden sector interaction thermalizes but not the connector, hidden sector freeze-out always occurs. This is different from the low mediator mass regime where in this case ({\em i.e.}~if $T'<T$) reannihilation always occurs. 

One should also emphasize the fact that one also gets a characteristic Mesa phase diagram in the more complex situation in which there are both a massless and a massive mediator. This applies in particular in the kinetic portal scenario, in the intermediate mass range $1$ GeV$\lesssim m_{DM} \lesssim m_Z/2$, see Appendix D.

To discuss the kinetic mixing portal case, we have considered a particularly simple model where the hidden sector consists of  a single particle charged under an unbroken $U(1)'$ gauge group, depending only on $m_{DM}$, on the kinetic mixing coupling and on the extra gauge coupling.
In addition to the relic density phase diagram, we have discussed in detail the phenomenology it may imply.
Direct detection is particularly interesting for this model because it involves a massless or light mediator, which implies that the cross section is inversely proportional to the recoil energy squared and is therefore strongly enhanced at low recoil energies.
{We have shown that the latest Xenon100 data exclude  the regimes of freeze-out (both in the hidden (III) and visible regimes (IV)) for candidates with $m_{DM}> \sim 5$~GeV (see Fig.9). The same data also probe ({\em i.e.} exclude) a large fraction of the parameter space corresponding to the reannihilation regime (II), but does not constrain freeze-in (regime I) so that, for the time being, the latter is  allowed for any DM mass. However, although the kinetic mixing parameter required to produce the observed DM abundance through freeze-in is very small, even this regime could be tested by the future Xenon-1T experiment. For one year (four years) of exposure, we have found that candidates in the range $50~\hbox{GeV}\lesssim m_{DM} \lesssim 140$~GeV ($50~\hbox{GeV}\lesssim m_{DM} \lesssim 600$~GeV) could be tested. As for the DAMA and CoGeNT data, they can be accounted for by considering values of the parameters which lay in the reannihilation regime, but which are excluded by the current Xenon-100 data, see Fig.8.} The characteristic $1/E_r^2$ recoil energy spectrum allows to distinguish this model from more standard DM models which, involving a mediator beyond the $\sim$~MeV scale, predict a cross section on nucleon that is independent of $E_r$. 

Other important constraints for the kinetic mixing setup are cosmological, related to the long range interaction caused by the light mediator. The most relevant constraint rests on  the ellipticity of galaxies, which puts an upper bound on $\alpha'$. There is also another stringent upper bound on $\alpha'$, which is related to the galactic magnetic fields and which one must satisfy if one wants DM to be present at the Sun location, so as to be probed using direct detection experiments. Both constraints turn out to be fully compatible with the freeze-in regime, but not with the other regimes, except for a DM mass larger  than a few hundreds GeV. We have shown how these constraints get considerably relaxed if instead of considering a massless mediator we make it slightly massive (still lighter than the DM mass scale, so that the relic densities we have obtained are unaffected).

The model we have considered for the Higgs portal consists of a scalar charged under a $U(1)'$ gauge group,
and depends only on $m_{DM}$, the $U(1)'$ gauge coupling and the Higgs portal interaction.
Here the 4 phases are all allowed by experimental data, even for DM masses above the GeV scale. Direct detection together with collider experiments could potentially cover the connector freeze-out phase and, optimistically, a part of the hidden sector freeze-out regime. 
The phase diagram we have derived can be consequently considered as the analytic prolongation (towards smaller Higgs portal coupling values) of the Higgs portal freeze-out regime {widely} considered in the literature. 

{\em A priori} such an analytic prolongation could be obtained with the same characteristic mesa shape for any DM candidate which belongs to a SM gauge singlet hidden sector, with (a) a visible sector which 
consists of particles that are in thermal equilibrium when they produce the DM (as in the case we considered here, where in the visible sector there are only SM particles), and (b) a primordial energy density in the hidden sector which is secondary. 
In this sense our results are representative of a large class of new DM models.

%%%%%%%%%%%%%%%%%%%%%%%%%%%%%%%%%%%%
\section*{Acknowledgements}
%%%%%%%%%%%%%%%%%%%%%%%%%%%%%%%%%%%%

We thank M.~Frigerio and E.~Masso for useful discussions, and Tongyan Lin, Timothy Cohen and Jay Wacker for useful comments. 
This work is supported by the FNRS-FRS, the IISN, the Belgian Science Policy (IAP VI-11), and the ARC ``Beyond Einstein''. 
TH thanks the Departamento de F\'isica Te\'orica (UAM-Madrid) and the IFT-Madrid for hospitality and the Comunidad de Madrid (Proyecto HEPHACOS S2009/ESP-1473).

\appendix
\def\thesection{Appendix \Alph{section}}
%\section{\label{appendix1}}

\section{Relevant  cross sections for the kinetic mixing portal}
\label{App0}
\bigskip
\noindent\underline{DM pair creation from SM fermions:}

This is the dominant process for DM creation through the kinetic mixing portal. It takes place through a photon or a Z in the s-channel:
\begin{eqnarray}
\sigma(f \bar{f} \rightarrow e' \bar e')=&&\displaystyle\sum_{f}N_f^C\frac{\pi
  q_{e'}^2\alpha\alpha'
  \hat{\epsilon}^2}{s}\sqrt{\frac{s-4m_{e'}^2}{s-4m_f^2}}\cdot\{\notag\\
&&(1+\frac{s^2-4m_f^2s-4m_{e'}^2s+16m_f^2m_{e'}^2}{3s^2}+\frac{4m_f^2+4m_{e'}^2}{s})\label{sigmaf}\\
&&\times
(q^2_f-\frac{g_V}{\cos\theta_W^2} q_fRe\frac{s}{S-m_Z^2-im_Z\Gamma_Z})\notag\\
&&+\frac{1}{4\cos\theta_W^4}\frac{1}{(s-m_Z^2)^2+m_Z^2\Gamma_Z^2}[(g_V^{2}-g_A^{2})m_f^2(4s+8m_{e'}^2)\notag\\
&&+(g_{V}^{2}+g_A^{2})(s^2+4sm_{e'}^2-8m_{e'}^2m_{f}^2+\frac{s^2-4m_f^2s-4m_{e'}^2s+16m_f^2m_{e'}^2}{3})]\} \notag,
\end{eqnarray}
where $\hat{\epsilon}=\epsilon\cos
\theta_W/\sqrt{1-\epsilon^2}$, and $g_V$, $g_A$, $N_f^C$ are
the V, A components and color factor of the
SM fermions with
$f=e,\mu,\tau,u,d,s,c,b,t$. 
For $m_{DM}>>m_Z$ the $\gamma$, $Z$ and $\gamma-Z$ interference terms involving charged $f$ particles, and the $f=\nu_{e,\mu.\tau}$ channels, contribute along the $8:2.8:-3.0:0.6$ proportions respectively.

\bigskip
\noindent\underline{Z-decay in DM pairs:}

A Z on-shell may decay into DM pairs, with a width given by:
\begin{equation}
\Gamma(Z \rightarrow e' \bar
  e')=\displaystyle\sum_{e'}\frac{ q_{e'}^2\alpha'\hat{\epsilon}^2}{3}m_Z\sqrt{1-\frac{4m_{e'}^2}{m_Z^2}}(1+\frac{2m_{e'}^2}{m_Z^2}).
\label{sigmaZ}
\end{equation}

\bigskip
\noindent\underline{DM/SM fermion scattering:}

In practice this t-channel process, which is dominated by photon exchange (Z contribution not shown), is negligible and is given for reference (see the discussion in Section 2):
\begin{equation}
\sigma(f e' \rightarrow f e')=\displaystyle\sum_{e,e'}\frac{\pi q_{f}^2 q_{e'}^2\alpha\alpha' \hat{\epsilon}^2}{s|P_{1\text{cm}}(m_f^2,m_{e'}^2)|^2}\int^{0}_{-4|P_{1\text{cm}}(m_f^2,m_{e'}^2)|^2}\left[\frac{(s-m_f^2-m_{e'}^2)^2}{t^2}+\frac{s}{t}+\frac{1}{2}\right]dt \,,
\label{sigmaft}
\end{equation}
with $|P_{1\text{cm}}(m_1,m_2)|^2=(s-m_1^2-m_2^2)^2/(4s)-m_1^2m_2^2/s$.
\bigskip

\noindent\underline{DM pair annihilation into hidden photons:}

\begin{equation}
\sigma(e' e' \rightarrow \gamma' \gamma')=\frac{4\pi(\alpha')^2 }{s}\left[\frac{2s^2+8m_{e'}^2s-16m_{e'}^2}{(s-4m_{e'}^2)s}\tanh^{-1}\Big(\frac{\sqrt{s-4m_{e'}^2}}{\sqrt{s}}\Big)-\frac{s+4m_{e'}^2}{\sqrt{s(s-4m_{e'}^2)}}\right] \,.\label{sigma1}
\end{equation}

\section{Reaction densities}
\label{App1}

In Eqs.~(\ref{generalboltzmann1}) and (\ref{generalboltzmann1HP})  the reaction densities (and accordingly the $\langle \sigma v \rangle$'s) are defined as
\begin{eqnarray}
\gamma(a\,b\leftrightarrow 1\,2) &=&  \iint d\bar{p}_a d\bar{p}_b %g_a g_b 
f_a^{eq}f_b^{eq} \iint d\bar{p}_1 d\bar{p}_2 (2\pi)^4 \delta^4(p_a+p_b-p_1-p_2) |{\cal M}|^2 \nonumber\\
&=&  \, \frac{T}{64~\pi^4} \int_{s_{min}}^{\infty} ds ~\sqrt{s}~\hat{\sigma}(s)~K_1\left(\frac{\sqrt{s}}{T} \right)\,.
\label{ScatRates}
\end{eqnarray}
We have defined $d \bar{p} \equiv 
d^3 p/((2 \pi)^3 2 E)$. 
Here $f_i^{eq}=(e^{E_i/T}\pm 1)^{-1}\simeq e^{-E_i/T}$ is the Maxwell-Boltzmann energy distribution,  
$|{\cal M}|^2$ is the amplitude squared summed over initial and final spins (with no averaging),
$s_{min}=\hbox{max}[(m_a+m_b)^2,(m_1+m_2)^2]$, and  the reduced cross section is defined by
\begin{equation}
\hat{\sigma}(a\,b\leftrightarrow 1\,2) = \frac{g_ag_b}{c_{ab}}\ \frac{2[
(s-m_a^2-m_b^2)^2 -4 m^2_a m^2_b]}{s} \ \sigma(a\,b\rightarrow 1\,2) \,,
\end{equation}
with $\sigma$ the particle physics cross section of Eqs.~(\ref{sigmaf}), $g_{a,b}$ the number of degrees of freedom of the particles 
$a,b$  and $c_{ab}$ a combinatorial factor equal to 2 (1) if $a$ and $b$ are identical (resp. different). 
As for the Hubble constant and the entropy density they are given by $H=1.67 \sqrt{g^{eff}_*} T^2/m_{Pl}$, with $g_*^{eff}=g_*^{SM}+g_*^{HS}$, and $s=(2 \pi^2/45)g_{*s} T^3$, with $g_{*s}=g_{*s}^{SM}+g_{*s}^{HS}$.

\section{Energy transfer Boltzmann equation}

It is convenient to rewrite Eq.~(\ref{Etransfereq}) as
\begin{equation}
\frac{d\rho'}{dt}+3H(\rho'+p')= g_1 g_2  \int \frac{d^3 p_1}{(2\pi)^3}   \frac{d^3 p_2}{(2\pi)^3} f_1(\vec p_1) f_2(\vec p_2) v_{Mol} \mathcal{E}(\vec p_1,\vec p_2) \,,
\end{equation}
where \cite{Gondolo:1990dk}
\begin{equation}
\mathcal{E}(\vec p_1,\vec p_2)\equiv\frac{1}{2E_1 2E_2 v_{Mol}}\left(\int \displaystyle\sum_{i=3}^{4} 
d^3 \bar{p}_i \cdot g_i |i\mathcal{M}_{1\,2 \leftrightarrow 3 \,4}|^2 (2\pi)^4 \delta^{(4)}(p_1+p_2-p_3-p_4) \Delta E_{tr}\right)\notag \,,
\end{equation}
with $v_{\text{Mol}}$ the so-called M\o ller velocity \cite{Gondolo:1990dk},
\begin{equation}
v_{Mol} = F/(E_1E_2)=\sqrt{[s-(m_1+m_2)^2][s-(m_1-m_2)^2]}/(2E_1E_2) \,.
\end{equation}
In these equations $f_i$ is the momentum distribution of the initial "i" particle, $p'$ is the pressure of the dark sector, $g_i$ is the number of effective degrees of freedom, $\Delta E_{tr}$ is the energy transferred in the process and $\mathcal{M}_{1\,2 \leftrightarrow 3 \,4}$ the usual invariant matrix element.
Note that
as long as $T'/T$ is not close to one, it is a good approximation to neglect the inverse processes because they transfer an energy to the hidden sector that is suppressed by a factor $\xi^4$ (one factor of $\xi^3$ from the hidden sector particle number density and one of $\xi$ from the energy transferred per scattering). We will nevertheless incorporate these processes to get results that are reliable for large $\xi$ values too.

Assuming a radiation dominated Universe in both sectors down to $T\sim1$~eV, {\em i.e.}~$p=\rho/3$, and  adopting Maxwell-Boltzmann momentum distribution one therefore obtains
\begin{equation}
\frac{d\rho'}{dt}+4H\rho'= g_1 g_2 \int \frac{d^3p_1}{(2\pi)^3} \frac{d^3 p_2}{(2\pi)^3} e^{-E_1/T}e^{-E_2/T} \,v_{Mol}  \mathcal{E}(\vec p_1,\vec p_2) \,,
\end{equation}

For the $e^+ e^- \rightarrow e'^+,e'^-$ pair production process
the Boltzmann equation integration is relatively straightforward because the energy transferred, $\Delta E_{tr}=E_1+E_2$, is independent of the momenta of final particles, so that $\mathcal{E}(\vec p_1,\vec p_2)$ is precisely $\sigma(\vec p_1,\vec p_2)\cdot \Delta E_{tr}$.
Since the cross section is Lorentz invariant it can be calculated simply in the center of mass frame,
which, defining $E_\pm=E_1\pm E_2$
 and $s=(E_1+E_2)^2-(\vec p_1+\vec p_2)^2$, gives
\begin{eqnarray}
\frac{d\rho'}{dt}+4H\rho'&= &g_1 g_2 \int \frac{d^3p_1}{(2\pi)^3} \frac{d^3 p_2}{(2\pi)^3} \cdot e^{-E_1/T}e^{-E_2/T} \sigma(\vec p_1,\vec p_2) v_{\text{Mol}} \Delta E_{tr}\cr
&=&\frac{g_1 g_2}{32\pi^4}  \int dE_+ dE_- ds \cdot e^{-E_+/T}\sigma(s) F E_+\cr
&=&\frac{g_1 g_2}{32\pi^4}  \int ds \int dE_+ E_+  e^{-E_+/T} \sigma(s) F \int dE_-  \,.
\end{eqnarray}
Integrating on the parameter space given by \cite{Edsjo:1997bg}
\[
\begin{pmatrix}
E_1\geq m_1\\
E_2\geq m_2\\
|\frac{\vec P_1 \cdot \vec P_2}{ P_1 P_2}|\leq 1
\end{pmatrix}
\Rightarrow
\begin{pmatrix}
s\geq (m_1+m_2)^2\\
E_+\geq \sqrt{s}\\
  -\frac{2F}{s}\sqrt{E_+^2-s}  \leq E_--E_+\frac{m_1^2-m_2^2}{s} \leq \frac{2F}{s}\sqrt{E_+^2-s}
\end{pmatrix}\,,
\] \,
and since $m_1=m_2=m$, one obtains the simplified form
\begin{eqnarray}
\frac{d\rho'}{dt}+4H\rho'&=& \frac{g_1 g_2}{32\pi^4}  \int ds \int dE_+ E_+  e^{-E_+/T}  \frac{4F}{s}\sqrt{E_+^2-s} \sigma(s) F \nonumber  \\
&=&\frac{g_1 g_2}{32\pi^4}  \int ds  \frac{4F}{s} s T K_2(\frac{\sqrt{s}}{T}) \sigma(s) F\nonumber \\
&=&\frac{g_1 g_2}{32 \pi^4}  \int ds\cdot \sigma (s) (s-4m^2)s T K_2(\frac{\sqrt{s}}{T}) \,,
\label{deltaE4}
\end{eqnarray}
which agrees with the result of other works, such as Ref.~\cite{Ciarcelluti:2008qk}. In terms of $\rho'/\rho$ and the variable $T$ this gives Eq.~(\ref{deltaE3}). 

\section{Interplay of decay and scattering processes for the kinetic mixing portal}

The interplay between decay and scatterings processes in the kinetic mixing portal is similar to that in the Higgs portal except for the important difference that, for the former, not all scatterings are mediated by the decaying particle (the $Z$ in this case). Indeed, for kinetic mixing in addition to the $Z\rightarrow e'\bar{e}'$ decay  and Z mediated $f \bar{f}\rightarrow e'\bar{e}'$ processes, there is also a contribution from $\gamma$, which is not suppressed at $T<<m_Z$. As a result, while for the Higgs portal the Higgs decay always dominates the DM production for $m_{DM}<m_h/2$, for the kinetic mixing this is true only within the range $\sim1\,\hbox{GeV}<m_{DM}<m_Z/2$.
This is due to the fact that, as we have seen before, DM production from the $\gamma$ contribution to $f \bar{f}\rightarrow e'\bar{e}'$  is infrared dominated, {\em i.e.}~it is enhanced at low temperatures (and maximum at $T\sim\hbox{Max}[m_{DM},m_f]$), while the production from the decay occurs at a temperature that is a few times smaller than $m_Z$ (before it becomes Boltzmann suppressed). 
It is useful to give a few more details about the interplay between decay and scattering processes that takes place in this intermediate mass range, and this for each regime.

\underline{Freeze-in}: In this regime the decay channel ({\em i.e.} the
resonant part of the scattering)  dominates both the energy transfer and the
DM production. The discussion is exactly the same as for the Higgs portal and
we will not repeat it. In particular, just in the same way as in Eq.~(\ref{YFIHP}), one has
\begin{equation}
Y=c \,\frac{n_Z^{eq} \Gamma(Z\rightarrow DM DM)}{s H}\Big|_{T=m_Z} \,.
\label{YFIGKM}
\end{equation}
with $c=3\pi/(2 K_2(1))\simeq 2.9$.
\begin{figure}[!t]
\centering
\includegraphics[height=7.5cm]{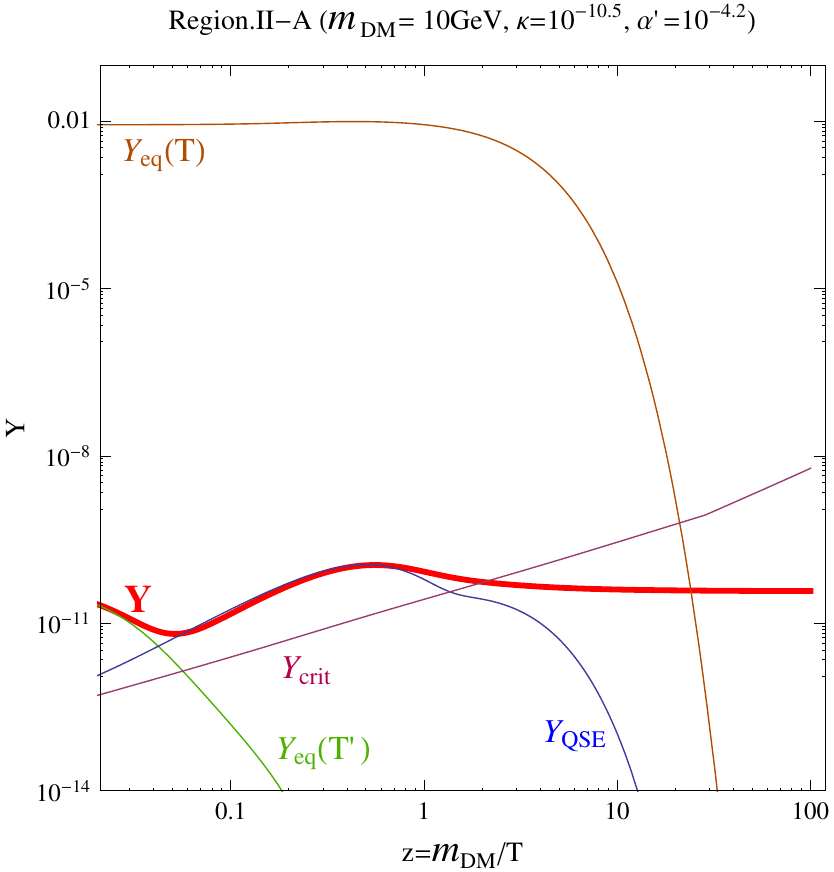}\includegraphics[height=7.5cm]{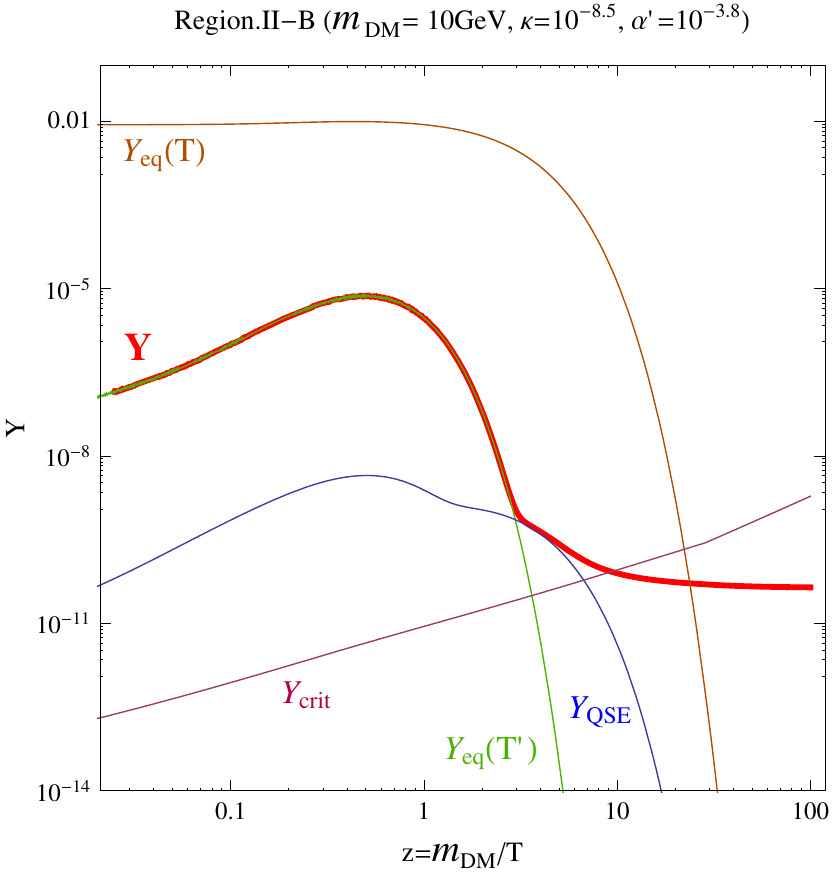}
\caption{Examples of evolutions of the DM number density as a function of $z$, in the reannihilation phase with $m_{DM}=10$~GeV. The first (second) panel corresponds to a case where $Y_{crit}$ intercepts $Y_{QSE}$ on the $Z$ decay bump ($\gamma$ mediated scattering bump). Both cases give the observed relic density. The first panel is obtained with $\kappa=10^{-10.5}$ and $\alpha'=10^{-4.2}$ whereas the second is obtained with $\kappa=10^{-8.5}$ and $\alpha'=10^{-3.8}$.}
\label{GKM-Ranni-10GeV}
\end{figure}

\underline{Reannihilation}: In this regime the situation is slightly more subtle than for the Higgs portal. In the case of the Higgs portal the QSE line is dominated by the decay contribution for $m_{DM}$ below $m_h/2$. In particular, just like the contribution from decay, the scattering process is cut off 
at low energies by the Higgs boson mass scale. Similarly, for kinetic mixing the $Z$ decay is cut off by the $Z$ mass scale but the $\gamma$ mediated $f \bar f\rightarrow e'\bar{e}'$ process is cut off only at $T\sim \hbox{Max}[m_{DM},m_f]$.
As a result $Y_{QSE}$ may display two bumps, one from the decay at $T\sim m_Z$  and one from the scattering at $T\sim m_{DM}$ (or at $T\sim m_e$ if $m_{DM}< m_e$). This is shown for two examples in Fig.~\ref{GKM-Ranni-10GeV}, both for $m_{DM}=10$~GeV. Therefore, to determine how reannihilation may occur, the key issue is to know when $Y_{QSE}$ crosses $Y_{crit}$ (at which point
 the relic density freezes). If this happens on the first bump, then the decay rate determines the relic density, as in the first panel of Fig.~\ref{GKM-Ranni-10GeV}, and the decay reannihilation equations of Section 6 apply. Conversely, if it occurs  on the second bump, as in the second panel of Fig.~\ref{GKM-Ranni-10GeV}, then the $\gamma$ mediated process determines the relic density and the scattering reannihilation equations of Section 3 apply. Actually, since $Y_{QSE}$ scales as $1/\alpha'$ while $Y_{crit}$ scales as $1/\alpha'^2$, large values of $\alpha'$ imply that the dominant contribution is from  scattering, whereas for small values of $\alpha'$ the dominant contribution is from Z decay. In Fig.~\ref{tableGKM-10GeVbis} we show, in addition to the possibilities described in the $m_{DM}=10$~GeV panel of Fig.~\ref{figure_tableGKM}, a line which delimitates these two reannihilation regimes, with IIA (IIB) reannihilation through Z decay processes (respectively $\gamma$ mediated processes).
\begin{figure}[!htb]
\centering
\includegraphics[height=7.5cm]{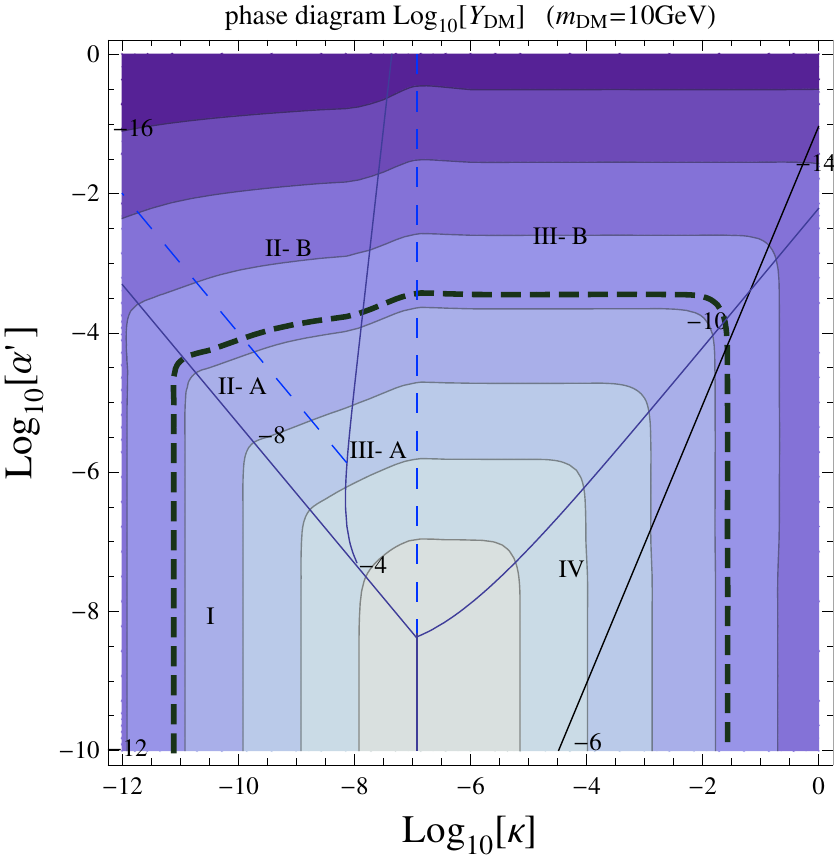}
\caption{Phase diagram for the kinetic mixing portal and $m_{DM}=10$~GeV separating explicitly the reannihilation regimes dominated by the decay (IIA) and by the $\gamma$ mediated scattering (IIB).}
\label{tableGKM-10GeVbis}
\end{figure}

\underline{Freeze-out regimes}:  For  kinetic mixing  the situation in the intermediate mass range is exactly similar to that of the Higgs portal, there is IIIA, IIIB and IV regimes. All cases the production is dominated by decay and, once both sectors have thermalized, there is standard freeze-out of the annihilation processes. If the case of hidden sector freeze-out, in which the connector does not thermalize, the transfer of energy is dominated by the decay channel.

\end{document}